\documentclass[preprint,11pt,a4paper]{elsarticle}
\usepackage{array}
\usepackage{times}
\usepackage{helvet}
\usepackage{courier}
\usepackage{amsmath}
\usepackage{amssymb}
\usepackage{subcaption}
\usepackage{color,soul}
\usepackage{mathtools}
\usepackage{multirow}
\usepackage[linesnumbered,ruled,vlined]{algorithm2e}
\usepackage{ctable}
\usepackage{flushend}
\usepackage{comment}

\usepackage{pifont}
\usepackage{wrapfig, blindtext}
\usepackage{url}

\usepackage{natbib}
 \bibpunct[, ]{(}{)}{,}{a}{}{,}%

\usepackage{cleveref}

\crefformat{subsection}{\S#2#1#3}
\crefformat{section}{\S#2#1#3}
\crefformat{subsubsection}{\S#2#1#3}
\crefname{section}{\S}{\S\S}

\SetKwComment{Comment}{$\triangleright$\ }{}

\usepackage[margin=1.0in]{geometry}

\usepackage{xcolor}
\newcommand*\circled[1]{\kern-2.5em%
  \put(0,4){\color{black}\circle*{13}}\put(0,4){\circle{12}}%
  \put(-3,0){\color{white}\bfseries\normalsize#1}~~}
\usepackage{enumitem}

\makeatletter
\def\ps@pprintTitle{%
 \let\@oddhead\@empty
 \let\@evenhead\@empty
 \def\@oddfoot{\centerline{\thepage}}%
 \let\@evenfoot\@oddfoot}
\makeatother

\newtheorem{exmp}{Example}[section]
\newcolumntype{L}[1]{>{\raggedright\let\newline\\\arraybackslash\hspace{0pt}}m{#1}}
\newcolumntype{C}[1]{>{\centering\let\newline\\\arraybackslash\hspace{0pt}}m{#1}}
\newtheorem{proposition}{Proposition}

\newtheorem{remark}{Remark}

\newcommand{\squishlist}{
 \begin{list}{$\bullet$}
  { \setlength{\itemsep}{0pt}
     \setlength{\parsep}{3pt}
     \setlength{\topsep}{3pt}
     \setlength{\partopsep}{0pt}
     \setlength{\leftmargin}{1.5em}
     \setlength{\labelwidth}{1em}
     \setlength{\labelsep}{0.5em} } }

\newcommand{\squishend}{
  \end{list}  }

\usepackage{color}

\newcommand{\bs}{\boldsymbol}
\begin{document}

\begin{frontmatter}

\title{An Iterative Security Game for \\Computing Robust and Adaptive Network Flows}

%----- First representation with emails on top -----------%
\author{Supriyo Ghosh$^\dag$\fnref{label1}}
\author {Patrick Jaillet$^\ddag$}

\address{$^\dag$IBM Research AI, Singapore 018983}
\address{$^\ddag$Department of Electrical Engineering and Computer Science, Massachusetts Institute of Technology, Cambridge, MA, 02139}
\author{supriyog@ibm.com\corref{cor1}}
\author{jaillet@mit.edu}
\fntext[label1]{Most of this work was done while the first author was with Singapore MIT Alliance for Research and Technology (SMART) Centre for Future Mobility (FM).}
\cortext[cor1]{Corresponding author}

%----- Second representation with emails in footnote -----------%
%\author[label1]{Supriyo Ghosh\corref{cor1}\fnref{ref1}}
%\ead{supriyog@ibm.com}
%\author[label2]{Patrick Jaillet}
%\ead{jaillet@mit.edu}
%\address[label1]{IBM Research AI, Singapore 018983}
%\address[label2]{Department of Electrical Engineering and Computer Science, Massachusetts Institute of Technology, Cambridge, MA, 02139}
%\fntext[ref1]{Most of this work was done while the first author was with Singapore MIT Alliance for Research and Technology (SMART) Centre for Future Mobility (FM).}

\begin{abstract}
The recent advancement in real-world critical infrastructure networks (e.g., transportation, energy distribution, water management, oil pipeline, etc.) has led to an exponential growth in the use of automated devices which in turn has created new security challenges. 
In this paper, we study the robust and adaptive maximum flow problem in an uncertain environment where the network parameters (e.g., capacities) are known and deterministic, but the network structure (e.g., edges) is vulnerable to adversarial attacks or failures. We propose a robust and sustainable network flow model to effectively and proactively counter plausible attacking behaviors of an adversary operating under a budget constraint. Specifically, we introduce a novel scenario generation approach based on an iterative two-player game between a defender and an adversary. We assume that the adversary always takes a best myopic response (out of some feasible attacks) against the current flow scenario prepared by the defender. On the other hand, we assume that the defender considers all the attacking behaviors revealed by the adversary in previous iterations in order to generate a new conservative flow strategy that is robust (maximin) against all those attacks. This iterative game continues until the objectives of the adversary and the administrator both converge. We show that the robust network flow problem to be solved by the defender is NP-hard and that the complexity of the adversary's decision problem grows exponentially with the network size and the adversary's budget value. We propose two principled heuristic approaches for solving the adversary's problem at the scale of a large urban network. Extensive computational results on multiple synthetic and real-world data sets demonstrate that the solution provided by the defender's problem significantly increases the amount of flow pushed through the network and reduces the expected lost flow over four state-of-the-art benchmark approaches. 
%The recent advancement in cyberphysical systems (e.g., transportation, water management, energy distribution system, etc.) has led to an exponential growth in the use of automated devices which in turn has created new security challenges. By manipulating cyberphysical components, a potential cyber/physical attacker can modify the capacities of multiple edges so as to disrupt the network of interest and reduce the amount of flow going through the network.
%, so as to proactively counter such adversarial failures
%Existing robust network flow models typically assume that the entire flow of an attacked edge gets lost. However, in many practical systems, the flow of an attacked edge could potentially be rerouted through adjacent edges with residual capacity. In order to address this feature, w
\end{abstract}

\end{frontmatter}

\newpage

\section{Introduction}
Network flow problems have been widely investigated from various points of views by many researchers and remain a central theme in operations research and computer science. \cite{ahuja1993network} provide a comprehensive study of theory, algorithms, and applications of network flow problems. Network flow problems have numerous applications in critical infrastructure network design and operations including urban transportation, water management, oil pipeline, energy distribution, and telecommunication systems.
Modern critical infrastructure networks have extensively installed a broad range of automated devices (e.g., sensors are installed at the road intersections for real-time traffic monitoring and intelligent traffic light management) to improve real-time operations. While the proliferation of automated devices in critical infrastructure networks provides many benefits (e.g., real-time monitoring, better sensing, data-centric planning, etc.) to the authorities, they are exposed to new security challenges, e.g., traffic light manipulation, tampering of sensors, destruction of transformers, etc. Several illustrations of cyber or physical attacks such as tampering of traffic monitoring sensors \citep{zetter2012hackers,reilly2015cybersecurity,cerrudo2014hacking}, manipulation of signal controllers \citep{ghena2014green,suzanne2014light} have been reported in recent past. By manipulating sensory components of an edge, a malicious attacker can either break the edge completely or modify its capacity. 

%In the context of transportation, an attacker can either set the traffic light to red or reduce the duration of green light (which is equivalent to capacity drop) by manipulating the sensors embedded in streets that feed data to traffic control systems.

There can be two principle ways to counter these adversarial disruptions and to develop resilient critical infrastructure networks: (a) Reactive approach -- the network administrator immediately dispatches available resources to recover the compromised edges for fast network restoration; and (b) Proactive approach -- the network administrator strategically circulates the flow to maximize the amount of flow that can be pushed to the destination node under worst-case adversarial attack. The adversarial attacks can also be countered using a combination of both the reactive and proactive approaches. This paper focuses on proactive and robust operation for resilient control of a critical infrastructure network where the network parameters (e.g., capacities) are known, but the network structure (e.g., edges) is vulnerable to adversarial attacks or failures. We next present a few real-world applications to motivate the study of robust and adaptive network flow problems with edge failures.

Let us consider a problem of a crude oil distribution network that links the production units to consumption centers via a number of intermediate pump stations. The traditional manual pipeline management methods do not consider that pipelines or pump stations may collapse. For example, if a pipeline segment is bombed or attacked during a war, it severely affects the entire oil distribution system as well as other industries. Therefore, it is crucial to manage crude oil distribution network in a way that is competent in dealing with such adversarial situation to reduce the shortage of crude oil at consumption centers possibly through rerouting but not rebuilding the network~\citep{bertsimas2013robust}. The challenge is to design a network flow strategy which utilizes the edge capacities efficiently under normal condition, but also preserve residual capacity to maximize the flow through network by rerouting flows under worst-case adversarial attacks. Another real-world motivating example is a supply-chain network that links source to sinks via several hubs that are connected by railways or roads. One can imagine a similar adverse scenario where a few segments may collapse due to natural disaster or attacks. Other motivating domains that anticipate robustness in flow solution include modern cyberphysical systems where the automated devices installed in nodes and edges are vulnerable to failures. %Note that the requirement of robustness arises in many other practical complex operation research models that use network flow as a sub-routine. 

In this paper, we consider the network flow problem in the framework of robust and adaptive optimization as the network structure is itself uncertain due to the possibility of (adversarial) edge failures. Robust optimization \citep{ben1998robust,bertsimas2003robust} is a broad research area that deals with problems where input data can vary within an uncertainty set and a robust solution should remain feasible for every realization of data within the uncertainty set. On the other hand, adaptive optimization models are used to address multi-stage decision-making problems under uncertainty. \cite{ben2004adjustable} provide a two-stage adaptive optimization model in which a decision is made before uncertainty is realized, followed by another set of decisions. \cite{atamturk2007two,poss2013affine} propose adaptive optimization models to solve network flow problems by considering different demand uncertainty sets. In contrast to these existing works on demand robustness, we consider network flow problems with edge failures that is motivated by the network interdiction problem.

%Our goal is to solve the robust network flow problem where the network administrator proactively plans to route a maximum amount of flow through the network by considering all the possible attacking behaviors of an adversary operating under a budget constraint. The robust network flow problem is motivated by 
%Another similar problem domain which has been solved effectively using game-theoretic models is strategic network design for critical infrastructure systems where the goal is to optimize a certain utility function by considering the possibility of edge failure \citep{laporte2010game, dziubinski2013network}.  or strategic network design problems for critical infrastructure bertsimas2016power,
%A classical network interdiction problem \citep{wood1993deterministic,cormican1998stochastic,sullivan2014exact} is represented through an attacker-defender model where either an interdictor places resources (e.g., security personnel) to inspect a subset of edges so as to detect and prevent a felonious entity carries illegal goods through the network, or an operator tries to maximize the flow through the network under worst-case attack.
In a classical network interdiction problem \citep{wood1993deterministic,cormican1998stochastic,sullivan2014exact}, a felonious entity carries illegal goods through the network and an interdictor places resources (e.g., security personnel) to inspect a subset of edges so as to detect and prevent the illegal activity. Many sequential and simultaneous game-theoretical models have been proposed to solve network interdiction problems \citep{washburn1995two,dahan2015network,guo2016optimal}. \cite{israeli2002shortest} propose a Benders decomposition approach to solve the complex two-stage max-min objective of shortest-path network interdiction problems where the master problem incrementally adds strategies (potential \emph{s-t} paths generated by slave solutions) to find an efficient interdiction strategy. In the similar spirit, \cite{guo2016optimal} recently propose a repeated game to solve the network interdiction problem and provide a column generation method to solve the repeated Stackelberg game with min-max objective. We also incrementally add potential attacks to the administrator's decision problem, but our third stage recourse objective that reroute flows through residual paths adds an extra layer of complexity. The network interdiction problems have also been extended to a three-stage problem called network fortification games
\citep{church2007protecting,scaparra2008bilevel,lozano2017backward}, in which the operator fortifies the network before the interdictor executes her action. 
However, different from the objectives in network interdiction or fortification games, our goal is to find a robust and adaptive flow strategy whereby an administrator is able to maximize the operational efficiency of the network under worst-case adversarial attacks possibly through flow rerouting.

Our work is motivated by the robust and adaptive network flow problem introduced by \cite{bertsimas2013robust}. For computing an adaptive maximum flow solution, they assume that the flow can be adjusted after edge failure occurred, but the adjusted flow is always bounded by the initial flow assigned to an edge. Furthermore, they propose a linear optimization model to approximately solve the adaptive maximum flow problem. However, we experimentally show that the proposed approximate solution performs poorly when the flow from an attacked edge is allowed to reroute through adjacent edges with residual capacity. In addition, to make the problem more realistic, we assume that the administrator faces a cost for routing flows through an edge, which increases the complexity of the optimization problem for computing an adaptive maximum flow solution.

Due to the aforementioned challenges, it is hard to formulate a tractable optimization problem for computing a robust and adaptive maximum flow solution using a two-stage robust model. Therefore, we treat the problem of computing a robust and adaptive maximum flow strategy as the result of a two-player iterative game between a network administrator and an adversary. 
We assume that the adversary is operating under a budget constraint (i.e., the number of attacked edges is bounded by a threshold value) and therefore, the adversary has a finite number of possible attacking choices. The assumption of budget constraint for the adversary is valid in many real-world applications and therefore, several existing works \citep{bertsimas2013robust,altner2010maximum,dahan2018probability} assume a budget constraint for the adversary to control the conservatism of robust solutions. As an example of practical consideration about such constraint, an adversary would need to be present physically within the geographical proximity in order to manipulate vehicle monitoring sensors which indirectly impact the traffic light timings \citep{cerrudo2014hacking}, implying a (human) resource budget constraint for the adversary. While the adversary's budget value can be learnt efficiently from previous attacking behaviors in case of repeated attacks, we also experimentally show that the initial estimation of the budget value can be done using sensitivity analysis (i.e., by observing the outcomes with varying budget value).

For the administrator, a feasible flow strategy should satisfy the flow conservation constraint at the nodes and the capacity constraint at the edges. As the strategy space of the administrator is extremely large and the attacker's strategy space grows exponentially with the budget, it is impossible to compute an equilibrium solution by considering all the pure strategies. To tackle such large strategy space, a common practice is to use an incremental strategy generation approach for choosing a small set of pure strategies which can be used to identify an equilibrium solution. \cite{jain2011double} propose a double oracle algorithm where both the players use best response against each other to incrementally generate their strategies and show that this iterative approach converges to a Stackelberg equilibrium. The final solution at the convergence is a mixed strategy over the incrementally generated pure strategies, which can readily be applied in a physical security scheduling domain as both attacker and defender execute their actions simultaneously. In contrast, we assume that the adversary is more powerful and has a perfect knowledge of the pure flow strategy chosen by the administrator. Therefore, our goal is to identify a robust pure strategy for the administrator that maximizes the worst-case adaptive flow value. %(e.g., security patrolling in airport or physical road network) 

In order to solve the iterative two-player game and to identify a robust pure strategy for the administrator, we propose a novel incremental strategy generation approach. The objective of the administrator is to maximize the ultimate flow that can be pushed to the terminal node while minimizing the total routing cost. On the other hand, the objective of the adversary is to identify an attack that minimizes the objective value of the administrator.
In each iteration of the game, the adversary who is restricted to a budget constraint, acts as a follower and optimally disrupts the current network flow strategy prepared by the network administrator. In turn, the administrator acts as a leader and generates a new network flow strategy which is robust (maximin) against all the attacking behaviors revealed in previous iterations. This iterative game continues until the objective values of the players converge to same value and they start repeating their previous actions. At that point, as the adversary repeats her previous attacks, the administrator has already considered these for generating the final flow strategy and therefore, it is a robust (maximin) response to the adversary's best response. In that sense, the iterative game has reached convergence.
Note that, unlike the solution approaches for traditional Stackelberg games (e.g., double oracle algorithm) that converge to a mixed strategy equilibrium, our solution converges to a robust pure flow strategy for the administrator. 

As the administrator model introduces additional constraints related to current attack on top of decision model from previous iteration, the objective value of the administrator reduces monotonically over the iterations. Due to more conservative flow, the adversary's ability to disrupt the flow strategy reduces over iterations and the objective value of the adversary increases. These two objective values are guaranteed to converge to the same value at some point. We show that the game converges to a maximin optimal flow solution for the administrator and therefore, the objective value of the administrator will at least be the value at which the game has converged for any realization of the feasible attacks.

We show that the robust and adaptive network flow problem to be solved by the administrator is NP-hard. Moreover, we empirically observe that the complexity of the adversary's decision problem grows exponentially with the network size and adversary's budget value. Therefore, we propose two principled heuristic approaches for solving the complex decision problem of the adversary at the scale of a large urban network. The first heuristic is an accelerated greedy approach where we identify one edge at a time in an incremental fashion so as to minimize the objective of the administrator for a given flow strategy, until the budget constraint of the adversary is exhausted. For the second heuristic, we partition the network into disjoint sub-networks and identify a set of edges to attack within the adversary's budget constraint by solving the corresponding sub-problems. We iteratively solve this process with random partitioning of the network and learn a set of best possible candidate edges to attack and finally, solve the adversary's decision problem to choose the best edges (within budget constraint) from these candidate edges. In each iteration of the game, we execute both heuristics and choose the one with better solution quality as the adversary's decision. 
By leveraging the computational effectiveness of the proposed heuristics, our solution approach can scale gracefully to large-scale problems while providing a consistent performance gain over four following benchmark approaches: (i) The administrator sends maximum flow through the network without considering any attacks; (ii) The administrator uses a myopic one-step reasoning against the adversary's behavior to choose a flow strategy; (iii) A robust maximum flow solution from \cite{bertsimas2013robust} where the administrator proactively computes a flow solution to improve the worst-case performance; and (iv) An approximate adaptive maximum flow solution employed from \cite{bertsimas2013robust}.

\emph{Our contributions.} The key contributions of this paper are as follows:
\begin{enumerate} [label=\protect\circled{\arabic*}]
\item We formally define the problem of computing a robust and adaptive maximum flow strategy for critical infrastructure networks by exploiting the fact that the flow of a compromised edge might be rerouted through adjacent edges with residual capacity. To solve the problem, we propose an iterative two-player game between a network administrator and an adversary which is referred as Traffic Network Flow Game (TNFG).
\item We develop novel optimization models to solve the decision problem of both players in each iteration of the game. The administrator's optimization model takes into account all the attacking strategies generated by the adversary in previous iterations and computes a robust flow strategy that maximizes the amount of flow pushed through the network in the worst-case over all the previous attacks.
The adversary's decision problem inspects the flow strategy generated by the administrator in the current iteration and generates an attacking strategy (out of the feasible attacks under a given budget constraint) to optimally disrupt the flow strategy.
\item We propose two novel heuristic approaches for solving the complex decision problem of the adversary at the scale of a large urban network. The first heuristic is an accelerated greedy approach that incrementally identifies the best edges to be attacked. The second heuristic is a network partitioning based approach that iteratively identifies a set of candidate edges in the network and then we solve the adversary's decision problem over these candidate edges. 
\item We provide extensive computational results on multiple synthetic and real-world benchmark data sets to demonstrate that our proposed solution approach scales gracefully to large-scale problems and significantly increases the amount of flow pushed through the network over the benchmark approaches. 
\end{enumerate}

\emph{Structure of the paper.} In \cref{related}, we elaborate on the relevant research. In \cref{problem}, we formally describe our problem by allowing the flow of an attacked edge to be rerouted through adjacent edges with residual capacity. In \cref{smallSol}, we demonstrate our proposed iterative two-player game between a network administrator and an adversary. We present the decision problem and optimization models for the adversary and the administrator in \cref{modelAdv} and \cref{modelAdmin}, respectively. In \cref{overallSol}, we provide the key iterative steps of our overall two-player game. In \cref{heuristics}, we describe two heuristic approaches to accelerate the solution process of the adversary's decision problem. In \cref{greedyHeuristic}, we present the accelerated greedy heuristic and in \cref{partitionHeuristic}, we describe the network partitioning based heuristic. We then provide the experimental setup and performance analysis of our proposed approach in \cref{allResults}. In \cref{smallResults}, we present the empirical results to verify the utility of our proposed solution approach on small-scale problem instances.  In \cref{largeResults}, we demonstrate the experimental results on large-scale problem instances by leveraging the computational effectiveness of the proposed heuristics. Finally, we provide concluding remarks in \cref{conclusion}.

\section{Related Work}
\label{related}
Given the practical importance of ensuring robustness in design and operation of critical urban infrastructure systems and evaluating the resilience of such systems, several game-theoretic models have been proposed to model the attacker-defender interactions \citep{manshaei2013game}. We summarize these contributions along four threads of research. \cref{related:1} summarizes existing research on applying security game models in critical infrastructure systems. \cref{related:2} recaps literatures on using Stackelberg games for ensuring physical security. \cref{related:3} summarizes literature in solving a relevant example of Stackelberg games called network interdiction problems, which is the motivation behind our robust and adaptive maximum network flow problems. Finally, in \cref{related:4}, we summarize relevant literatures in robust and adaptive optimization, and their applications on network flow problems.  

\subsection{Security Games in Critical Infrastructure Systems} \label{related:1}
Our work closely resembles the broader research theme of network security games in critical infrastructure systems. Security games, that are used to model the attacker-defender interactions in a network, have been employed for strategic network design \citep{laporte2010game, dziubinski2013network} of critical infrastructure systems (e.g., railways and defense) where the goal is to optimize a certain utility function by considering the possibility of edge failure. 
Furthermore, security games are widely used to examine the vulnerability of nodes and edges of urban networks including transportation \citep{baykal2014infrastructure}, shipment of hazardous material \citep{szeto2013routing} and communication networks \citep{gueye2012towards}. \cite{gueye2012game} use a game-theoretic approach to quantify the trade-off between vulnerability and security cost for supply-demand networks. They model an operator who transmits a feasible flow through the network and an attacker, who faces a cost for attacking an edge, seeks to maximize the amount of lost flow by disrupting only one edge and show that the game converges to a mixed-strategy Nash equilibrium.  

Several recent research works \citep{dahan2018probability, he2012game,ma2011game,wu2018security,wu2018signaling} have designed simultaneous attacker-defender network flow games in which the adversary is allowed to disrupt multiple edges within a fixed budget constraint so as to identify critical and vulnerable edges in the network. The vulnerability assessment methods for complex urban networks have also been designed using network flow models that are built upon the \emph{max-flow} and \emph{min-cut} theorem \citep{assadi2014minimum, dwivedi2013maximum}.
\cite{dahan2015network} and \cite{dahan2018probability} combine network flow models within a simultaneous game framework to learn the attacker-defender interaction. The simultaneous game efficiently captures the strategic uncertainty on the adversary's behavior, and the game has a mixed strategy Nash equilibrium if both the adversary and administrator are restricted to a fixed budget constraint. However, they did not consider the option to reroute flows through adjacent edges with residual capacity. 

We employ network flow models to represent the decision problems of both the administrator and the adversary, but our approach differs from the simultaneous game methods mentioned in this section as we propose an iterative game to identify a pure robust flow strategy by considering the fact that the flow of an attacked edge can be rerouted through other paths with residual capacity.

\subsection{Stackelberg Games in Physical Security} \label{related:2}
In this section, we summarize existing research on a relevant class of iterative security game that focusses on identifying optimal Stackelberg strategies \citep{tsai2010urban, kar2017trends}. These leader-follower based sequential games have been applied successfully in many real-world applications ranging from security patrolling \citep{pita2008deployed, brown2014streets} to wild-life protection \citep{fang2016deploying} to opportunistic crime \citep{zhang2016using} to cyber security \citep{sinha2015physical}. The fundamental concept behind these problems is to efficiently allocate security personnel to defend adversarial events. However, solving these large normal-form Stackelberg games are practically infeasible and therefore, sophisticated methods are needed to speed up the solution process by exploiting specific domain structure. 

To identify a randomized patrolling strategy for large-scale infrastructure protection, \cite{jain2010security} propose a combination of column generation and branch-and-bound algorithm that exploits the network flow representation of the problem and iteratively generates tighter bounds using fast and efficient algorithms. In the similar direction, \cite{guo2016optimal} develop a column and constraint generation algorithm to approximately solve the network security game. \cite{jain2011double} employ a double oracle algorithm to solve the network security game. On the contrary to enumerating the entire exponential sized strategy space for the players, the proposed double oracle algorithm generates the oracle for the attacker and the defender by adding their pure strategy best response in each iteration and finally compute an equilibrium solution for the restricted game consisting of pure strategy oracles of both the players. The final solution at the convergence is a mixed strategy that can be applied in a physical security scheduling domain as both players execute their actions simultaneously. 

However, these approaches cannot be readily employed to solve our problem as we have exponential sized strategy space and for each pair of attacker and defender strategy, we need to solve an optimization model (due to the fact that the flow can be rerouted through other paths in case of adversarial attacks) to compute the payoff value.
Furthermore, we assume that the adversary is more powerful and has a perfect knowledge of the flow strategy chosen by the administrator and therefore, we need to identify a robust pure strategy for the administrator.

\subsection{Network Interdiction and Fortification Games} \label{related:3}
Robust and adaptive network flow problems are motivated by the network interdiction problem which is an example of Stackelberg games. In classical network interdiction problems, a malicious entity carries illegal goods through the network and a security agency deploy a set of interdictors to prevent the illegal activity. Due to its practical importance in defense and drug enforcement, a wide variety of research papers have addressed both deterministic \citep{wood1993deterministic} and stochastic \citep{cormican1998stochastic} network interdiction problems. Several research papers \citep{wollmer1964removing, ratliff1975finding, ball1989finding} propose methodologies for identifying $n$ ``most vital" edges in the network which can recommend a deterministic interdiction strategy. For solving the maximum flow network interdiction problem in Euclidean space where the interdictors can be placed anywhere in the network, \cite{sullivan2014exact} provide a sequence of integer programs to iteratively identify the upper and lower bounds on solution quality and show that these bounds are convergent. 
Many sequential and simultaneous game-theoretical models have also been proposed to solve the network interdiction problem. \cite{washburn1995two} propose a two-player zero-sum game to compute a probabilistic edge selection strategy for the interdictors. \cite{bertsimas2016power} introduce an arc-based and a path-based formulation to solve the sequential network interdiction game where the interdictor iteratively chooses a pure strategy and then the other player routes a feasible flow through the network. In contrast to these two stage min-max games, we consider a three stage robust maximum flow problem in which the administrator can modify the flow in the third stage possibly through rerouting.
%On the other hand, in maximum flow network interdiction problems, an operator tries to maximize the flow through network against worst-case attacks by learning most critical and vulnerable edges.

The network interdiction problems have also been extended to three stage defender-attacker-defender (DA\underline{D}) setting which are often referred as network fortification games. In general network fortification games, the first stage (the defender fortifies the network) and second stage (the attacker executes her attack plan) decision variables are binary valued and the third stage is represented by a recourse function. \cite{church2007protecting} solve the interdiction median problem with fortification (IMF) for facility protection. They convert the three-level problem into a single-level mixed integer program (MIP) by enumerating all the possible attack plans. \cite{scaparra2008exact} reformulate the three-level problem as a single-level maximal covering problem with precedence constraints. \cite{scaparra2008bilevel} design a bilevel program for IMF and solve it with efficient implicit enumeration algorithm, which was further extended by \cite{cappanera2011optimal} for allocating resources in a shortest path network.
\cite{brown2006defending} take the dual of third stage problem to combine second and third stage problems of the DA\underline{D} model and solve the resulting problem using Benders decomposition. \cite{smith2007survivable} combine the second and third stage problems into a bilinear problem and solve the resulting problem using cutting-plane approach. In a similar vein, \cite{alderson2011solving,alderson2013sometimes,alderson2015operational} use DA\underline{D} model in the context of critical infrastructure resilience in which the components of the network can either be fortified or rebuild after realization of the attack and solve these three stage problems using decomposition approaches after combining the second and third stage problems using duality mechanism. \cite{lozano2017backward} recently study the fortification games in which the first and second stage decision variables are binary valued, but the third stage recourse function can take any general form. They propose to iteratively refine the samples from third stage solution space so as to solve the bilinear interdiction problems which are then added as cuts in the first stage defender model. For our problem setting, the first stage decision is to identify a feasible flow strategy represented by continuous variables and the second stage decision variables for the attacker are only binary valued. We combine the second and third stage problems using duality mechanism and iteratively generate samples from second stage solution space (i.e., attack plans) to improve the first stage defender solution.

%\cite{prince2013three} use DA\underline{D} model for procurement optimization problem under uncertainty with non-convex third stage recourse function. They convert the non-convex recourse problem into a large scale shortest path MIP problem. 
%\cite{brown2006defending} study  for protecting critical component in electric power grid. 

In network interdiction problem, the goal is to determine the worst-case scenario, assuming the decision maker is in a position to act after the realization of uncertainty. The operator can either fortify the network components partially against the worst-case scenario or rebuild some components after the failure is realized. Our work differs from this thread of research as we are interested in proactively identifying those flow decisions that are robust against any plausible attacks.

\subsection{Robust and Adaptive Optimization} \label{related:4}
The last thread of research which is complementary to our work presented in this paper is on robust and adaptive optimization. Robust optimization is a broad research area and the solution methodology varies for different problem settings \citep{bertsimas2011theory, bertsimas2018data}. In a robust optimization framework, the input data of a problem can vary within an uncertainty set and a robust solution should remain feasible for every realization of the data within the uncertainty set. 
Several studies have considered network flow problems within the robust optimization framework \citep{bertsimas2003robust,el1998robust,ben1998robust}.
The design of a robust optimization solution varies with different objective functions such as worst-case \citep{vorobyov2003robust,ghosh2016robust}, regret \citep{ahmed2013regret} and risk sensitive \citep{adulyasak2015models} objectives. 
The design of a robust optimization solution and the level of conservatism also varies with different types of uncertainty sets such as ellipsoidal \citep{el1998robust,ben2000robust,bertsimas2004robust}, polyhedral \citep{bertsimas2003robust,bertsimas2004price} and interval \citep{li2008multiobjective,ghosh2019improving} uncertainty sets. Our proposed iterative game and incremental strategy generation approach is designed for robust network flow problem with worst-case objective and interval uncertainty set.
%Our proposed iterative game and incremental strategy generation approach can be used for robust optimization with worst-case objective and interval uncertainty set in the context of network flow problems. 

Adaptive optimization models are recently introduced to address multi-stage decision-making problems under uncertainty. \cite{ben2004adjustable} provide a two-stage adaptive optimization model where a decision is made before uncertainty is realized which is followed by another set of decisions. They show that the adaptive counterpart of this optimization problem is NP-hard in general and therefore, several approximation methods have been proposed subsequently to tackle this problem~\citep{ben2004adjustable,bertsimas2011theory,bertsimas2010power}. Adaptive two-stage robust optimization models have been proposed to solve network flow problems by considering different demand uncertainty sets. \cite{atamturk2007two} present a model of the first-stage robust decisions for network flow problems with demand uncertainty using an exponential number of constraints and show that the corresponding separation problem is NP-Hard.
\cite{mattia2013robust} consider a robust network loading problem with dynamic routing in which the demand is represented using polyhedral uncertainty set, and solve it using branch-and-cut algorithm.
\cite{poss2013affine} solve adaptive network design problem with polyhedral demand uncertainty by introducing the concept of affine routing.
On the contrary to these existing works on demand robustness, we consider network flow problems with edge failures.
%\cite{ordonez2007robust}

Our work is motivated by the robust and adaptive network flow problem introduced by \cite{bertsimas2013robust}. They assume that the flow at an edge can be adjusted after edge failure occurred, but the adjusted flow should be bounded by the initially assigned flow value. They further propose a linear optimization model to solve the complex three stage adaptive network flow problem quickly and approximately. However, we experimentally show that the proposed approximate solution performs poorly when the flow from an attacked edge is allowed to reroute through adjacent edges with residual capacity. Moreover, we also assume that the administrator faces a cost for routing flows through an edge, which increases the complexity of the optimization problem for computing a robust adaptive maximum flow solution.

\section{Problem Formulation}
\label{problem}
We begin with a formal definition of the Traffic Network Flow Game (\textbf{TNFG}). Let $G=\langle {\mathcal V}, {\mathcal E}\rangle $ denote an urban network, where ${\mathcal V}$ symbolizes the set of nodes and $s,t\in {\mathcal V}$ represent the source and terminal node, respectively. ${\mathcal E}$ denotes the set of edges, where $e_{ij} \in {\mathcal E}$ represents a directed edge from node $i$ to node $j$. We also denote an edge simply as $e\in {\mathcal E}$ and assume that it has a finite capacity $U_e$ (corresponding to the maximum amount of traffic that can be sent through the edge). We introduce an artificial edge $e_{ts}$ with infinite capacity between the terminal node and the source node. In addition, we introduce the following notations:
\squishlist
\item $\delta^+_v$: The set of incoming edges to node $v$. %For the source node $s$, $\delta^+_s$ includes the artificial edge $e_{ts}$.
\item $\delta^-_v$: The set of outgoing edges from node $v$. %For the terminal node $t$, $\delta^-_t$ includes the artificial edge $e_{ts}$.
\item $\Lambda_v$: The set of all possible simple paths (i.e., without any internal cycles) from node $v$ to terminal node $t$.
\item $\Lambda_{\sigma} \!=\! \bigcup_{v\in \sigma} \Lambda_v$: The set of all possible simple paths from one of the node of set $\sigma$ to terminal node $t$.
\item $\Gamma$: Budget for the adversary. That is to say, a maximum of $\Gamma$ edges can be attacked.
\squishend

A flow scenario $x$ is a function $x: {\mathcal E} \rightarrow \mathbb{R}_{\geq 0}$ that assigns a non-negative flow value $x_e$ to each edge $e\in {\mathcal E}$ such that the following capacity and flow conservation constraints~\eqref{eq:flowconv} are ensured: 
\begin{equation}
	\label{eq:flowconv}
	\begin{aligned}
		\sum_{e\in \delta^+_v} x_e - \sum_{e\in \delta^-_v} x_{e} = 0  \hspace{0.85in} \forall v \in {\mathcal V}\setminus \{s\} \\
		0 \leq x_e \leq U_e \hspace{1.7in} \forall e \in {\mathcal E} 
	\end{aligned}
\end{equation}
Let ${\mathcal X}$ denote the set of all possible flow scenarios that satisfy the flow conservation constraints~\eqref{eq:flowconv}. An attack $\mu$ is a function $\mu: {\cal E} \rightarrow \{0,1\}$, where $\mu_e$ is set to 1 if the edge $e\in {\mathcal E}$ is attacked by the adversary and 0 otherwise. The set of possible attacks that satisfy the budget constraint of the adversary is denoted by $\Psi$.
\begin{equation}
	\label{eq:advscen}
	\begin{aligned}
		\Psi := \left.\Big\{  \mu= (\mu_e)_{e\in {\mathcal E}} \in \{0,1\}^{|{\mathcal E}|}  \right\vert \sum_{e \in {\mathcal E}} \mu_e \leq \Gamma\Big \} 
	\end{aligned}
\end{equation}

We now describe the traffic network flow game (TNFG) in the context of critical infrastructure networks where the flow of an attacked edge might be rerouted through adjacent edges with residual capacity. In the crude oil pipeline application, the initial pipeline is built to optimally transmit a planned flow. Once this structure is modified due to adversarial failures, it is feasible to reroute the flows to optimally utilize the remaining pipes. After a failure occurred, flow through an edge can be adjusted to maximize the system efficiency, but the adjusted flow cannot exceed the initial flow, for which the pipeline is initially designed. We consider a more generic setting, where the initial flow might permit residual capacity in some of the edges, and therefore, the adjusted flow in an edge can utilize the residual capacity only for rerouting flows from an attacked edge. This assumption is motivated by the simple properties of liquid flow in which due to the pressure in pipes, additional flows from attacked edges should naturally be rerouted through adjacent edges with residual capacity.

We first provide the additional setup needed to describe the problem. If the adversary attacks an edge $e\in {\mathcal E}$ (i.e., $\mu_e=1$), then the capacity of the edge is modified from $U_e$ to $m_e$. We set the value of $m_e$ to 0, if the edge is completely blocked. We assume that the administrator faces routing cost for transporting the flow through edges. Let $p_e$ denote the cost for routing one unit of flow through edge $e\in {\mathcal E}$ and the reward for successfully getting one unit of flow to the destination node is assumed to be $1$. 
Moreover, we assume that the administrator faces an additional $p_e$ cost for rerouting one unit of flow through edge $e$. For example, in the oil distribution pipeline application, the initial flow allocation cost (e.g., sum of required capacities of installed pipelines) remains unchanged after disruption, and the flow rerouting (e.g., utilizing residual capacity) incurs additional operational cost.
If the sum of initial routing cost and rerouting cost (due to attacks) for pushing one unit of flow to the destination node exceeds $1$, then an optimal strategy is to send zero flow through the network. To avoid such trivial situation, we assume that the value of $p_e$ is always upper bounded by $\frac{1}{2L}$, where $L$ represents the maximum number of edges in a source to destination path. The cost for routing one unit of flow through the artificial edge $p_{e_{ts}}$ is set to 0.

In this setting, given an initial flow $x$ and a resulting attack $\mu$, we can compute the maximum adaptive value of $x$ for the administrator after rerouting flows, $M(x,\mu)$, by solving the following linear optimization (LO) model \eqref{eq:flow1}-\eqref{eq:flow5}, where $y_e$ denotes the resulting amount of flow going through edge $e\in {\mathcal E}$ and $z_e$ represents the amount of additional flow that is being rerouted through edge $e$ due to the disruption from the attack $\mu$. 
The objective function \eqref{eq:flow1} of the LO model computes the trade-off between maximizing the ultimate flow that can be pushed to the terminal node (which is equivalent to the flow $y_{(t,s)}$ of the artificial edge $e_{ts}$) and minimizing the total rerouting cost of the additional flow from the attacked edges\footnote{As the input flow $x$ is fixed, the initial flow routing cost (i.e., $\sum_{e  \in {\mathcal E}} p_e x_e$) is constant in the objective function \eqref{eq:flow1}.}. %It should be noted that, herein and throughout the rest of the paper \emph{we fix the value of $W$ to 1 (i.e., $p_e \leq \frac{1}{2L}$)}.
{ \small
\begin{align}
	M(x,\mu) = & \max \hspace{2pt} \big\{  y_{(t,s)} - \sum_{e  \in {\mathcal E}} p_e z_e \big \} - \sum_{e \in {\mathcal E}} p_e x_e& \label{eq:flow1}\\
	&\hspace{10pt}\text{s.t. }  \sum_{e\in \delta^+_v} y_e - \sum_{e\in \delta^-_v} y_{e} \geq 0  & \forall v \in {\cal V}\setminus \{s\} \label{eq:flow2} \\
	& \hspace{0.33in} y_e \leq x_e & \forall e \notin \Lambda_{\sigma_{\mu}} \label{eq:flow3}\\
	& \hspace{0.33in} y_e \leq (1-\mu_e) U_e + \mu_e m_e & \forall e \in \Lambda_{\sigma_{\mu}} \label{eq:flow4} \\
	& \hspace{0.33in} z_e \geq y_e - x_e & \forall e \in \Lambda_{\sigma_{\mu}} \label{eq:flow6} \\
	& \hspace{0.33in} y_e \geq 0; z_e \geq 0 & \forall e \in {\cal E} \label{eq:flow5}
\end{align}}
As a portion of the flow might be lost due to the attack, we employ a weaker notion of flow conservation at the nodes using constraints~\eqref{eq:flow2} to avoid infeasibility issues. Let $\sigma_{\mu}$ denote the set of source nodes of the attacked edges for attack $\mu$. As indicated earlier, $\Lambda_{\sigma_{\mu}}$ represents the set of all possible paths from any of the nodes of set $\sigma_{\mu}$ to the terminal node. If an edge $e$ does not belong to any of these paths, then constraints~\eqref{eq:flow3} ensure that the value of $y_e$ is bounded by the initial flow value $x_e$. For all the other edges which belong to one of the paths in $\Lambda_{\sigma_{\mu}}$, there can be two possibilities: (a) If the edge $e$ is attacked, then we can send maximum $m_e$ amount of flow through the edge; and (b) If the edge is not attacked, then we can send maximum $U_e$ amount of flow through edge $e$. We combine these two possibilities and represent them using constraints~\eqref{eq:flow4}. Finally, constraints~\eqref{eq:flow6}-\eqref{eq:flow5} insure the amount of rerouted flow for edge $e$, $z_e=\max(0,y_e-x_e)$.

\begin{figure}[!htb]
	\centering
	\includegraphics[width=0.8\linewidth]{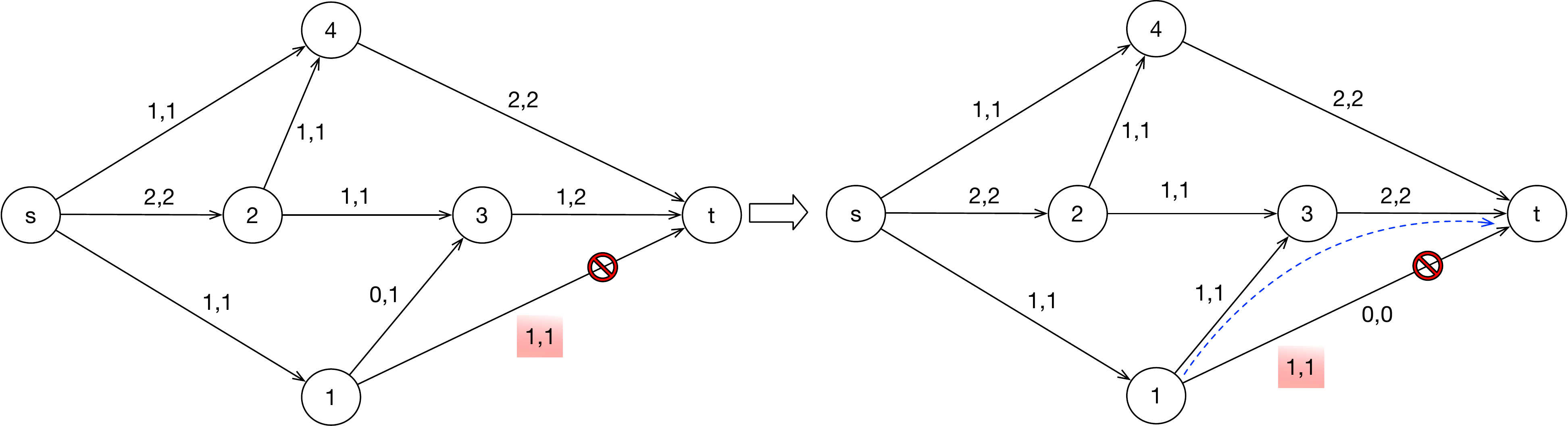}
	 \caption{ Illustration of resulting flow after an attack in a network.}
	 \label{fig:map1}
	 %\vspace{-0.1in}
\end{figure}

\begin{exmp}\label{exp1}
We provide an illustrative example in Figure~\ref{fig:map1} to better explain the constraints in the LO model. We consider a network with six nodes and nine edges. We show an initial flow solution $x$ in the first network, where each pair of numbers represents the flow and the capacity for that particular edge. Assume that the adversary can attack a maximum of one edge in the network. Assume the adversary attacks the edge $e_{1t}$ and the resulting flow in that edge is set 0. As the flow of edge $e_{1t}$ can only be rerouted through edge $e_{13}$ and $e_{3t}$, the resulting flows for all the other edges remain the same. The resulting flow after the attack is shown in the second network where the blue dotted line shows the augmented path through which the flow of the attacked edge is being rerouted.
\end{exmp}

Given an initial flow strategy $x$ chosen by the administrator, the adversary observes the flow and executes an attack that optimally disrupts the flow $x$. Therefore, for a given flow $x$, an optimal attack $\mu(x)$ and the objective of the adversary (also referred as the adaptive value of $x$) $AV(x)$ are defined as follows: 
\begin{equation}
	\label{eq:probadv}
	\begin{aligned}
		\mu(x) \in \arg\min_{\mu \in \Psi} M(x,\mu) \\
		AV(x) = \min_{\mu \in \Psi} M(x,\mu)
	\end{aligned}
\end{equation}

Finally, the goal of the network administrator is to identify a robust flow strategy which has the maximum adaptive value. Therefore, the objective of the administrator is defined as follows:
\begin{equation}
	\label{eq:probadm}
	\begin{aligned}
		%\max_{x\in {\cal X}} \min_{\mu \in \Psi} M(x,\mu) 
		\max_{x\in {\cal X}} AV(x) = \max_{x\in {\cal X}} M(x,\mu(x)) 
	\end{aligned}
\end{equation}

\begin{proposition}
The problem of computing a robust and adaptive maximum flow strategy to be solved by the administrator from expression~\eqref{eq:probadm} is strongly NP-hard.
\end{proposition}
\textbf{Proof.} The adaptive maximum flow problem introduced by \cite{bertsimas2013robust} is a related problem where the routing cost is assumed to be 0 and one can adjust the flow solution after every realization of edge failure by assuming that the adjusted flow is bounded by the initial flow assigned to an edge. \cite{bertsimas2013robust} show that the adaptive maximum flow problem is strongly NP-hard by reducing it to a classical network interdiction problem \citep{wood1993deterministic}. Our robust and adaptive maximum flow problem can be reduced to the adaptive maximum flow problem if we set the routing cost to 0 and restrict the adjusted flow for an edge to be upper bounded by the initial flow assigned to it. Due to this trivial reduction to the adaptive maximum flow problem which is strongly NP-Hard, our problem is also strongly NP-Hard.
$\blacksquare$

\section{Solution Approach} \label{smallSol}
In this section, we describe an iterative sequence of best-response plays between the administrator and an adversary to solve the problem defined in \cref{problem}. In the first iteration of the game, the administrator assumes that no attack is planned in the system and computes a flow solution that maximizes the amount of incoming flow to the destination node. Next, the adversary finds an optimal attack to disrupt the flow solution computed by the administrator. In the subsequent iterations, the administrator considers all the attacks revealed by the adversary in previous iterations and finds a robust (maximin) strategy against those attacks. The adversary always computes a best myopic disruption against the flow solution revealed by the administrator in the current iteration. This iterative process continues until the objectives of both the players converge to the same value. We now describe the details of the optimization problems to be solved by the adversary and by the administrator during each iteration of the game.
%At that point, as the adversary repeats her previous attack, the final flow solution would be a robust (maximin) response to the adversary's best possible attack. 
% So, the resulting objective value of this iterative process will provide a lower bound on the objective value for any possible attack from $\Psi$.

\subsection{Optimization Problem for the Adversary} \label{modelAdv}
Once the administrator reveals a flow scenario $\bar{x}$, the adversary solves the problem \eqref{eq:probadv} to generate an attack that optimally disrupts the flow $\bar{x}$ by considering the fact that the administrator can reroute flow from an attacked edge through other paths with residual capacity. 
We now provide an alternative formulation of the problem \eqref{eq:probadv} so as to mathematically represent the notion of rerouting paths. The set of expressions~\eqref{eq:cadv1}-\eqref{eq:cadv8} illustrates the alternative decision problem for the adversary. The objective function \eqref{eq:cadv1} corresponds to finding an attack $\mu$ that minimizes the resulting objective of the administrator, represented by the inner maximization problem. 
Constraints~\eqref{eq:cadv2}-\eqref{eq:cadv3} ensure flow conservation and capacity constraints on the flow variables $y$.
Let $\rho_e$ denote the set of edges, which if attacked, can have a portion of their flows rerouted through edge $e$, i.e.,
$ \rho_e = \{(u,v)=e' \in {\mathcal E} | e \in \Lambda_u\}. $
Let $\pi_e$ denote a variable which is set to 0 if none of the edges in the set $\rho_e$ is attacked, and otherwise the value of $\pi_e$ is set to 1. Constraints~\eqref{eq:cadv5} and \eqref{eq:cadv7} set the value of $\pi$. Constraints~\eqref{eq:cadv4} enforce that the flow passing through an edge $e$ is bounded by the given flow $\bar{x}_e$ if none of the edges in $\rho_e$ is attacked (i.e., $\pi_e=0$). 
Finally, as the inner objective function minimizes the rerouting cost, constraints~\eqref{eq:cadv6} are sufficient alone to accurately compute the value of $z_e$.% as $\max (0, y_e - \bar{x}_e)$. 
\begin{align}
	\min_{\mu \in \Psi} \big\{ \max & \hspace{5pt}  y_{(t,s)} - \sum_{e\in {\cal E}} p_e\cdot z_e \big \} \hspace{0.5in} & \label{eq:cadv1}\\
	\hspace{10pt}\text{s.t. } & \sum_{e\in \delta^+_v} y_e - \sum_{e\in \delta^-_v} y_{e} \geq 0  \hspace{0.1in} & \forall v \in {\cal V}\setminus \{s\} \label{eq:cadv2} \\
	&  y_e \leq (1-\mu_e) U_e + \mu_e m_e \hspace{0.1in} & \forall e\in {\cal E} \label{eq:cadv3}\\
	& y_e \leq (1-\pi_e)\bar{x}_e +\pi_e \cdot U_e \hspace{0.1in} & \forall e\in {\cal E} \label{eq:cadv4}\\
	& \pi_e \leq \sum_{e' \in \rho_e} \mu_{e'} \hspace{0.1in} & \forall e\in {\cal E} \label{eq:cadv5} \\
	& z_e \geq y_e - \bar{x}_e  \hspace{0.1in} & \forall e \in {\cal E}\label{eq:cadv6} \\
	& \pi_e \in \{0,1\} & \forall e \in {\cal E} \label{eq:cadv7} \\
	& y_e \geq 0; z_e\geq 0 & \forall e \in {\cal E} \label{eq:cadv8} 
\end{align}
\begin{proposition}\label{obs4}
The integrality constraints~\eqref{eq:cadv7} can be relaxed to $0\leq \pi_e \leq 1$, without compromising the feasibility or optimality of the optimization problem \eqref{eq:cadv1}-\eqref{eq:cadv8}. 
\end{proposition} 
{\bf Proof:} As $\bs{\mu}$ variables are binary, constraints~\eqref{eq:cadv5} enforce that the value of $\pi_e$ is set to 0 if none of the edges in set $\rho_e$ is attacked. On the other hand, if some of the edges in set $\rho_e$ are attacked, then the excessive flow of those edges might be rerouted through edge $e$ and therefore, the actual flow through edge $e$ can take any value between $\bar{x}_e$ and $U_e$. As $\pi_e$ is used by constraints~\eqref{eq:cadv4} to only enforce an upper bound on the flow variable $y_e$, even with continuous relaxation, $\pi_e$ will take a value of 1 to maximize the inner objective of expression~\eqref{eq:cadv1}.
$\blacksquare$

Unfortunately, the adversary's optimization model cannot be solved directly as a linear program due to the minimax function in objective \eqref{eq:cadv1}. However, as the inner maximization problem is a linear program, we can convert the entire problem into a minimization problem by taking dual of the inner problem. The dual problem, which is quadratic nature, is compactly shown in Table~\ref{alg:cadv}. $\alpha, \beta, \gamma, \zeta, \omega$ and $\eta$ represent nonnegative dual price variables for constraints \eqref{eq:cadv2}-\eqref{eq:cadv7}, respectively. We set $\mu$ as variable in the dual problem and incorporate the domain constraints~\eqref{eq:cadvdual5} of $\mu$ to represent the optimization problem of the adversary.

%we first obtain the Lagrangian function, $L$ by relaxing constraints \eqref{eq:cadv2}-\eqref{eq:cadv7} using price variables $\alpha, \beta, \gamma, \delta, \omega$ and $\eta$, respectively.
%\begin{equation}
%	\label{eq:clargange}
%	\begin{aligned}
%		L(\alpha, \beta, \gamma, \delta,\eta,\omega) = & - y_{(t,s)} + \sum_{e\in {\cal E}} p_e z_e+ \sum_{ v\in {N}\setminus s} \alpha_v \big [\sum_{e\in \delta^-_v} y_e - \sum_{e\in \delta^+_v} y_{e}\big ] + \sum_e \delta_e  \big [ \pi_e - \sum_{e' \in \rho_e} \mu_{e'} \big ] \\
%		& + \sum_{e\in {\cal E}} \gamma_e \big [ y_e - \bar{x}_e -\pi_e (U_e -\bar{x}_e) \big ] + \sum_e \eta_e \big [ \pi_e -1 \big ] + \sum_e \omega_e \big [ y_e - z_e - \bar{x}_e \big ]\\ 
%		& + \sum_{e\in {\cal E}} \beta_e \big [ y_e - U_e + \mu_e(U_e-m_e)\big ]  
%	\end{aligned}
%\end{equation}

%We can now write the optimization problem for the adversary by using the dual problem from the Lagrangian function $L$. Specifically, we set $\mu$ as variable in the dual problem and incorporate the domain constraints~\eqref{eq:cadvdual5} of $\mu$ to represent the optimization problem of the adversary. The optimization problem, which is quadratic nature, is compactly shown in Table~\ref{alg:cadv}.
\begin{table}[!htb]
	\centering
	\begin{tabular}{|c|}
		\hline
		\begin{minipage}{0.97\textwidth}
			{	
			\begin{flalign}
				\min_{\alpha, \beta, \gamma, \zeta,\eta, \omega ,\mu} & \hspace{5pt}  \sum_{e \in {\cal E}} \beta_e U_e - \sum_{e \in {\cal E}} \beta_e \mu_e (U_e-m_e) + \sum_{e \in {\cal E}} \gamma_e \bar{x}_e + \sum_{e \in {\cal E}}\zeta_e \sum_{e' \in \rho_e} \mu_{e'} + \sum_{e \in {\cal E}} \eta_e + \sum_{e \in {\cal E}} \omega_e \bar{x}_e  \label{eq:cadvdual1}\\
				\hspace{10pt}\text{s.t. } & \beta_e + \gamma_e + \omega_e + \alpha_v -\alpha_w \geq 0 \hspace{2.4in} \forall e = (v,w) \label{eq:cadvdual2}\\
				& \zeta_e +\eta_e - \gamma_e (U_e - \bar{x}_e) \geq 0 \hspace{2.8in} \forall e\in {\cal E}\label{eq:cadvdual3}\\
				& \omega_e \leq p_e \hspace{4.0in} \forall e\in {\cal E} \label{eq:cadvdual4} \\
				& \sum_{e \in {\cal E}} \mu_e \leq \Gamma \label{eq:cadvdual5} \\
				& \alpha_t =1 ; \alpha_s =0 \label{eq:cadvdual6} \\
				& \alpha_v , \beta_e , \gamma_e, \zeta_e, \eta_e, \omega_e \geq 0; \mu_e \in \{0,1\} \label{eq:cadvdual7}
			\end{flalign}}
		\end{minipage} \\
		\hline
	\end{tabular}
	\caption{{\sc AdversaryProblem}($G,U,p,\bs{x}, \Gamma$)}
	\label{alg:cadv}
\end{table}

The objective function \eqref{eq:cadvdual1} -- redefined by minimizing the problem \eqref{eq:cadv1}-\eqref{eq:cadv8} over all the variables except for $\mu$ -- would be concave in $\mu$. As a concave function over a compact domain achieves the optimal solution at an extreme point of the feasible region, we can relax the binary variables $\mu$ to continuous ones (see e.g., the proof of Lemma 6 of \citeauthor{bertsimas2013robust}~\citeyear{bertsimas2013robust}). However, if $\mu$ variables are binary, then the bilinear terms in the objective function \eqref{eq:cadvdual1} can be represented as \emph{special ordered set} (SOS) constraints and these constraints are efficiently implemented by existing mixed-integer optimization solvers like CPLEX or Gurobi. 
%In other words, the objective function \eqref{eq:cadvdual1} computes a point-wise minimum of linear functions of $\mu$ and therefore, concave in $\mu$ (see \citeauthor{boyd2004convex} \citeyear{boyd2004convex}).

\subsection{Optimization Problem for the Administrator} \label{modelAdmin}
In a fully secured network, the optimal strategy for the network administrator is to adopt one of the \emph{max-flow min-cost} solutions~\citep{ford1956maximal}. However, in case of an adversarial attack, the \emph{max-flow min-cost} solution is not necessarily an optimal strategy. 
The administrator's goal is then to identify a robust flow strategy that maximizes the objective value under worst-case attack, as indicated in problem \eqref{eq:probadm}. However, as the strategy space of the adversary increases with network size and budget value, we use an incremental approach and consider the attacks revealed by the adversary in previous iterations. In the last iteration of the game, as the adversary repeats her previous attacks as the best response, the administrator at this point has already generated a robust flow strategy against these worst-case attacks without considering the entire large strategy space of the adversary.

Given a set of $K$ attacks (the set of indices for these attacks is denoted by ${\cal K}$) revealed by the adversary in previous $K$ iterations, the administrator generates a flow strategy that maximizes the minimum objective value over $K$ attacks. Let $\hat{\mu}^k$ denote the attack $k$ and ${\hat{x}}$ denote the robust flow strategy against these attacks. Even though the initial flow decision ${\hat{x}}$ is chosen, it may lead to different outcomes under different attacks and the actual amount of flow that reaches the destination will vary depending on the attack. So, we introduce ${\hat{y}}^k$ variables to represent the actual flow and ${\hat{z}}^k$ variables to denote the additional rerouted flow under the attack $k$. $\hat{\pi}^k_e$ is a given input to the administrator's decision model which is set to 0 if none of the edges in edge set $\rho_e$ (from which the flow can be rerouted to edge $e$) is attacked under attack $k$ and otherwise, it is fixed to 1:
\begin{equation}
\begin{aligned}
\hat{\pi}^k_e = \min(1, \sum_{e' \in \rho_e} \hat{\mu}^k_{e'}) \hspace{0.3in} \forall k\in {\cal K}, e\in {\mathcal E} \label{computePi}
\end{aligned}
\end{equation}

% for an initial robust flow $\bs{\hat{x}}$, actual resulting flow $\bs{\hat{y}}$, and additional rerouted flow $\bs{\hat{z}}$, can be
%\begin{equation}
%	\begin{aligned}
%		\label{cobjflow2}
%		\max & \hspace{0.05in} \lambda \\
%		\textbf{s.t.} &  \hspace{0.05in}  \lambda \leq \hat{y}^k_{(t,s)} -\sum_{e \in {\cal E}}  p_e \big[ \hat{x}_e + \hat{z}^k_e \big ]  \hspace{0.3in} \forall k \in {\cal K}
%	\end{aligned}
%\end{equation}

We show the entire linear optimization problem for the administrator compactly in Table~\ref{alg:piterflow}. 
The goal of the administrator is to maximize the worst-case objective value over $K$ attacks.
\begin{equation}
	\begin{aligned}
		\label{cobjflow1}
		\max \min_{k \in {\cal K}} & \hspace{0.05in} \hat{y}^k_{(t,s)} -\sum_{e \in {\cal E}}  p_e \big[ \hat{x}_e + \hat{z}^k_e \big ] 
	\end{aligned}
\end{equation}
The objective function \eqref{cobjflow1} can easily be linearized using expression~\eqref{piterflow1}-\eqref{piterflow2}, where we maximize the proxy variable $\lambda$ that represents the worst-case objective value over $K$ attacks.
Constraints~\eqref{piterflow3} compute the value of $\hat{z}^k_e$. Constraints~\eqref{piterflow4}-\eqref{piterflow7} enforce the flow preservation and capacity constraints on the flow variables ${\hat{x}}$ and ${\hat{y}}^k$.
Finally, constraints~\eqref{piterflow8} enforce that the actual flow going through edge $e$ under attack $k$ is bounded by the initial flow $\hat{x}_e$ if none of the edges in set $\rho_e$ is attacked and otherwise, the upper bound is set to the capacity of the edge, $U_e$.

\begin{table}[t]
	\centering
	\begin{tabular}{|c|}
		\hline
		\begin{minipage}{0.8\textwidth}
			{	
			\begin{flalign}
				\max & \hspace{0.05in} \lambda & \label{piterflow1}\\
				s.t. & \hspace{0.05in} \lambda \leq \hat{y}^k_{(t,s)} -\sum_{e \in {\cal E}}  p_e \big[ \hat{x}_e + \hat{z}^k_e \big ] & \forall k\in {\cal K} \label{piterflow2}\\
				& \hat{z}^k_e \geq \hat{y}^k_e - \hat{x}_e & \forall k\in {\cal K}, e \in {\cal E} \label{piterflow3}\\
				& \sum_{e\in \delta^+_v} \hat{x}_e - \sum_{e\in \delta^-_v} \hat{x}_{e} = 0 & \forall v \in {\cal V}\setminus \{s\} \label{piterflow4}\\
				& \hat{x}_e \leq U_e  & \forall e \in {\cal E} \label{piterflow5} \\
				& \sum_{e\in \delta^+_v} \hat{y}^k_e - \sum_{e\in \delta^-_v} \hat{y}^k_{e} \geq 0 &  \forall k\in {\cal K}, v \in {\cal V}\setminus \{s\} \label{piterflow6}\\
				& \hat{y}^k_e \leq (1-\hat{\mu}^k_e)U_e + \hat{\mu}^k_e\cdot m_e &  \forall k\in {\cal K}, e \in {\cal E} \label{piterflow7}\\
				& \hat{y}^k_e \leq (1-\hat{\pi}^k_e)\hat{x}_e + \hat{\pi}^k_e\cdot U_e &  \forall k\in {\cal K}, e \in {\cal E} \label{piterflow8}\\
				& \lambda\geq 0; \hat{x}_e\geq 0; \hat{y}^k_e \geq 0; \hat{z}^k_e \geq 0   &  \forall k\in {\cal K}, e \in {\cal E} \label{piterflow9}
			\end{flalign}}
		\end{minipage} \\
		\hline
	\end{tabular}
	\caption{{\sc AdministratorProblem}($G,U,p, \bs{\hat{\mu}}$)}
	\label{alg:piterflow}
\end{table}

\subsection{Overall Solution Approach} \label{overallSol}
In this section, we put together the optimization problems delineated in \cref{modelAdv} and \cref{modelAdmin}. To better understand the overall framework of our proposed solution approach for solving the TNFG, we explain the key iterative steps in Algorithm~\ref{algo2}. We essentially formulate it as a leader-follower game where the administrator is the leader and the adversary is the follower. In that sense, the adversary is more powerful and can take decisions after observing the decision of the administrator. As a result, the administrator needs to take robust decisions by considering all possible attacking behaviors of the adversary. 

{\centering
\begin{algorithm}[!htb]
	\textbf{Initialize:} $ \bs{X}\leftarrow \{\},\bs{\mu} \leftarrow \{\}, \mu_{old}\leftarrow \bs{0}, V^L\leftarrow 0, V^U \leftarrow \infty $ \;
	\While{$( V^U \neq V^L )$}{
		$\bs{\tilde{x}}, V^U\leftarrow$ {\sc AdminstratorProblem}$(G,U,{p},\bs{\mu})$ \Comment*[r]{\scriptsize Robust flow solutions against $\bs{\mu}$}
		$\bs{x}_p \leftarrow \bs{\tilde{x}} \cap \bs{X}$ \Comment*[r]{\scriptsize $\bs{x}_p$ contains previously executed flows}
		$\bs{x}_c \leftarrow \bs{\tilde{x}} \setminus \bs{x}_p$ \Comment*[r]{\scriptsize $\bs{x}_c$ contains flows which are not executed yet}
		\If{$|\bs{x}_c| = 0$}{
			%\Return $\bs{\tilde{x}}, \bs{\mu}$ 
			\Return $\bs{\tilde{x}}, \bs{\mu}$ \Comment*[r]{\scriptsize Process converges}% and administrator executes one flow from $\bs{\tilde{x}}$} 
		}
		\Else{
			$x^* \leftarrow Random(\bs{x}_c)$ \Comment*[r]{\scriptsize Randomly pick one flow from $\bs{x}_c$ and execute it} 
			$\bs{X} \leftarrow \bs{X} \cup x^*$ \Comment*[r]{\scriptsize Add flow $x^*$ into flow strategy pool $\bs{X}$} 
		}
		$\tilde{\mu}, V^L \leftarrow$ {\sc AdversaryProblem}($G,U,{p},x^*,\Gamma$) \Comment*[r]{\scriptsize Generate best attack against $x^*$}
		\If{$\tilde{\mu} \neq \mu_{old}$}{
			$\bs{\mu} \leftarrow \bs{\mu} \cup \tilde{\mu}$ \Comment*[r]{\scriptsize Adversary executes $\tilde{\mu}$ and add it to attacking solution pool $\bs{\mu}$} 
			$\mu_{old} \leftarrow \tilde{\mu}$ \Comment*[r]{\scriptsize Update the old attack}   
		}
		\Else{
			%\Return $x^*, \bs{\mu}$ 
			\Return $x^*, \bs{\mu}$ \Comment*[r]{\scriptsize Process converges}% and administrator executes flow $x^*$} 
		}
	}
	\Return $x^*, \bs{\mu}$
	\caption{\sc{SolveTNFG}($G, U,p, \Gamma$)}
	\label{algo2}
\end{algorithm}}

Let $\bs{X}$ and $\bs{\mu}$ store all the previously generated flow strategies and attacks respectively, and both are initialized as empty set. $\mu_{old}$ keeps track of the attack from the last iteration and it is initialized to no attack. Let $V^L$ and $V^U$ denote the lower and upper bound of the game which are initialized to 0 and $\infty$, respectively.
In each iteration of the game, the administrator solves the optimization problem from Table~\ref{alg:piterflow} to compute the optimal robust flow strategy against the previously generated attack pool $\bs{\mu}$. Note that the administrator's optimization problem from Table~\ref{alg:piterflow} can have multiple optimal solutions. For example, in the first iteration, the administrator essentially solves a \emph{max-flow min-cost} problem which can have multiple flow solutions with optimal objective value. Let, $\bs{\tilde{x}}$ denote the set of optimal robust flow solutions computed by the administrator in the current iteration. We update the upper bound of the game $V^U$ to the objective value of the administrator. From these flow solutions, we identify the flows, $\bs{x}_c$ which are not executed before. If all the solutions of $\bs{\tilde{x}}$ are executed before, then the process terminates as any solution from $\bs{\tilde{x}}$ would be a robust flow against any plausible attack from $\Psi$. Otherwise, we randomly pick one flow $x^*$ from $\bs{x}_c$ and execute it. We also add this new flow $x^*$ into the pool of flow solutions $\bs{X}$.

Once the administrator generates a flow $x^*$ and executes it, the adversary computes a best myopic attack $\tilde{\mu}$ to optimally disrupt the flow $x^*$ by solving the optimization model from Table~\ref{alg:cadv}. In addition, the lower bound of the game $V^L$ is updated to the adversary's objective value. If the adversary does not repeat the attack from previous iteration (i.e., $\mu_{old}$), then we add the new attack $\tilde{\mu}$ into the attack pool $\bs{\mu}$ and update $\mu_{old}$ to the new attack $\tilde{\mu}$. On the contrary, if the adversary repeats the same attack from previous iteration, then the process converges as $x^*$ is the robust flow scenario against $\tilde{\mu}$. Specifically, the objective values of both the players would converge at this point and we can guarantee that if the administrator executes the flow $x^*$, then the lower bound on her objective value would be the value at which the objectives of the players converged. It should be noted that as the administrator's optimization problem can have uncountable number of solutions with optimal objective value, a situation might arise where although the objectives of both the players converge to the same value, they can come up with new strategies in subsequent iterations that lead to same objective value. However, as the game value does not change from this point onward, we terminate the process if the lower and upper bound of the game converge. On the other hand, if the optimization problems of the players are solved sub-optimally (e.g., see \cref{heuristics} for large-scale problems), then the lower and upper bound of the game might not converge even though the players start repeating their previous strategies. To tackle these situations, we enforce both the termination conditions in Algorithm~\ref{algo2}. 

\begin{proposition}
The proposed iterative game converges to a maximin equilibrium and produces a robust maximum flow strategy for the administrator.
\end{proposition}
\textbf{Proof.} As the adversary has a finite number of attacks (i.e., $|\Psi |$ = $|{\cal E}| \choose \Gamma$), the proposed iterative game is guaranteed to converge. We begin the proof by showing that if the players start repeating their previous strategies and their respective decision problems are solved optimally, then the lower and upper bound of the game would converge. Let us suppose the game converges after $K$ iterations and the set of $K-1$ attacks computed by the adversary in previous iterations is denoted by $\bs{\mu}$. Let $\tilde{x}$ represent the flow strategy of the administrator and $\tilde{\mu} \in \bs{\mu}$ denote the attack from the last iteration. As the adversary repeats one of her previous attacks in the final iteration, the objective of the adversary for the final iteration, $V^L = M(\tilde{x},\tilde{\mu})=\min_{\mu \in \bs{\mu}} M(\tilde{x},\mu)$, where $M(\tilde{x},\tilde{\mu})$ represents the adaptive value of the flow $\tilde{x}$ under attack $\tilde{\mu}$ which can be computed using the LO model \eqref{eq:flow1}-\eqref{eq:flow5}. Similarly, as the administrator maximizes the minimum adaptive flow value over all the attacks in $\bs{\mu}$, the objective value for the final iteration is $V^U =\min_{\mu \in \bs{\mu}} M(\tilde{x},\mu)=V^L$.

Let $V^L=V^U=V$. From the adversary's optimized solution, since $\forall \mu\in \Psi, M(\tilde{x},\mu) \geq V$, we know $\min_{\mu\in \Psi} M(\tilde{x},\mu) \geq V$.
On the other hand, from the administrator's optimized solution, since $\forall x\in \mathcal{X}, \min_{\mu\in \bs{\mu}} M(x,\mu) \leq V$, we know $ \min_{\mu\in \Psi} M(x,\mu) \leq V$. By combining these two inequalities, we have $\forall x\in \mathcal{X}, \min_{\mu\in \Psi} M(\tilde{x}, \mu) \geq \min_{\mu\in \Psi} M(x, \mu)$, which concludes the proof that the administrator's final flow strategy $\tilde{x}$ is maximin optimal.
$\blacksquare$

If Algorithm~\ref{algo2} ends up considering all the attacks (i.e., $K=|\Psi|$), then the proposed iterative approach will be slower than solving the original problem of identifying a flow strategy that maximizes the minimum adaptive flow value by considering all the attacks in $\Psi$. However, it is reasonable to expect that only some attacking choices are better to execute in practice than others. For example, if all the incoming edges to a node is attacked, then none of the outgoing edges from that node will be selected, or if the edge with maximum capacity in a source to destination path is attacked, then other edges in that path are less likely to be attacked. Our proposed approach provides a principle way to identify those crucial attacks. In fact, for all the problem instances in our experiment, we observe that the proposed iterative game converges within 20 iterations.

\section{Heuristics for Large-scale Networks}\label{heuristics}
The runtime complexity of our iterative game approach increases with the network size and adversary's budget value (refer to \cref{runtimeSmall}). The key reason behind this behavior is that we need to solve a complex quadratic program from Table~\ref{alg:cadv} for the adversary's decision problem in each iteration of the game. The adversary's decision problem seeks to identify $\Gamma$ best edges to attack so as to minimize the resulting objective of the administrator for a given flow strategy. In case of ``s-t planar graphs", this problem can be solved in polynomial time \citep{wollmer1964removing} if the entire flow of an attacked edge is assumed to be lost. However, if the flow of an attacked edge is allowed to reroute through other paths, then the resulting flow at an edge after attack is not upper bounded by the initial flow value and therefore, the existing polynomial time algorithms cannot be employed to solve our problem. However, for a given attack, we can precompute the set of edges through which the additional flow might be rerouted and compute an upper bound on the resulting flow assigned to the edges. Therefore, the utility of a given attack can be computed by solving a \emph{max-flow min-cost} problem (in polynomial time) on a network whose edge capacities are set to the upper bounds. However, to compute an optimal attack, we need to solve such polynomial time algorithms for $|{\cal E}| \choose \Gamma$ times using an exhaustive search which is practically intractable for large networks.

In this section, we provide two novel heuristic approaches to efficiently solve the complex optimization problem of the adversary. 
The first heuristic is developed using an accelerated greedy approach and the second heuristic is a network partitioning based optimization approach. We now describe the details of the two heuristic approaches for solving the adversary's decision problem.

\subsection{Greedy Approach} \label{greedyHeuristic}
The greedy approach incrementally identifies the set of edges to be attacked by the adversary so as to disrupt the network for a given flow strategy. We begin with an alternative formulation of the optimization model \eqref{eq:flow1}-\eqref{eq:flow5} to evaluate the utility of an attack $\mu$ for a given flow scenario ${\bar{x}}$. The alternative LO model is compactly shown in Table~\ref{alg:simulation}. The value of ${\pi}$ is computed using equation~\eqref{computePi} and is given as an input to the LO model. 
%The objective is to route maximum amount of flow to the terminal node while minimizing the total amount of rerouting cost. First two sets of constraints enforce the flow preservation and capacity constraints. The last set of constraints compute the amount of rerouted flow $z_e$ through the edge $e$. 
The only differentiating constraints from LO model \eqref{eq:flow1}-\eqref{eq:flow5} are the third set of constraints which use $\pi$ to ensure the upper bound on the flow variables $y$.
It should be noted that the solution of the LO model can be obtained in polynomial time by solving a \emph{max-flow min-cost} problem on a modified network where the capacity of an edge $e$ is set to 0 if $\mu_e = 1$, $\bar{x}_e$ if $\pi_e = 0$ and otherwise it remains $U_e$.

\begin{table}[!htb]
	\centering
	\begin{tabular}{|c|}
		\hline
		\begin{minipage}{0.75\textwidth}
			{	
			\begin{flalign}
				\max & \hspace{5pt}  y_{(t,s)} - \sum_{e\in {\cal E}} p_e\cdot z_e & \nonumber\\
				\hspace{10pt}\text{s.t. } & \sum_{e\in \delta^+_v} y_e - \sum_{e\in \delta^-_v} y_{e} \geq 0  & \forall v \in {\cal V}\setminus s \nonumber \\
				&  y_e \leq (1-\mu_e) U_e + \mu_e m_e  & \forall e\in {\cal E} \nonumber\\
				& y_e \leq (1-\pi_e)\bar{x}_e +\pi_e \cdot U_e  & \forall e\in {\cal E} \nonumber \\
				& z_e \geq y_e - \bar{x}_e   & \forall e \in {\cal E} \nonumber \\
				& y_e \geq 0 ; z_e \geq 0 & \forall e \in {\cal E} \nonumber 
			\end{flalign}}
		\end{minipage} \\
		\hline
	\end{tabular}
	\caption{{\sc IdentifyFlow}($G,U,{p}, {\bar{x}},\mu$)}
	\label{alg:simulation}
\end{table}

Algorithm~\ref{algo5} provides the details of the greedy algorithm. We start with an empty attacked edge set $\mu$. Let $E$ denote an edge set that initially contains all the edges. We first compute the objective value $O$ for the given administrator's flow strategy ${\bar{x}}$ without executing any attacks in the network. In each iteration, we calculate the utility $g_e$ for adding an edge $e \in E$ in the current attack set $\mu$ by employing the LO model from Table~\ref{alg:simulation} and compute the marginal gain in the adversary's objective for attacking the edge $e$. Then we add the best edge $e^*$ (that provides maximum marginal gain) into $\mu$ and remove it from the candidate edge set $E$. This process continues until the budget for the adversary is exhausted.

{\centering
\begin{algorithm}[!htb]
	\textbf{Initialize: } $\mu \leftarrow \{\}; it \leftarrow 0; E\leftarrow {\cal E}$ \;
	$O \leftarrow$ {\sc IdentifyFlow}($G, U, {p},  {\bar{x}}, \mu$) \Comment*[r]{\scriptsize Compute objective for the given flow ${\bar{x}}$}
	\Repeat{$(|\mu| \geq \Gamma) $}{
		$it \leftarrow it +1$ \;
		$g_e, {\hat{x}} \leftarrow$ {\sc IdentifyFlow}($G, U, {p},  {\bar{x}}, \mu\cup \{e\}$) $\hspace{0.2in} \forall e\in E$\; 	
		$g_e \leftarrow O-g_e\hspace{0.5in} \forall e\in E$ \Comment*[r]{\scriptsize Compute marginal gain for edge $e$}	
		$e^* \leftarrow \arg\max\limits_{e\in E} g_e$ \Comment*[r]{\scriptsize Choose edge $e^*$ with highest marginal gain}	
		$O \leftarrow O+g_{e^*}$  \Comment*[r]{\scriptsize Update objective value for current attack $\mu$}
		$\mu \leftarrow \mu \cup \{e^*\}$ \Comment*[r]{\scriptsize Update current attack $\mu$}
		$E \leftarrow E - \{e^*\}$ \Comment*[r]{\scriptsize Update candidate edge set $E$}		 
	}
	\Return $\mu$\\
	\caption{\sc{Greedy}($G = <{\cal V}, {\cal E}>, U,{p}, {\bar{x}}, \Gamma$)}
	\label{algo5}
\end{algorithm}}

\begin{figure}[!htb]
	\centering
	\begin{subfigure}{0.49\textwidth}
		\includegraphics[width=\textwidth]{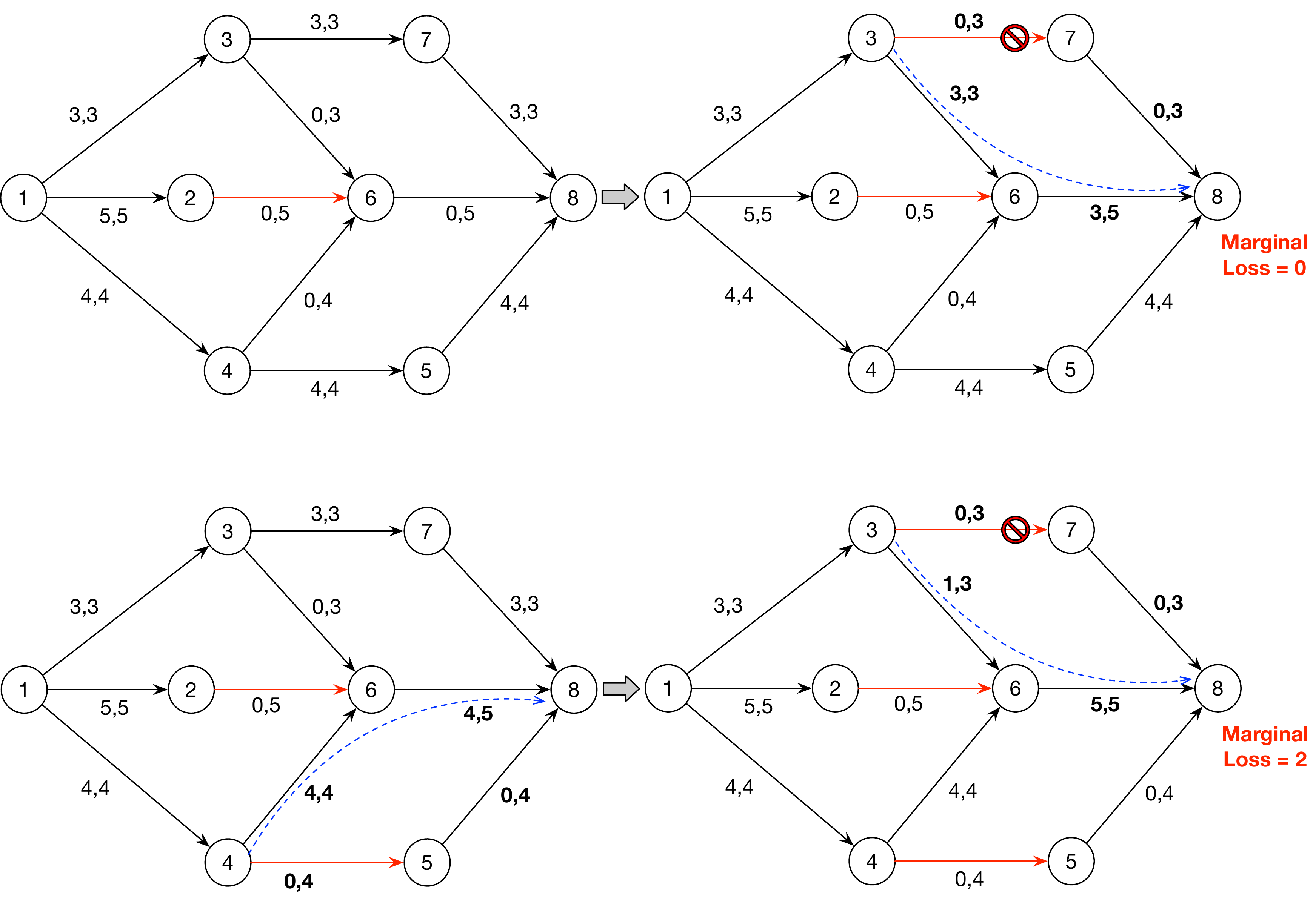} \caption{}
	\end{subfigure} 
	\vline \vspace{0.1in}
	\begin{subfigure}{0.49\textwidth}
		\includegraphics[width=\textwidth]{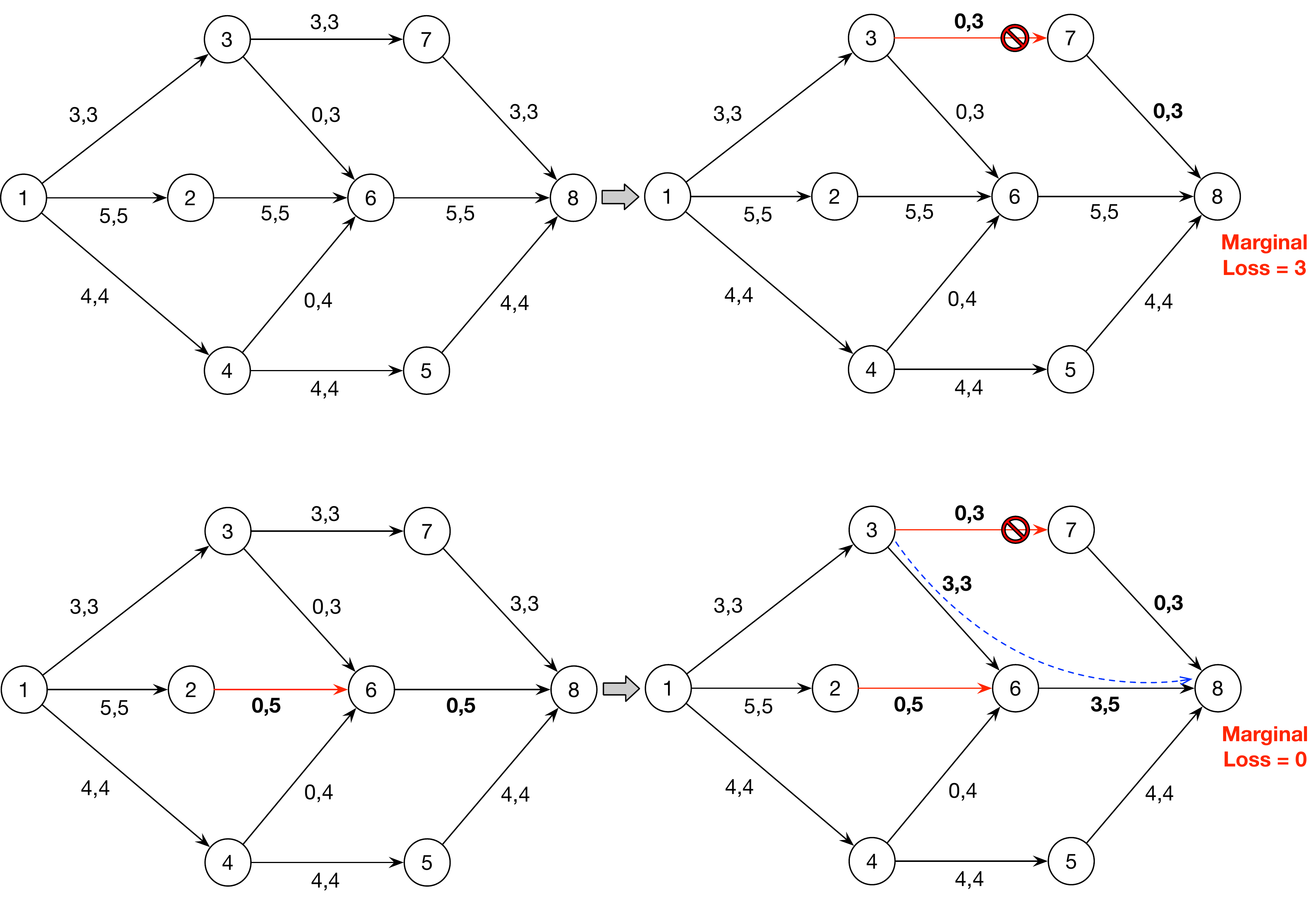} \caption{}
	\end{subfigure} 
	 \caption{Adversary's decision problem is (a) neither sub-modular; (b) nor super-modular.}
	 \label{fig:map6}
\end{figure}

Although the LO model from Table~\ref{alg:simulation} can be solved in polynomial time, the greedy approach needs to solve it $(\Gamma \times |{\cal E}|)$ times and therefore, it is not suitable for problems with large number of edges. In case of sub-modular objective function, lazy greedy algorithms \citep{minoux1978accelerated} can be employed to accelerate the solution process. However, the following two remarks show that the adversary's decision problem is neither sub-modular nor super-modular.
\begin{remark}\label{exp6}
\textbf{The adversary's decision problem is not sub-modular.} A function is monotone sub-modular if the marginal gain for adding an element into the subset is always higher than adding the same element into its superset. Figure~\ref{fig:map6}(a) illustrates that the adversary's decision problem does not exhibit this property. The network has 8 nodes and 11 edges and the pair of numbers in each edge represents the flow and capacity of the corresponding edge. Let us assume that the unit cost for routing the flow is 0 for all the edges. If we add the edge $e_{37}$ in the attacked edge set which only contains edge $e_{26}$, the marginal gain in objective remains 0 as the entire flow can be rerouted through the augmented path $\{e_{36}, e_{68}\}$. In contrast, if we add the edge $e_{37}$ in the superset which contains $e_{26}$ and $e_{45}$, then the marginal gain in objective is 2, as the residual capacity of $e_{68}$ is now shared by the rerouted flow from both $e_{37}$ and $e_{45}$. As the marginal gain for adding edge $e_{37}$ into superset is higher, the adversary's decision problem is not sub-modular.
\end{remark}
\begin{remark}\label{exp7}
\textbf{The adversary's decision problem is not super-modular.} A function is referred as monotone super-modular if the marginal gain for adding an element into the subset is always lower than adding the same element into its superset. Figure~\ref{fig:map6}(b) illustrates that the adversary's decision problem is not super-modular. We employ the same network with 8 nodes and 11 edges. In this example, if we add the edge $e_{37}$ in the attacked edge set as the first element, then the entire flow is lost and therefore, the gain in the adversary's objective is 3. In contrast, if we add the edge $e_{37}$ in the superset which contains edge $e_{26}$, the entire flow can be rerouted to the terminal node and the marginal gain in objective is 0. Hence, the adversary's decision problem is clearly not super-modular, as the marginal gain for adding edge $e_{37}$ into superset is lower.
\end{remark}

Even though the adversary's decision problem is neither sub-modular nor super-modular, the following remark shows that the upper bound on the marginal gain for attacking an edge can be precomputed which provides us the basis to develop an accelerated greedy algorithm.

\begin{remark}\label{obs5}
The maximum amount of lost flow for attacking an edge $e$ is bounded by the flow assigned to that edge, $\bar{x}_e$ if the adversary's budget is 1 (i.e., $\Gamma = 1$).
\end{remark}

As we incrementally elect one edge at a time for the greedy approach, we exploit the observation from remark~\ref{obs5} to develop an accelerated greedy algorithm which is shown in Algorithm~\ref{algo6}. 
Let $E$ denote the set of candidate edges and $\mu$ be the set of attacked edges which is initialized as an empty set. We first compute the objective value $g^0$ for the given flow strategy ${\bar{x}}$ without considering any attacks. In each iteration, we keep an upper bound $B_e$ on the marginal gain for edge $e$ which is initialized to the given flow value $\bar{x}_e$. We also introduce a set $\hat{E}$ (initialized as an empty set) that stores the edges for which the marginal gain has already been computed in the current iteration. We iteratively select the edge $e^*$ with maximum upper bound and add it to the edge set $\hat{E}$. Then we employ the LO model from Table~\ref{alg:simulation} to compute the objective value and marginal gain for adding the edge $e^*$ in the current attacking set $\mu$ and update the upper bound $B_{e^*}$ with the marginal gain value. In addition, as the flow values for the edges change due to modified attacks, we store the updated flow strategy ${\hat{x}}$. If the marginal gain for edge $e^*$ is equal or greater than the upper bound for all the unexplored edges (i.e., $e\in E\setminus \hat{E}$), then the best edge for the current iteration is identified. We then select the edge $e^{**}$ with highest marginal gain, insert it to the set of attacked edges $\mu$ and remove it from the candidate edge set $E$. Finally, we update the flow strategy ${\bar{x}}$ with the modified flow strategy ${\hat{x}}$. Therefore, in each iteration, the accelerated greedy algorithm is able to identify the best edge without executing the LO model for $|E\setminus \hat{E}|$ times in comparison to the greedy algorithm, which significantly reduces the runtime. This iterative edge selection process continues until the adversary's budget is exhausted.

{\centering
\begin{algorithm}[!htb]
	\textbf{Initialize: } $\mu \leftarrow \{\}; it \leftarrow 0; E\leftarrow {\cal E}$ \;
	$g^0, {\hat{x}} \leftarrow$ {\sc IdentifyFlow}($G, U, {p}, {\bar{x}}, \mu$) \Comment*[r]{\scriptsize Compute objective for the given flow ${\bar{x}}$}
	\Repeat{$(|\mu| \geq \Gamma) $}{
		$it \leftarrow it +1$ \;
		$B_e \leftarrow \bar{x}_e \forall e\in E$ \Comment*[r]{\scriptsize Update upper bounds with the new flow ${\bar{x}}$ from last iteration}
		$\hat{E}\leftarrow \{\}$ \;
		\Repeat{True}{
			$e^* \leftarrow \arg\max\limits_{e\in E} B_e$ \Comment*[r]{\scriptsize Choose the edge $e^*$ with highest upper bound}
			$\hat{E} \leftarrow \hat{E} \cup \{e^*\}$ \Comment*[r]{\scriptsize Update the list of visited edges}
			$g^{it}, {\hat{x}} \leftarrow$ {\sc IdentifyFlow}($G, U, {p}, {\bar{x}}, \mu\cup \{e^*\}$)\;
			$B_{e^*} \leftarrow g^{it-1} - g^{it}$ \Comment*[r]{\scriptsize Update the upper bound for edge $e^*$}
			\If{$B_{e^*} \geq B_e , \forall e \in E\setminus \hat{E}$ }{
				$e^{**} \leftarrow \arg\max\limits_{e \in \hat{E}} B_e$ \Comment*[r]{\scriptsize Choose the edge $e^{**}$ with highest marginal gain}
				$\mu \leftarrow \mu \cup \{e^{**}\}$ \Comment*[r]{\scriptsize Update the current attacking edge set}
				$E \leftarrow E - \{e^{**}\}$ \Comment*[r]{\scriptsize Update the candidate edge set $E$ }	
				${\bar{x}} \leftarrow {\hat{x}}$ \Comment*[r]{\scriptsize Update the flow scenario ${\bar{x}}$ according to new attack $\mu$}
				$\textbf{Break}$\;
			}			
		}
	}
	\Return $\mu$\\
	\caption{\sc{AcceleratedGreedy}($G = <{\cal V}, {\cal E}>, U, {p}, {\bar{x}}, \Gamma$)}
	\label{algo6}
\end{algorithm}}

\subsection{Network Partitioning Based Optimization} \label{partitionHeuristic}
As the adversary's decision problem grows exponentially with the number of edges in the network, for our second heuristic, we first identify a small subset of candidate edges that are highly likely to be present in an optimal solution and then the adversary's optimization problem from Table~\ref{alg:cadv} is solved by only considering the small set of candidate edges. To identify the candidate edges, we propose a network partitioning based iterative approach. In each iteration, the network is partitioned into disjoint sub-networks and those sub-problems are solved independently to elect $\Gamma$ best edges to attack. 
 
We begin the discussion with a random network partitioning approach that is compactly shown in Algorithm~\ref{algo3}. Let us consider a network with $N (=|{\cal V}|)$ nodes. We first randomly sample $(\frac{N}{2}-1)$ nodes that excludes the source and terminal node\footnote{In case of multiple source nodes, we create an artificial source node and connect it to all the original source nodes. Similarly, in case multiple terminal nodes, we create an artificial terminal node and connect all the original terminal nodes to the artificial node. During the network partitioning process, we ensure that all the original source nodes are kept in the first sub-network and all the original terminal nodes are kept in the second sub-network.}. All the randomly sampled nodes along with the source node are kept in the first sub-network and the remaining nodes are kept in the second sub-network. In addition, an artificial terminal node $\hat{t}$ for first sub-network and an artificial source node $\hat{s}$ for the second sub-network are created. For each edge $e (=\{u,v\})$ in the network, we carry out the following operations:
\begin{itemize}
%\squishlist
\item If both $u$ and $v$ lie in the first sub-network, we create a directed edge between them.
\item If both $u$ and $v$ lie in the second sub-network, we create a directed edge between them.
\item If $u$ lies in first sub-network and $v$ lies in the second sub-network, then we create two edges, one from node $u$ to node $\hat{t}$ in the first sub-network and another one from node $\hat{s}$ to node $v$ in the second sub-network. The flow and capacity values for both the newly introduced edges are directly taken from edge $e$\footnote{It should be noted that due to our edge reconstruction, there might be multiple parallel edges from one node of the first sub-network to the terminal node or from the source node of second sub-network to another node. In that case, we replace all the parallel edges with same source and destination node with a single edge whose flow and capacity values are computed as the sum of flows and capacities of all the parallel edges.}.
\item If $u$ lies in the second sub-network and $v$ lies in first sub-network, then we remove edge $e$ from the sub-problems.
%\squishend
\end{itemize}

\begin{figure}[!htb]
	\centering
	\includegraphics[width=0.98\linewidth]{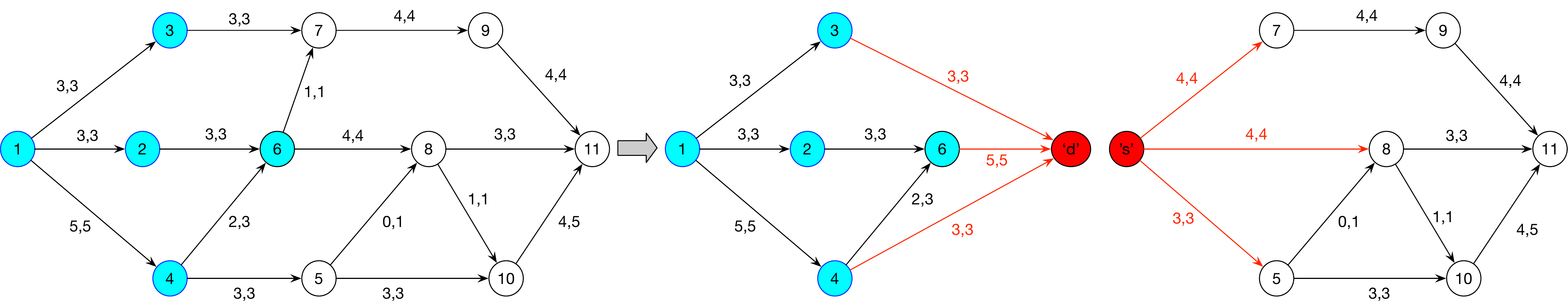} 
	 \caption{Illustration of partitioning approach.}
	 \label{fig:partition}
	 %\vspace{-0.1in}
\end{figure}

{\centering
\begin{algorithm}[!htb]
	${\cal V}^s \leftarrow$ {\sc RandomSample} (${\cal V} \setminus \{s,t\}, |{\cal V}|/2$) \Comment*[r]{\scriptsize Randomly sample half of the nodes}
	${\cal V}^s \leftarrow {\cal V}^s \cup \{s\} \cup \{\hat{t}\}$ \Comment*[r]{\scriptsize ${\cal V}^s$ represents the set of nodes in the first sub-network}
	${\cal V}^t \leftarrow  {\cal V} \setminus \{{\cal V}^s \cup \{s\}\} \cup \{\hat{s}\}$ \Comment*[r]{\scriptsize ${\cal V}^t$ represents the set of nodes in the second sub-network}
	${\cal E}^s \leftarrow \{\}$ \Comment*[r]{\scriptsize ${\cal E}^s$ represents the set of edges in the first sub-network}
	${\cal E}^t \leftarrow \{\}$ \Comment*[r]{\scriptsize ${\cal E}^t$ represents the set of edges in the second sub-network}
	\For{$e=\{u,v\} \in {\cal E}$}{
		\If{$u,v \in {\cal V}^s$}{
			${\cal E}^s \leftarrow {\cal E}^s  \cup e$ \Comment*[r]{\scriptsize Add edge $e$ directly if both $u$ and $v$ lie in first sub-network}
		}
		\If{$u,v \in {\cal V}^t$}{
			${\cal E}^t \leftarrow {\cal E}^t  \cup e$ \Comment*[r]{\scriptsize Add edge $e$ directly if both $u$ and $v$ lie in second sub-network}
		}
		\If{$(u \in {\cal V}^s) \land (v \in {\cal V}^t)  $}{
			${\cal E}^s \leftarrow {\cal E}^s  \cup \{u,\hat{t}\}$ \Comment*[r]{\scriptsize Create an edge between $u$ and $\hat{t}$ in first sub-network}
			${\cal E}^t \leftarrow {\cal E}^t  \cup \{\hat{s},v\}$ \Comment*[r]{\scriptsize Create an edge between $\hat{s}$ and $v$ in  second sub-network}
		}		
	}
	\Return $G^s = <{\cal V}^s, {\cal E}^s>, G^t = <{\cal V}^t, {\cal E}^t>$\\
	\caption{{\sc NetworkPartitioning}($G = <{\cal V}, {\cal E}>$)}
	\label{algo3}
\end{algorithm}}
\begin{exmp}\label{exp7}
Figure~\eqref{fig:partition} illustrates our random network partitioning approach on a small synthetic network. The network has 8 nodes and 16 edges. The numbers associated with each directed edge represent the corresponding flow and capacity values. The blue colored nodes represent the chosen nodes for the first sub-network and other nodes are kept in the second sub-network. We create an artificial terminal node `$d$' for the first sub-network and an artificial source node `$s$' for the second sub-network. Finally, we replace each edge that connects the two sub-networks with two artificial edges (shown in red color), one sinks to the artificial terminal node and other one originates from the artificial source node. If multiple edges share the same source and destination node, we replace them with a single artificial edge with cumulative flow and capacity values.
\end{exmp}

Algorithm~\ref{algo4} describes the key steps for the network partitioning based heuristic approach. Let $\mu^{\cal E}$ denote the set of potential candidate edges, which is initialized as an empty set. In each iteration, we randomly partition the network into two sub-networks (i.e., $G^s, G^t$), compute the modified initial flows for two sub-networks (i.e., $\bar{x}^s, \bar{x}^t$), and solve the adversary's decision model for both the sub-networks independently with a budget of $\frac{\Gamma}{2}$ for each sub-problem. Then we insert the resulting attacked edges from both the sub-problems to the candidate edge set $\mu^{\cal E}$. This iterative process continues until a predetermined number of iterations is completed or the cardinality of the candidate edge set reaches a given threshold value $\Delta$. Finally, we solve the adversary's optimization problem from Table~\ref{alg:cadv} to identify the best $\Gamma$ edges to attack from the candidate edges $\mu^{\cal E}$. Specifically, we manually set the value of $\mu_e$ to 0 if the edge $e$ does not belong to the candidate edge set (i.e., $e\notin \mu^{\cal E})$. This problem is computationally less expensive as the search space reduces from $|{\cal E}|$ to $|\mu^{\cal E}| (\ll |{\cal E}|)$. 

It should be noted that, in case of extensively large networks, even the sub-problems might become intractable if we partition the network into two sub-networks. In such scenarios, our approach can be extended to recursively partition the sub-networks into even smaller networks and then the decision problem can be solved independently for all the small networks to compute the set of candidate edges. In addition, as the network size increases, the value of the threshold parameter $\Delta$ needs to be reduced accordingly for solving the final decision problem over the candidate edges.

{\centering
\begin{algorithm}[!htb]
	\textbf{Initialize: } $\mu^{\cal E} \leftarrow \{\}; it \leftarrow 0$\Comment*[r]{\scriptsize Initialize candidate edge set $\mu^{\cal E}$ as an empty set}
	\Repeat{$(|\mu^{\cal E}| \geq \Delta) \lor (it \geq M)$}{
		$it \leftarrow it +1$ \;
		$\{G^s, G^t\} \leftarrow$ {\sc NetworkPartitioning($G$)} \Comment*[r]{\scriptsize Randomly partition the network}
		$\mu^s \leftarrow $ {\sc AdversaryProblem}($G^s,U, {p}, {\bar{x}^s},\frac{\Gamma}{2}$) \Comment*[r]{\scriptsize Solve first sub-network problem} 
		$\mu^t \leftarrow $ {\sc AdversaryProblem}($G^t,U, {p}, {\bar{x}^t},\frac{\Gamma}{2}$) \Comment*[r]{\scriptsize Solve second sub-network problem} 
		$\mu^{\cal E} \leftarrow \mu^{\cal E} \cup \mu^s \cup \mu^t$ \Comment*[r]{\scriptsize Add the new set of potential edges to $\mu^{\cal E}$}
	}
	${\mu} \leftarrow $ {\sc AdversaryProblem}($\{G, \mu^{\cal E} \},U, {p}, {\bar{x}},\Gamma$) \Comment*[r]{\scriptsize  Solve the problem over $\mu^{\cal E}$ edges}
	\Return ${\mu}$\\
	\caption{{\sc PartitioningHeuristic}($G = <{\cal V}, {\cal E}>, U, {p}, {\bar{x}}, \Gamma$)}
	\label{algo4}
\end{algorithm}}

\section{Empirical Results} \label{allResults}
In this section, we demonstrate the performance of our two-player iterative game approach on a set of synthetic and real-world datasets. We perform experiments on a 3.4 GHz Intel Core i7 machine with 16GB DDR3 RAM and all the linear optimization models are solved using IBM ILOG CPLEX optimization Studio V12.7.1. \cref{smallResults} provides the performance analysis of our optimal solution strategy from \cref{smallSol} on a wide range of small scale problem instances. \cref{largeResults} presents the performance of our approach on a set of large scale problem instances where the adversary's decision problem in each iteration is solved using heuristics from \cref{heuristics} for computational efficiency.
We compare the performance of our approach against the following four well-known state-of-the-art benchmark approaches: 
\begin{enumerate}
\item Max-Flow solution (\textbf{MF}): In this approach, we assume that the administrator is not aware of any attacks and sends the optimal (i.e., \emph{max-flow min-cost} solution) flow through the network. 
\item One step planning (\textbf{OSP}): This is a one-stage game model where the administrator first computes a \emph{max-flow min-cost} solution, then the adversary identifies an optimal attack to disrupt that solution and finally, the administrator finds an optimal flow solution for the network which is damaged according to the attack revealed by the adversary. This one-stage game model is applicable to the settings where the administrator takes a myopic view and assumes that the attacker cannot observe the flow sent by the administrator. Therefore, from the administrator's perspective, the best policy for the adversary is to attack the initial \emph{Max-Flow} solution which can be computed if the network structure and edge capacities are known to her. 
\item Robust flow solution (\textbf{RF}): For this approach, we compute a robust flow solution where the entire flow of an attacked edge is assumed to be lost. \cite{bertsimas2013robust} propose an optimization model for computing a robust flow solution by assuming that a maximum of $\Gamma_v$ incoming edges to node $v$ can fail. We modify the optimization model to ensure that a maximum of $\Gamma$ edges in the network can be attacked. The details of the modified optimization model is provided in \ref{sec:appendixA}.  
\item Approximate adaptive maximum flow solution (\textbf{AAMF}): For this approach, our goal is to compute an adaptive maximum flow solution where the flow can be adjusted after the edge failure occurred. We employ an optimization model from \cite{bertsimas2013robust} which provides an approximate solution to the adaptive maximum flow problem. The details of the approximate adaptive maximum flow solution is provided in \ref{sec:appendixB}. 
\end{enumerate}
Finally, we refer to our iterative game approach as robust and adaptive maximum flow (\textbf{RAMF}) solution. To ensure fairness in comparison, we assume that the adversary always acts as a follower and optimally disrupts the resulting flow solution for all the five approaches\footnote{For all the experiments, we set the modified capacity of an edge, $m_e$ to 0 if the edge is attacked by the adversary.}.
Since our goal is to optimize the network flow solution under adversarial conditions, we consider two crucial and complementary performance metrics for comparison: 
\begin{itemize}
\item \textbf{Objective} -- The objective value of the administrator, which we want to maximize, is computed as the difference between the amount of flow pushed to the terminal node and the total cost of routing and rerouting the flow through the network; and 
\item \textbf{Lost flow} -- The difference between the flow sent from the source node and the amount of flow reached to the terminal node. The amount of lost flow is an important indicator in many practical applications, e.g., these can correspond to congestion in case of urban transportation or introduce additional pressure in pipes in case of crude oil distribution application.
\end{itemize}
The trade-off between the objective value and the amount of lost flow can be captured using the routing cost parameter. If the routing cost is very high, then the optimal solution is to send zero flow through the network. On the other hand, if the routing cost is 0, then the optimal solution can have a large amount of lost flow. Therefore, we generate the routing costs randomly from a given range and report both the objective value and the amount of lost flow as performance metrics in the experiments.

\subsection{Empirical Results on Small-scale Data Sets}\label{smallResults}
In this section, we provide the following key comparison results of our \emph{RAMF} approach with the four benchmark approaches on small problem instances:
\begin{enumerate}
\item Sensitivity results with respect to lost flow and objective value on a set of synthetic networks by varying three tunable input parameters: (a) the number of nodes; (b) the edge density which controls the number of edges; and (c) budget value of the adversary, $\Gamma$.
\item The runtime analysis and convergence results of our \emph{RAMF} approach.
\item Performance of our proposed heuristic approaches against the optimal solution.
\item Behavioral insights from the strategies generated by the administrator and by the adversary.
\item Performance analysis on a real-world benchmark data set called SNDlib \citep{orlowski2010sndlib}.
\end{enumerate}

\subsubsection{Sensitivity Results over Different Settings of Input Parameters} \label{smallResultsSensitive}
We now demonstrate the performance comparison on a set of synthetic problem instances. We generate random connected directed networks by varying the number of nodes and the edge density. The capacities of the edges are drawn randomly from the range of 1 to 20 and the unit costs for transporting flows, $p$ are generated randomly from the range of 0.01 to 0.1. In the default setting of experiments, we use networks with 20 nodes and 0.8 edge density, and the adversary's budget value is set to 5. For each input setting, we generate 5 random problem instances and report the average lost demand and percentage gain in the objective value. 
Let $U^{max}$ denote the maximum capacity of an edge (i.e., $U^{max} = \max_e U_e $)\footnote{As the capacities are drawn randomly from the range of 1 to 20, we set the value of $U^{max}$ to 20.}. Then, the maximum loss in the objective value due to adversarial attacks is bounded by $(U^{max}\times \Gamma)$. So, the percentage gain of our approach against a benchmark approach (e.g., \emph{MF}) is computed as the ratio between the difference in objectives and the upper bound on the marginal gain. 
\begin{align}
\text{\%Gain over \emph{MF}} \!=\! \frac{(\text{Obj of \emph{RAMF} - Obj of \emph{MF}})\!\times \! 100}{U^{max}\times \Gamma}  \label{percentGain}
\end{align}

\begin{figure}[!htb]
	\centering
	\begin{subfigure}{0.325\textwidth}
		\includegraphics[width=\textwidth]{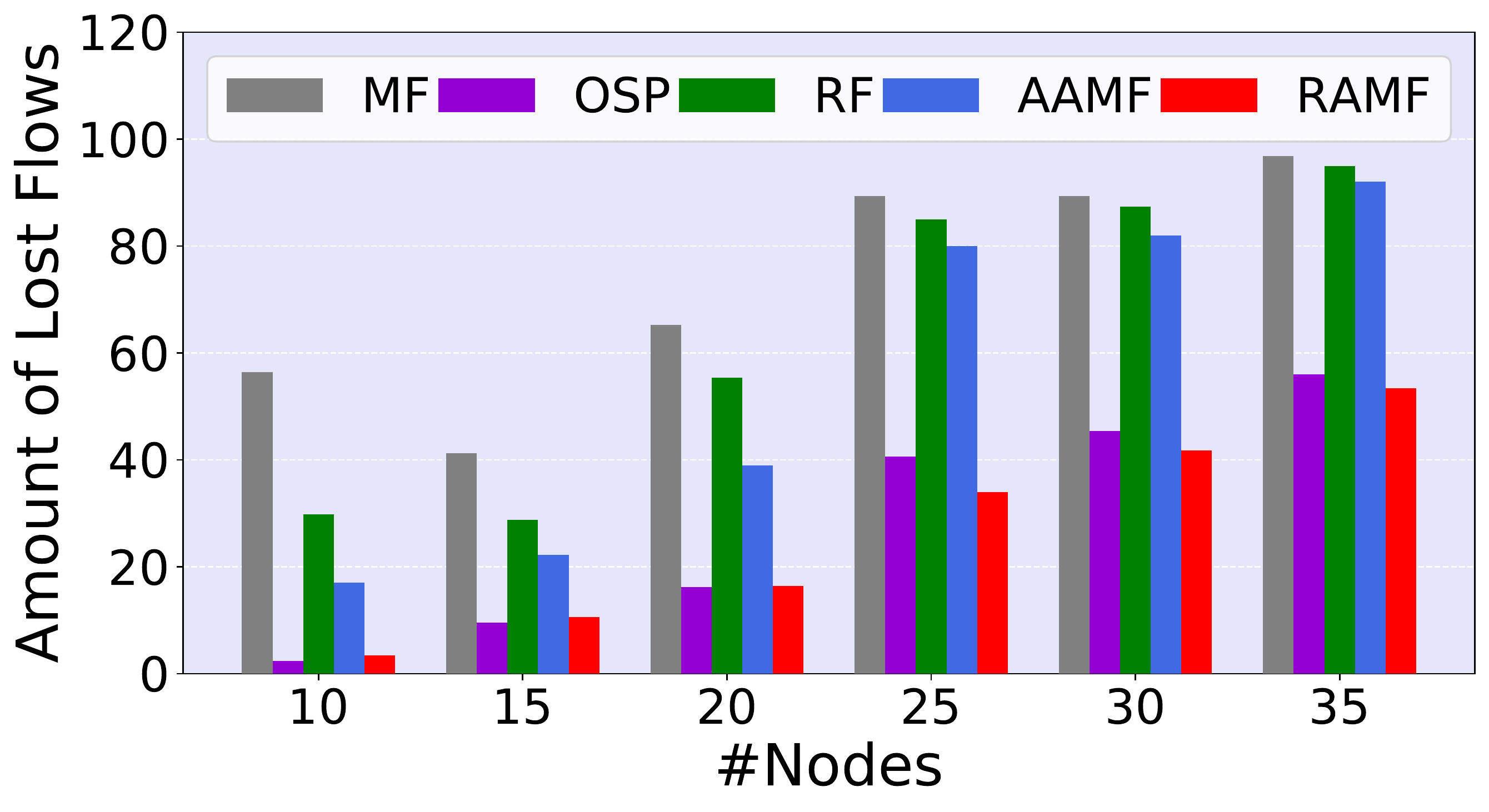} \caption{}
	\end{subfigure} 
	\begin{subfigure}{0.325\textwidth}
		\includegraphics[width=\textwidth]{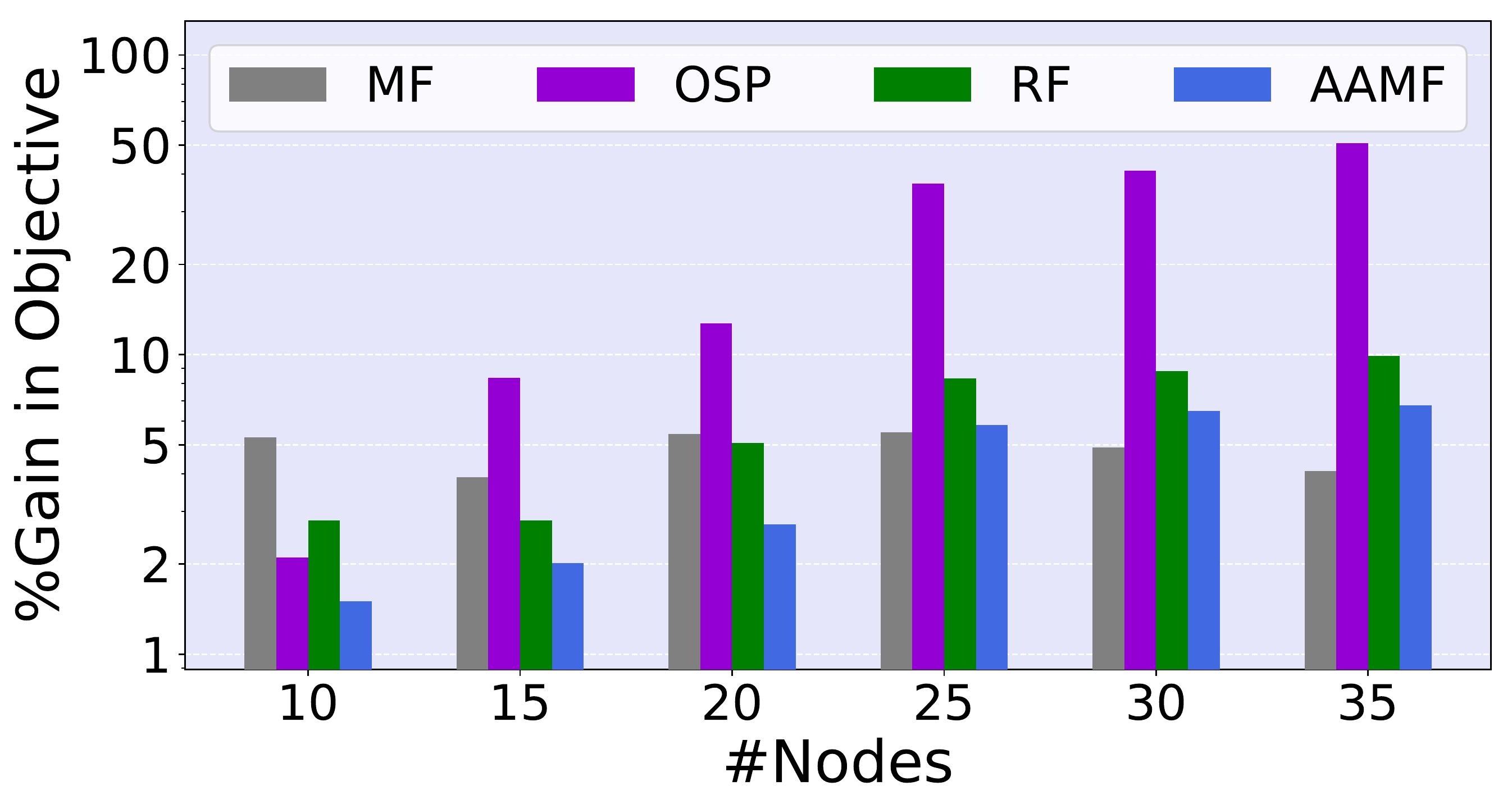} \caption{}
	\end{subfigure} 
	\begin{subfigure}{0.325\textwidth}
		\includegraphics[width=\textwidth]{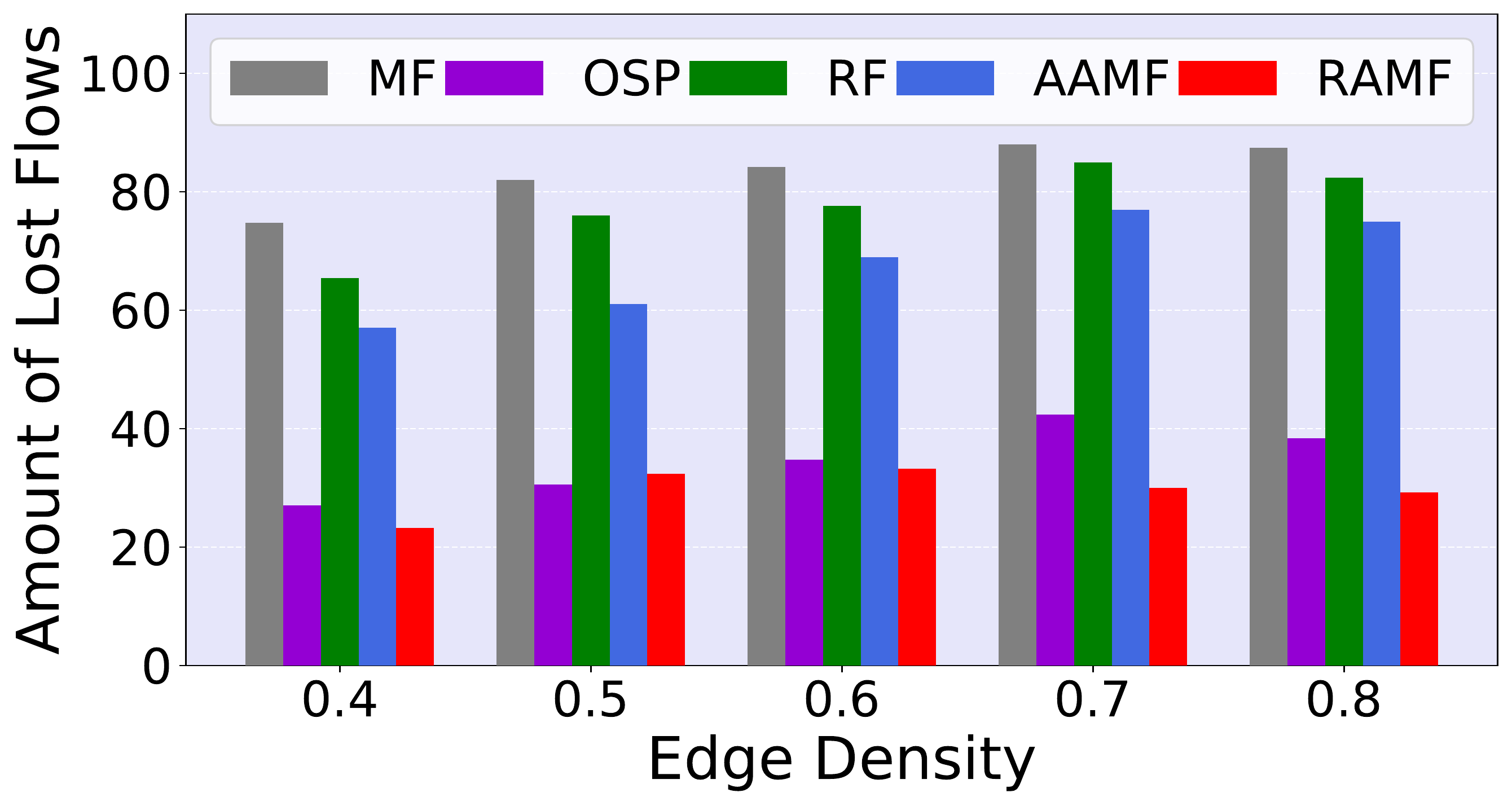} \caption{}
	\end{subfigure} 
	\begin{subfigure}{0.325\textwidth}
		\includegraphics[width=\textwidth]{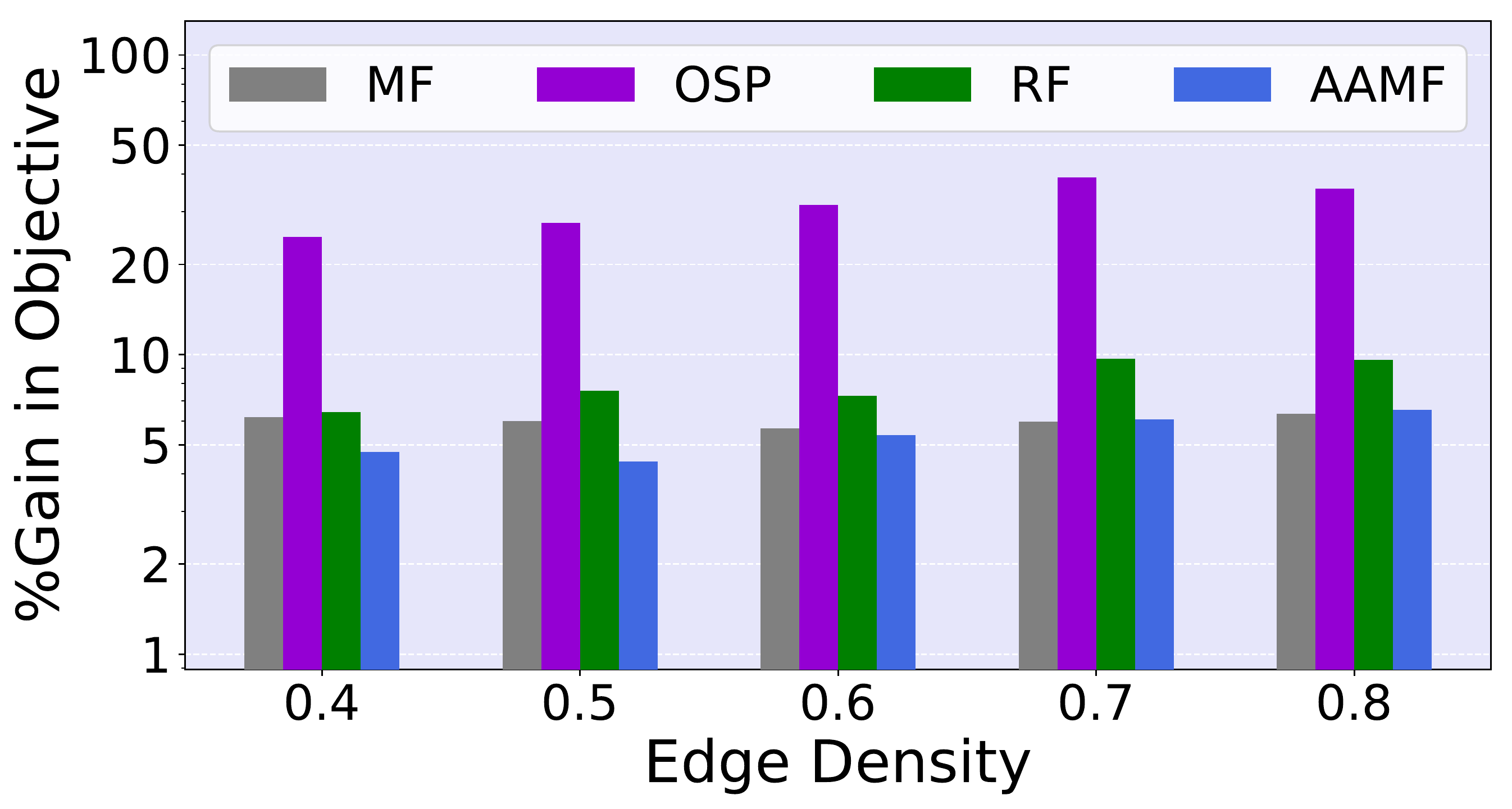} \caption{}
	\end{subfigure} 
	\begin{subfigure}{0.325\textwidth}
		\includegraphics[width=\textwidth]{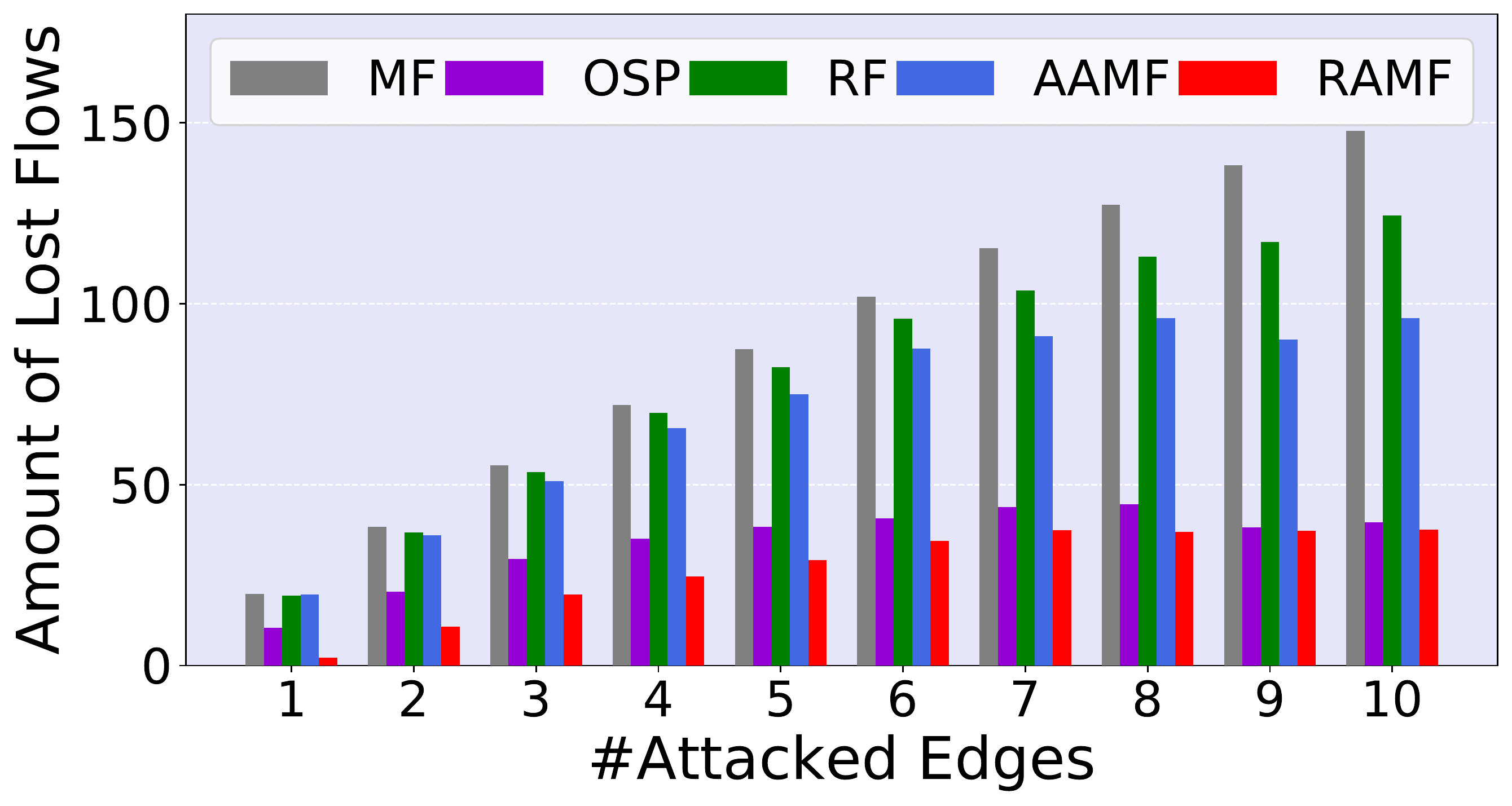} \caption{}
	\end{subfigure} 
	\begin{subfigure}{0.325\textwidth}
		\includegraphics[width=\textwidth]{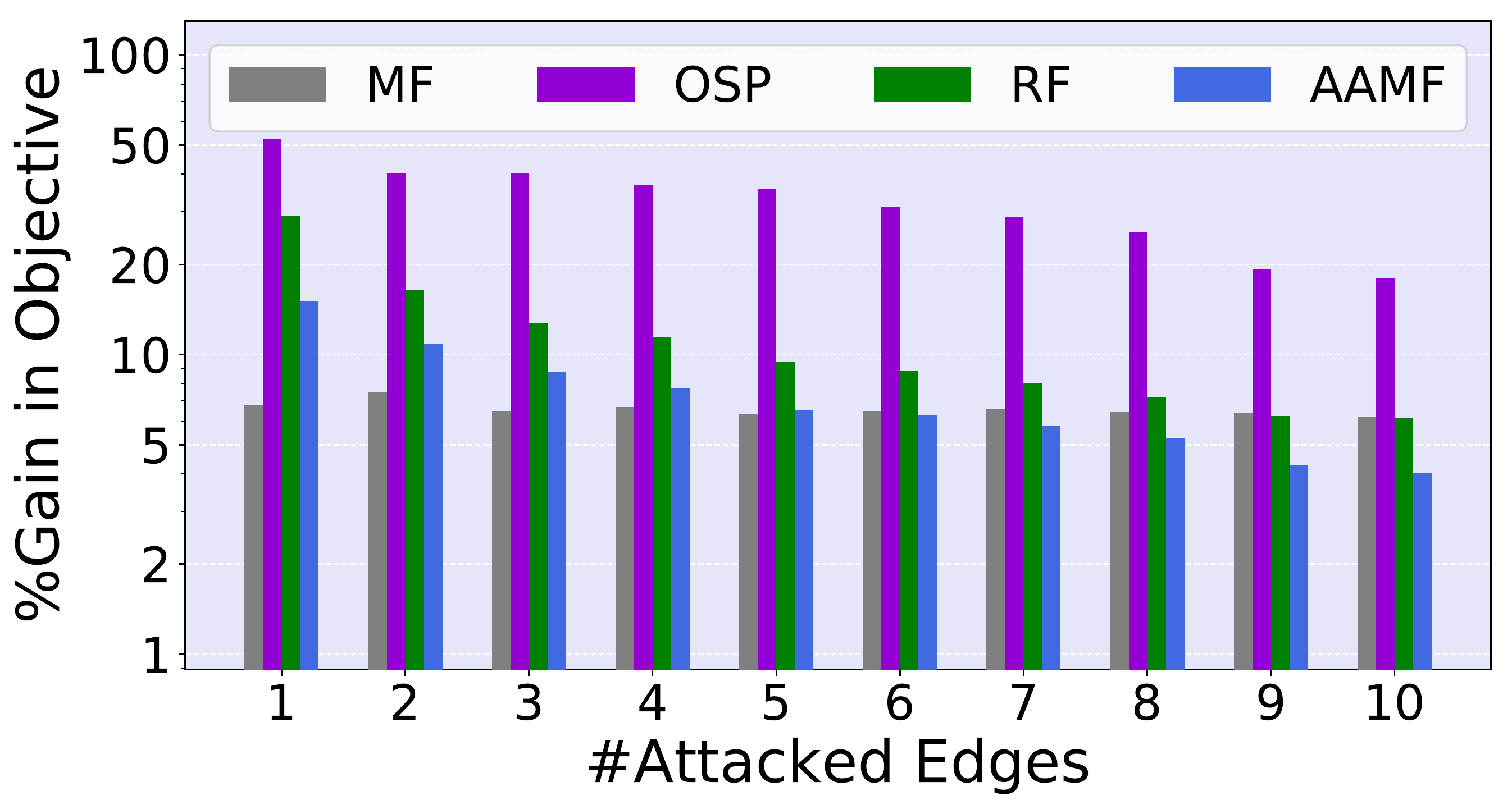} \caption{}
	\end{subfigure} 
	 \caption{Effect of number of nodes on (a) Lost flow and (b) Objective value; Effect of number of edges on (c) Lost flow and (d) Objective value; Effect of the value of $\Gamma$ on (e) Lost flow and (f) Objective value.}
	 \label{fig:smallResult}
	 \vspace{0.1in}
\end{figure}

Figure~\ref{fig:smallResult}(a) shows the net amount of lost flow due to attacks, where we vary the number of nodes in the X-axis. As expected, the amount of lost flow is significantly high for the \emph{MF} approach, as it ignores the adversary's attacking behavior. The amount of lost flows for \emph{RF} and \emph{AAMF} are also significantly higher than our \emph{RAMF} approach. The \emph{OSP }approach is more conservative and therefore, the lost flow for the \emph{OSP} is relatively lower. The lost flow for our \emph{RAMF} approach is almost always lower than all the four benchmarks. 
Figure~\ref{fig:smallResult}(b) delineates the percentage gain in the objective value of the administrator in a logarithmic scale. Our approach always outperforms all the four benchmark approaches. The average percentage gains in the objective value for our approach over \emph{MF}, \emph{OSP}, \emph{RF} and \emph{AAMF} are 4.8\%, 25.4\%, 6.3\%  and 4.2\%, respectively. 

Figure~\ref{fig:smallResult}(c) shows the net amount of lost flow for all the approaches, where we vary the edge density from 0.4 to 0.8 in the X-axis. We observe a consistent pattern that the lost flows for \emph{MF}, \emph{RF}, \emph{AAMF} approaches are at least two times higher than \emph{RAMF} approach in all the cases. While the lost flow for the \emph{OSP} approach is almost similar for edge density 0.4 to 0.6, our \emph{RAMF} approach performs better when the number of edges increases. 
Figure~\ref{fig:smallResult}(d) delineates the percentage gain in the objective value in a logarithmic scale. The gap between objectives and the gain for our approach remain consistent in all the settings. As expected, \emph{AAMF} always outperforms \emph{RF} approach due to the flow adjustment. On an average, the percentage gains in objective for our approach over \emph{MF}, \emph{OSP}, \emph{RF} and \emph{AAMF} approaches are 6\%, 31.8\%, 8.1\% and 5.4\%, respectively.

Figure~\ref{fig:smallResult}(e) exhibits net amount of lost flow for all the approaches, where we vary the adversary's budget from 1 to 10 in the X-axis. The amount of lost flow for \emph{MF} approach increases monotonically from 20 to 150 as we increase the adversary's budget, whereas the lost flow for the \emph{RAMF} approach is always bounded by 40. Moreover, we observe that the number of lost flow for our \emph{RAMF} approach remains consistent when the value of $\Gamma$ goes beyond 7. Such sensitivity analysis results can be used for initial estimation of the adversary's budget value.
Figure~\ref{fig:smallResult}(f) demonstrates the percentage gains in the objective value. As the upper bound on marginal gain (i.e., $U^{max}\times \Gamma$) increases with adversary's budget, the percentage gain in objective over \emph{OSP}, \emph{RF} and \emph{AAMF} approaches reduces monotonically with the value of $\Gamma$. Although the gain of our approach over the \emph{MF} approach remains consistent for different values $\Gamma$, the net difference between the objectives increases monotonically. Therefore, these results clearly indicate that the performance of our approach improves gradually if the adversary becomes stronger. 

%In a nutshell, we observe that the \emph{RAMF} approach reduces the lost flow significantly over \emph{MF}, \emph{RF} and \emph{AAMF} approaches. On the other hand, the percentage gains in the objective value for the \emph{RAMF} approach over all the benchmarks are always positive and the gain is significantly high over the \emph{OSP} approach. Therefore, we can conclude that among five approaches, only our \emph{RAMF} approach is able to maintain the right trade-off between the two performance metrics. 

\subsubsection{Convergence Results} \label{convergenceSmall}
\begin{figure}[!htb]
	\centering
	\begin{subfigure}{0.31\textwidth}
		\includegraphics[width=\textwidth]{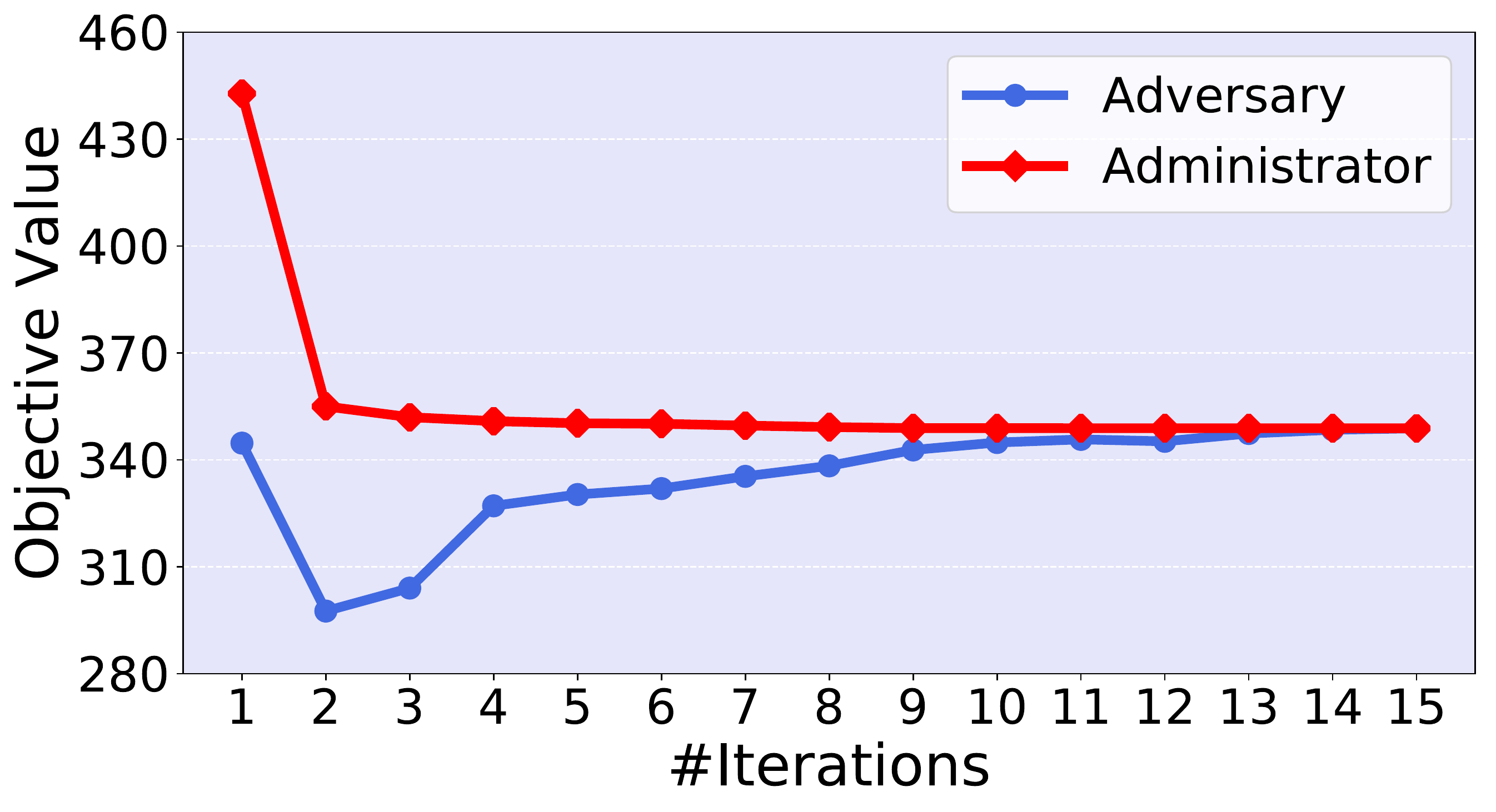} \caption{}
	\end{subfigure} 
	\begin{subfigure}{0.31\textwidth}
		\includegraphics[width=\textwidth]{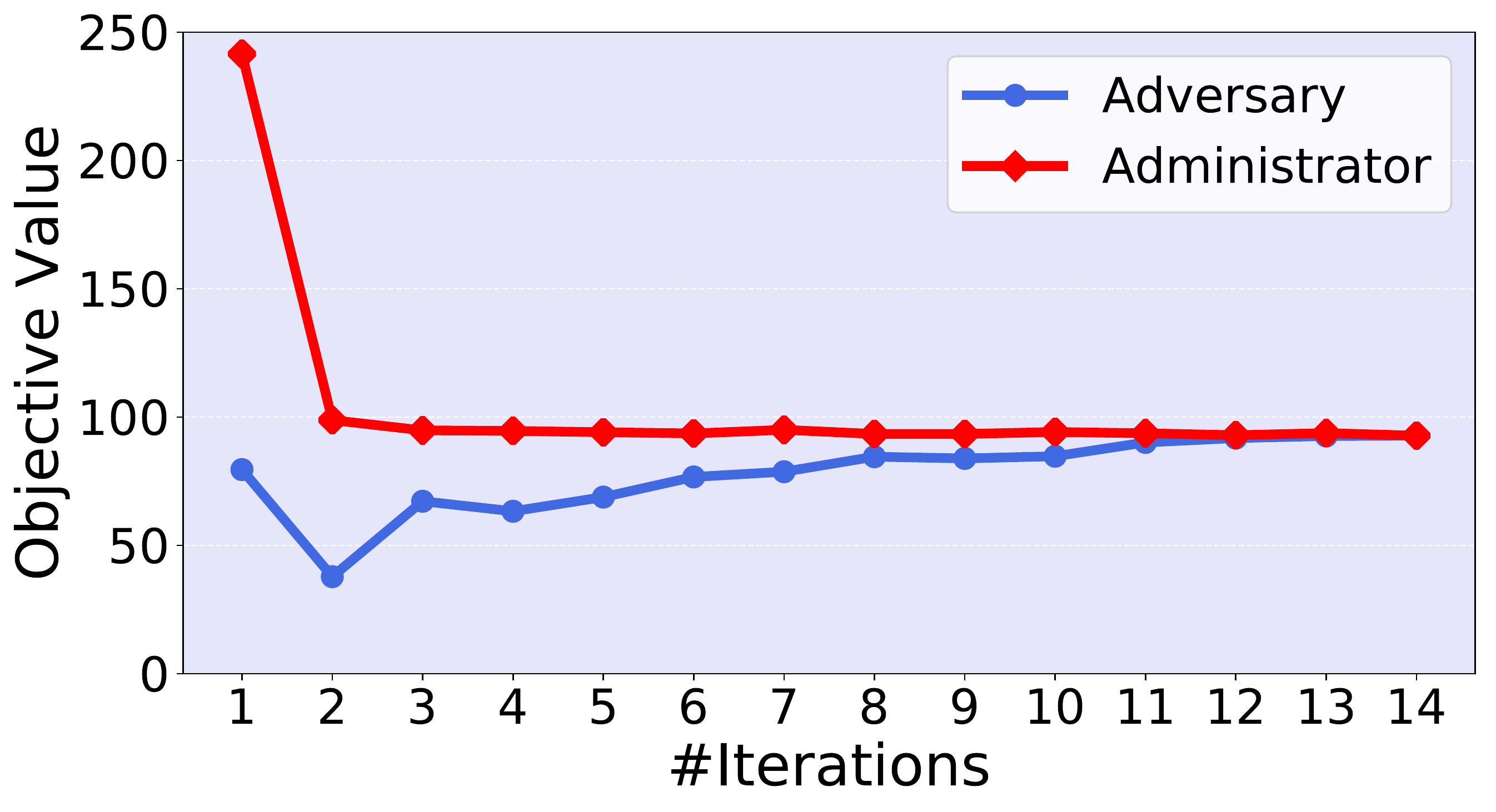} \caption{}
	\end{subfigure} 
	\begin{subfigure}{0.35\textwidth}
		\includegraphics[width=\textwidth]{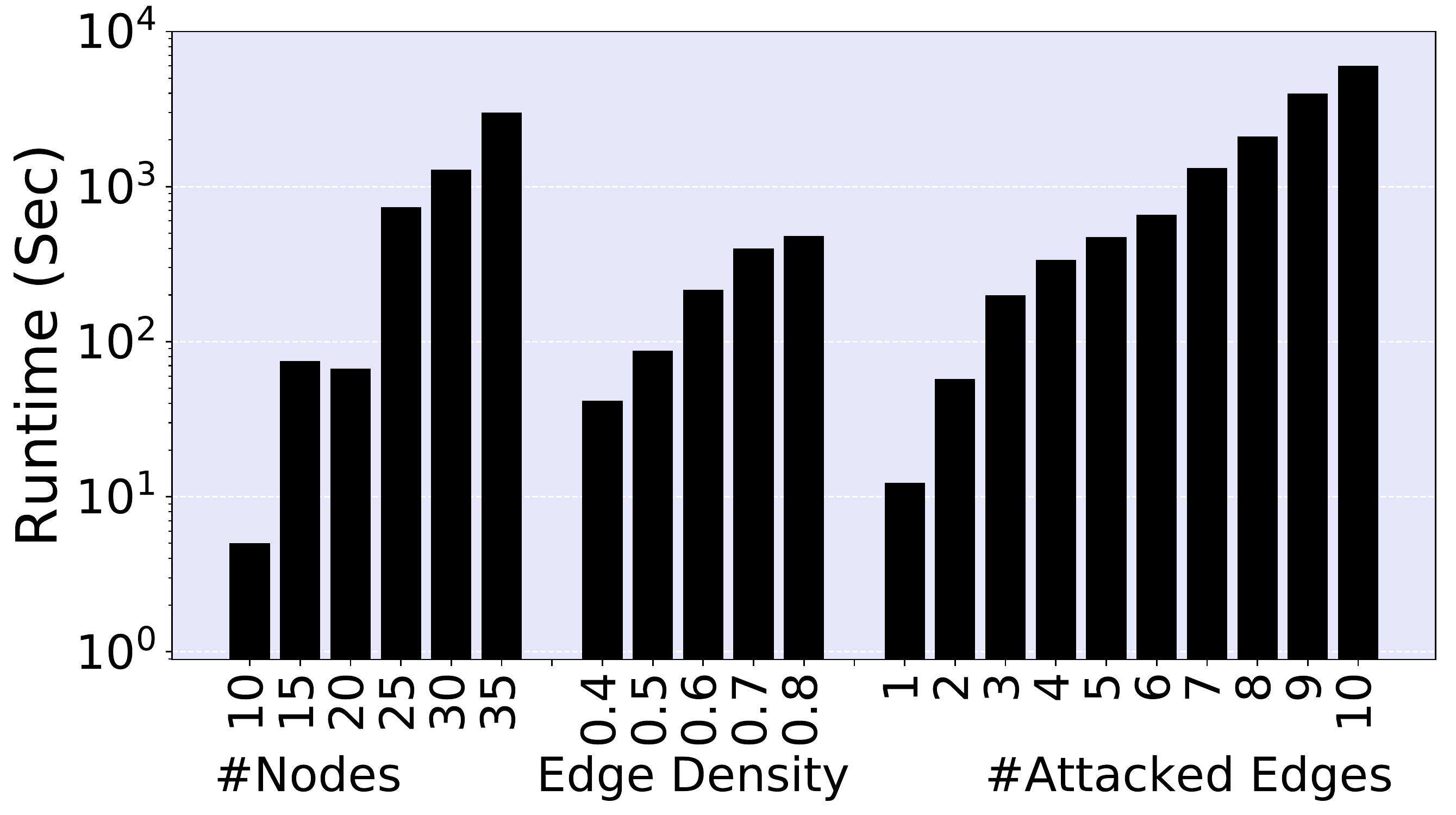} \caption{}
	\end{subfigure} 
	 %\vspace{-0.1in}
	 \caption{ Convergence of the proposed iterative game on problems with (a) 35 nodes and (b) $\Gamma$=10; (c) Runtime results for varying number of nodes, edges and adversary's budget.}
	 \label{fig:smallConvergence}
	 %\vspace{-0.1in}
\end{figure}

Figure~\ref{fig:smallConvergence}(a) shows the convergence of our proposed iterative game on a problem with 35 nodes and edge density 0.4, where the adversary's budget is set to 5. The X-axis represents the iteration number of the game and the Y-axis denotes the objective value obtained by the players. As expected, the objective for the administrator reduces monotonically over the iterations. As the administrator generates more conservative solution over the iterations, the adversary's ability to disrupt the flow strategy reduces. The objective values converge to 349 after 15 iterations. So, we can claim that the administrator's objective will at least be 349 (for any attacks from $\Psi$) if the resulting robust flow strategy is executed. Figure~\ref{fig:smallConvergence}(b) delineates objectives of the players in each iteration of the game on another problem instance with 20 node, edge density 0.8 and adversary's budget as 10, which converges after 14 iterations. We experimentally observe that the game converges within 20 iterations for all the other problem instances.

\subsubsection{Runtime Performance} \label{runtimeSmall}
As all the benchmark approaches provide solution from single-step optimization model, the runtimes for them are always lower than our iterative \emph{RAMF} approach. Therefore, we only demonstrate the runtime performance of our \emph{RAMF} approach for different settings of network and adversary's budget value.
Figure~\ref{fig:smallConvergence}(c) presents the runtime for the \emph{RAMF} approach in seconds in a logarithmic scale. 
As shown clearly, the runtime increases monotonically as the size of the network (in terms of both the number of nodes and edges) increases. Furthermore, the quadratic optimization problem for the adversary from Table~\ref{alg:cadv} becomes more computationally expensive as we increase the value of $\Gamma$ and therefore, the runtime increases monotonically as the adversary becomes stronger. %As the computational complexity of our approach increases gradually with all the three input tunable parameters, we only show the performance on small-scale problems in this section. 

\subsubsection{Performance of Heuristic Approaches} \label{heuristicPerformanceSmall}
\begin{figure}[!htb]
	\centering
	\begin{subfigure}{0.48\textwidth}
		\includegraphics[width=\textwidth]{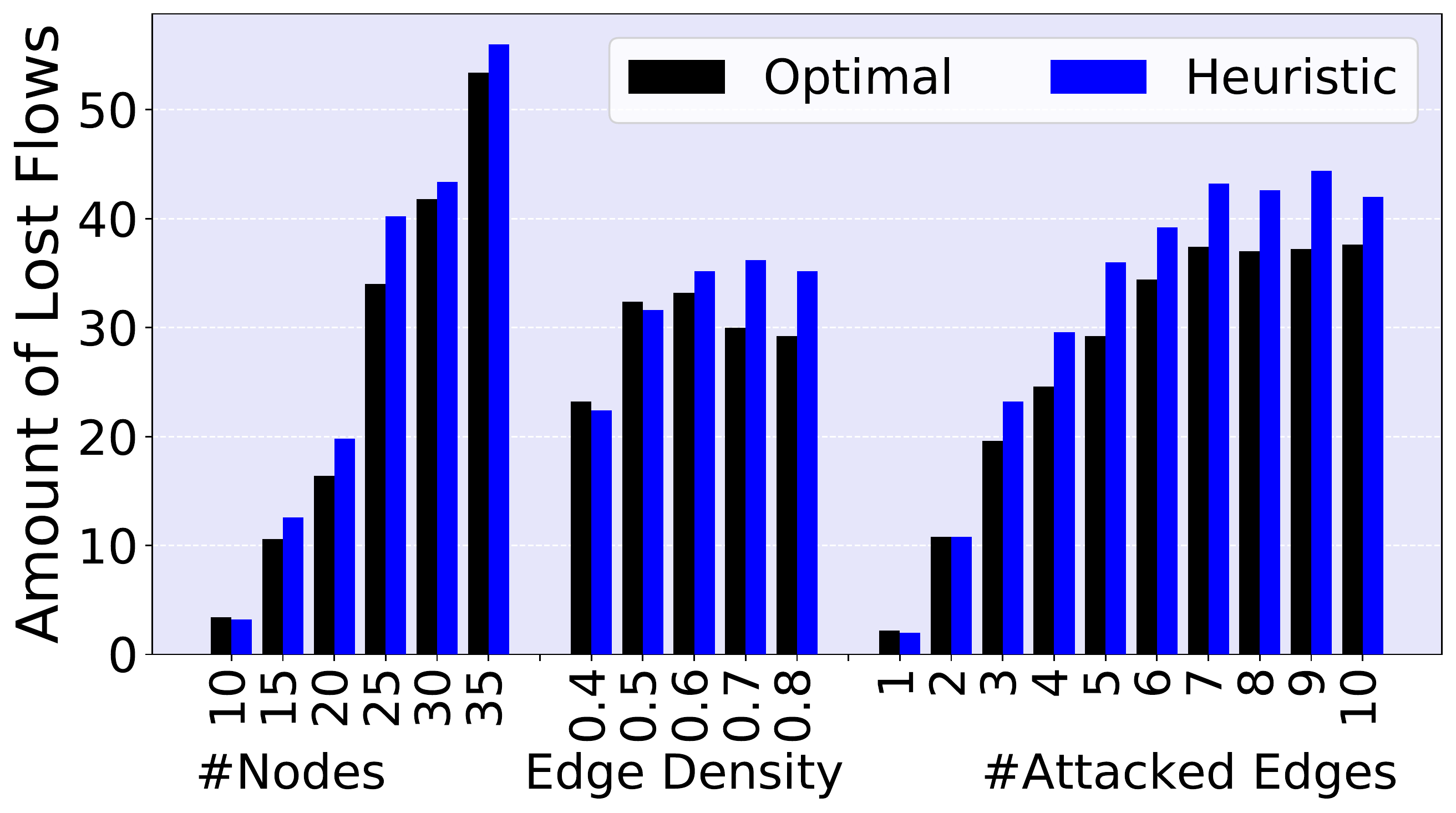} \caption{}
	\end{subfigure} 
	\hspace{0.15in}
	\begin{subfigure}{0.48\textwidth}
		\includegraphics[width=\textwidth]{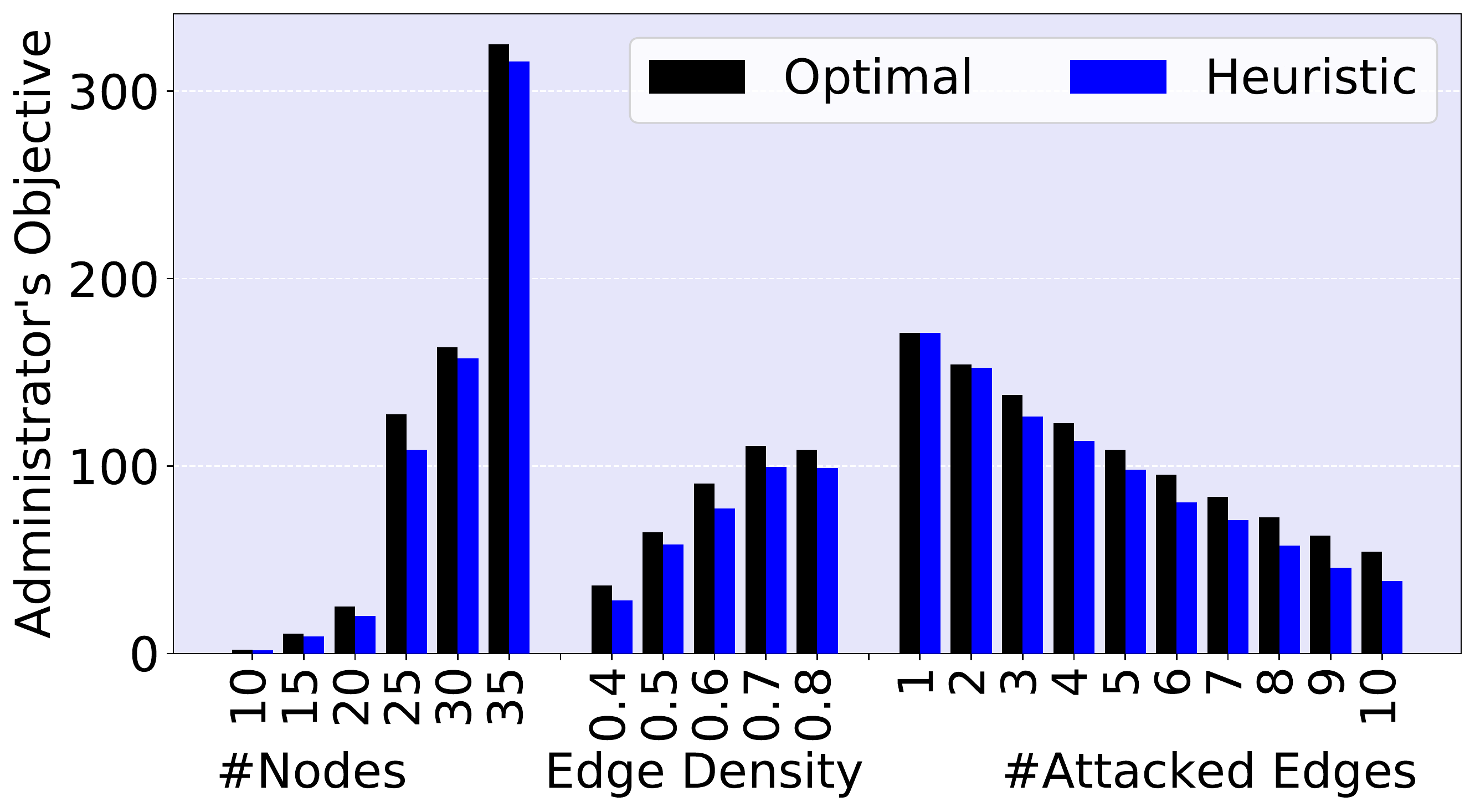}\caption{}
	\end{subfigure} 
	 \caption{ Performance comparison between the proposed heuristic approaches and optimal solution on (a) Lost flow and (b) Objective value. }
	 \label{fig:smallHeuristicGap}
\end{figure}

To assess the performance of two proposed heuristic approaches against the optimal solution on small-scale problem instances, we solve the adversary's decision problem using heuristics in each iteration of the game. In the last iteration, when we have the final flow solution, we employ the optimal attack from \cref{modelAdv} to disrupt the final flow and compare its performance with the optimal flow solution (where in each iteration of the game, the adversary executes her optimal attack) results presented in \cref{smallResultsSensitive}. Figure~\ref{fig:smallHeuristicGap} demonstrates the net objective value and lost flows comparison. On an average, the objective value for our heuristic based solution is 6.8\%, 9.7\% and 9.4\% far from the optimal solution for varying nodes, edges and adversary's budget value setting, respectively. 

\begin{figure}[!htb]
	\centering
	\begin{subfigure}{0.48\textwidth}
		\includegraphics[width=\textwidth]{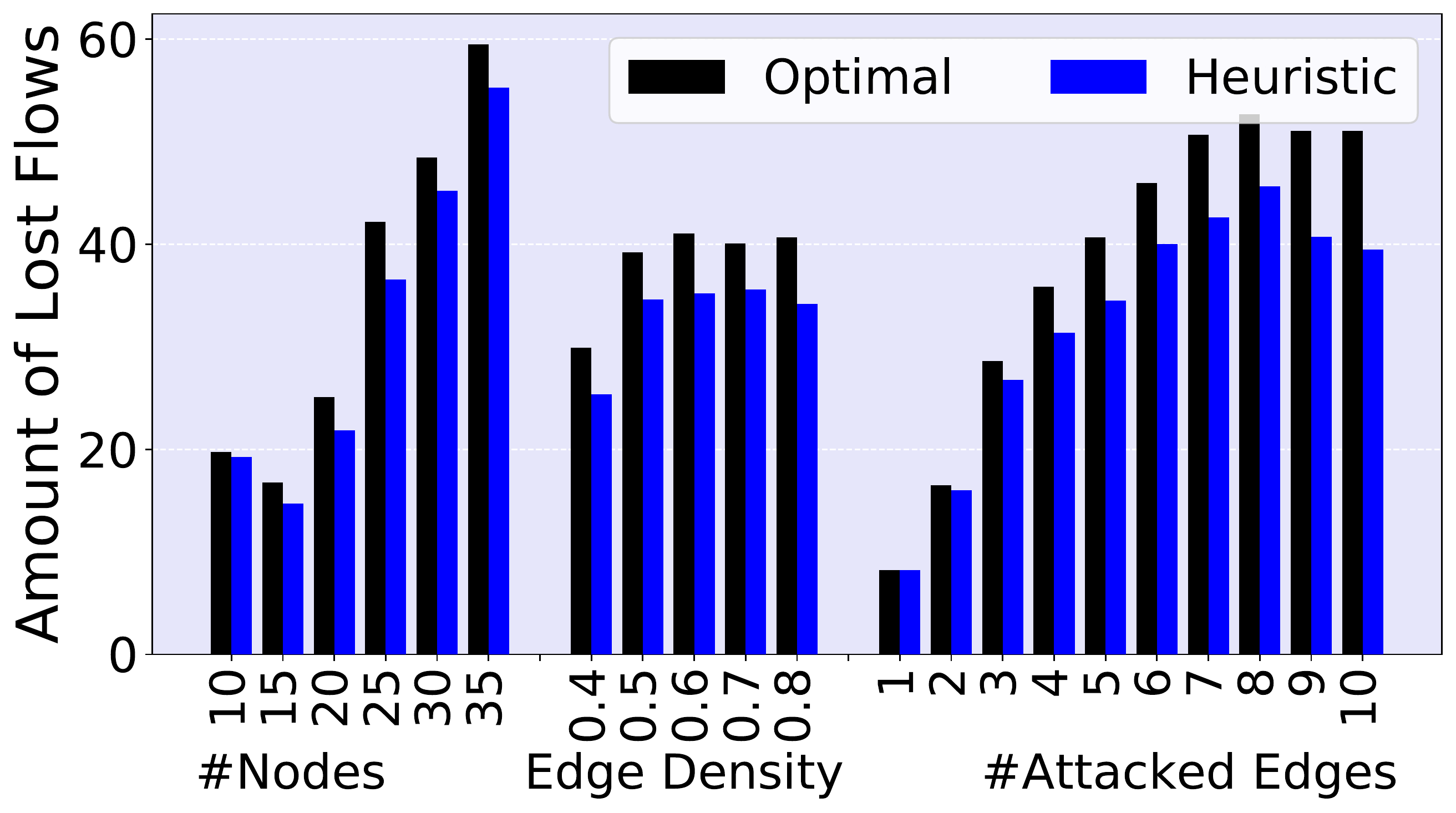} \caption{}
	\end{subfigure} 
	\hspace{0.15in}
	\begin{subfigure}{0.48\textwidth}
		\includegraphics[width=\textwidth]{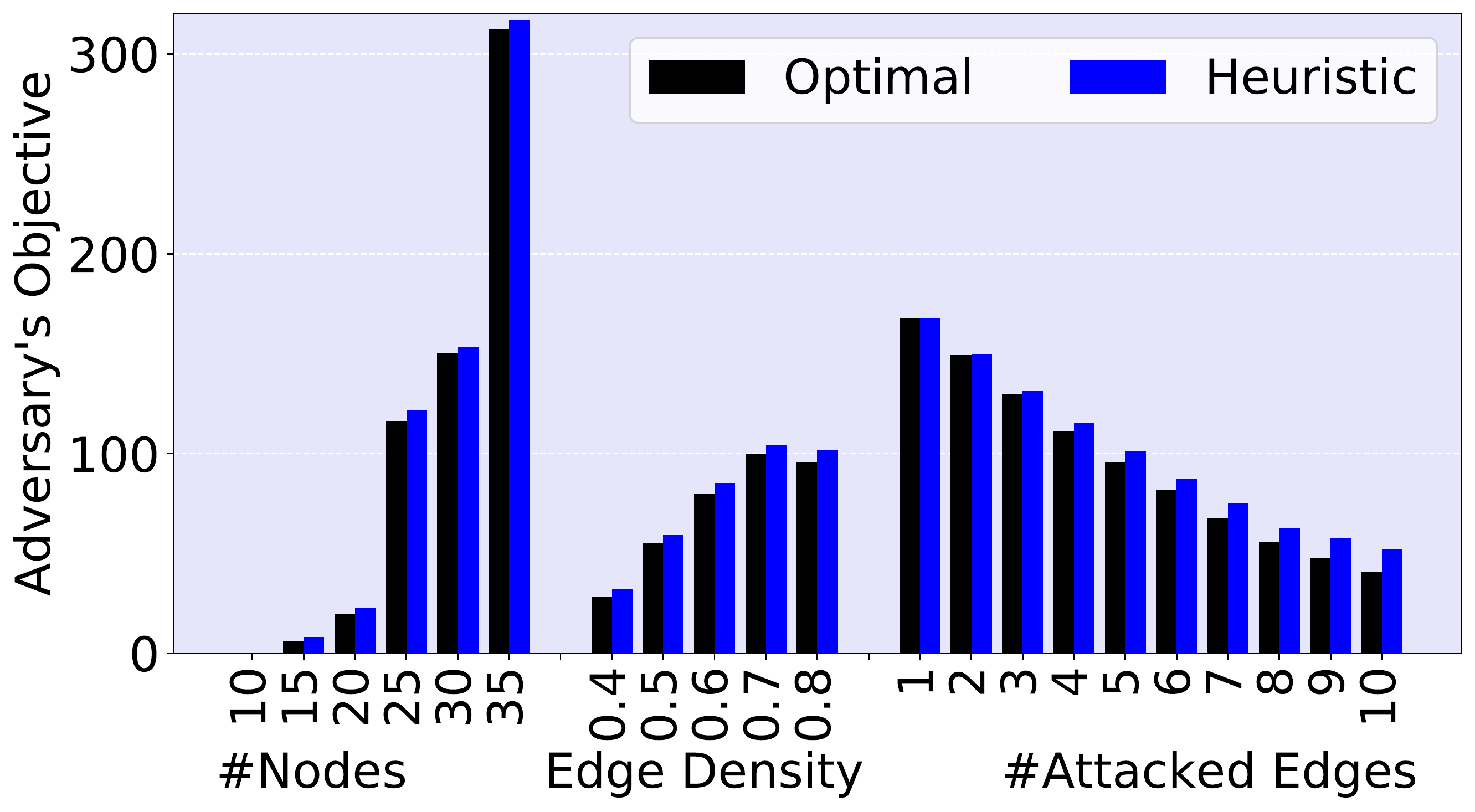}\caption{}
	\end{subfigure} 
	 \caption{ Optimality gap for the proposed heuristic approaches on (a) Lost flow and (b) Adversary's objective value. }
	 \label{fig:smallHeuristicGapOpt}
\end{figure}

We now present another set of experimental results to assess the optimality gap for the heuristic solutions of the adversary. For these experiments, in each iteration of the game, for a given flow solution by the administrator, the adversary computes and executes her optimal attack using the decision model from \cref{modelAdv}. In addition, for the same flow strategy given by the administrator, we compute the adversary's attack using both heuristic approaches and choose the solution with lower objective value. Figure~\ref{fig:smallHeuristicGapOpt} exhibits the average solution quality comparison of the heuristic approaches and optimal attacks in each iteration of the game. It should be noted that a lower objective value and higher amount of lost flows are better for the adversary. On an average, the adversary's objective values for heuristics are 3.2\%, 4.5\% and 5.1\% worse than the optimal solution for varying nodes, edges and adversary's budget value setting, respectively.

\subsubsection{Behavioral Insights for the Administrator and Adversary Strategies}
\begin{figure}[!htb]
	\centering
	\begin{subfigure}{0.48\textwidth}
		\includegraphics[width=\textwidth]{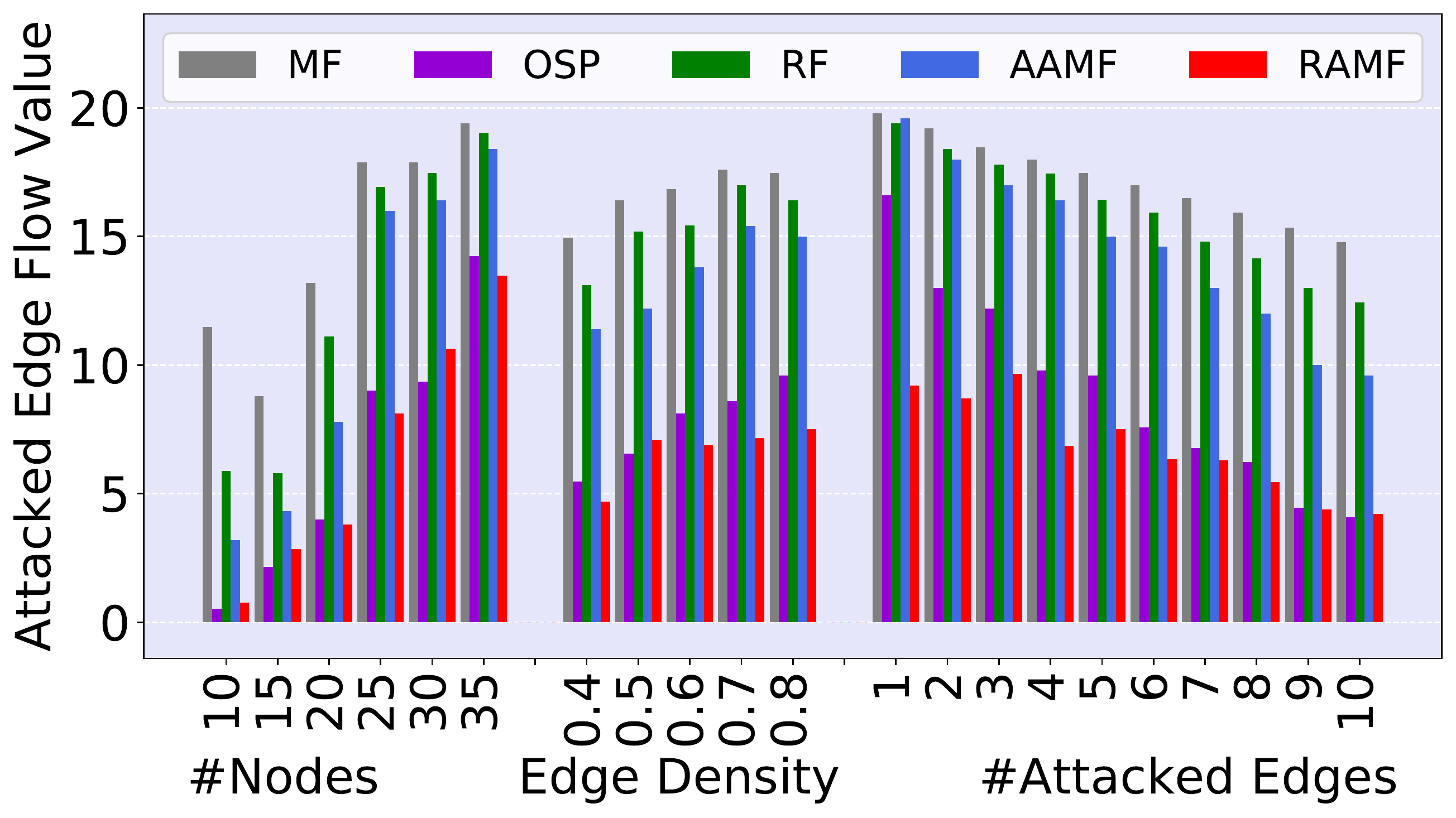} \caption{}
	\end{subfigure} 
	\hspace{0.15in}
	\begin{subfigure}{0.48\textwidth}
		\includegraphics[width=\textwidth]{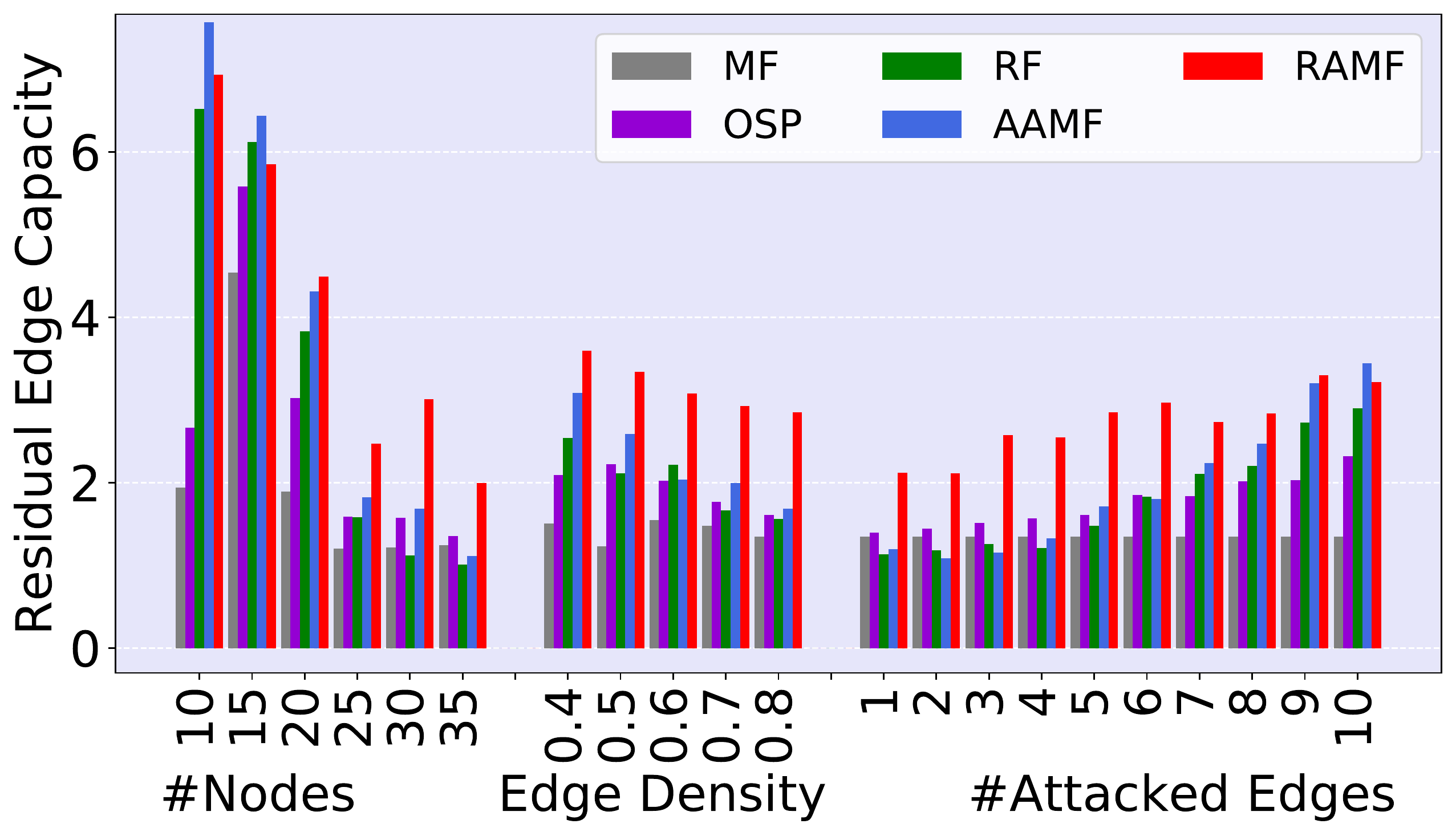}\caption{}
	\end{subfigure} 
	 \caption{ Behavioral insights from solutions: (a) Amount of flow assigned to edges attacked by the adversary's strategy; and (b) Amount of residual capacity allocated to intermediate edges by the Administrator's flow strategy. }
	 \label{fig:behaviorInsights}
\end{figure}

We now provide behavioral insights generated from the strategies of the adversary and the administrator for different approaches. Figure~\ref{fig:behaviorInsights}(a) demonstrates the average amount of initial flow assigned to edges that are being attacked by the adversary's strategy. We observe that MF, AAMF and RF solutions transmit higher amount of flow through critical high capacity edges and the adversary can gain significantly by attacking those critical edges. The OSP solution assigns a lower amount of flow in all the edges that leads to a significantly lower objective value. Our RAMF solution is able to find intelligent flow strategies that transmit high amount of flow through the edges, whose flow can be rerouted if attacked, and therefore, the adversary has more incentive in attacking edges with less flow than those important edges with higher flow. On an average, the flow value assigned to attacked edges by RAMF approach is 60\%, 8\%, 56\% and 50\% less than the MF, OSP, AAMF and RF approaches, respectively. 
To further investigate the nature of the flow solutions generated by the administrator, we compute the residual capacity for the intermediate edges (i.e., except for the edges connected to source or terminal node) in which the administrator has assigned a portion of flow. Figure~\ref{fig:behaviorInsights}(b) depicts the average residual capacity left by the administrator's initial flow solution for intermediate edges. In contrast to other benchmark approaches, our RAMF solutions diversify the flow assigned to intermediate edges to maximize the flow reaching the terminal node under nominal condition and leave enough residual capacity so as to reroute flows in case of adversarial attacks. 

\subsubsection{Performance analysis on SNDlib Data Set}\label{sec:SNDlib}
In this section, we provide performance comparison results on instances from the Survivable Network Design Library [SNDLib] \citep{orlowski2010sndlib}. SNDLib database consists of 26 problem instances. The capacities of edges, $U$ are directly taken from the database\footnote{For some instances, the capacities for all the edges are stated as 0 in the SNDlib database. For those instances, we set the capacities to randomly drawn values between 500 to 1000.}. The unit edge costs for routing flows, $p$ are randomly drawn from the range of 0.01 to 0.1. In the default setting of our experiments, we set the adversary's budget to 5. However, due to small number of edges in some instances, we observe that the flow reaching the terminal node is always 0 for all the approaches, if we allow the adversary to attack 5 edges. So, we reduce the adversary's budget accordingly for those instances.

\begin{table*}[!htb]
\small
\begin{center}
\begin{tabular}{>{}m{1.55cm} >{\centering}m{0.35cm} >{\centering}m{0.35cm} >{\centering}m{0.25cm} >{\centering}m{0.78cm} >{\centering}m{0.78cm} >{\centering}m{0.78cm} >{\centering}m{0.78cm} >{\centering}m{0.9cm} >{\centering}m{0.52cm} >{\centering}m{0.52cm} >{\centering}m{0.52cm} >{\centering}m{0.8cm} c}
\hline
\multicolumn{4}{>{\centering}m{3.5cm}|}{Problem Instance} & \multicolumn{5}{>{}m{4.5cm}|}{Number of Lost Flow} & \multicolumn{4}{>{\centering}m{3.5cm}|}{\%Gain in Objective \ of RAMF over} & \multicolumn{1}{>{\centering}m{1.0cm}}{Runtime (sec)} \\
\hline
{\footnotesize Name} & $|{\mathcal V}|$ & $|{\mathcal E}|$ & \multicolumn{1}{>{\centering}m{0.25cm}}{$\Gamma$} & {\footnotesize MF} & {\footnotesize OSP} & {\footnotesize RF} & {\footnotesize AAMF} & \multicolumn{1}{>{\centering}m{0.9cm}}{{\footnotesize RAMF}} & {\footnotesize MF} & {\footnotesize OSP} & {\footnotesize RF} & \multicolumn{1}{>{\centering}m{0.8cm}}{{\footnotesize AAMF}} & {\footnotesize RAMF} \\
\specialrule{.1em}{.05em}{.05em} 
{\small abilene} & 14 & 22 & 2 & 1828 & 1602 & 1796 & 1692 & 1602 & 2.2 & 65.4 & 1.9 & 0.9 & 0.2 \\
{\small atlanta} & 17 & 31 & 3 & 28711 & 5000 & 12000 & 12000 & 7966 & 3.9 & 5.8 & 1.8 & 1.3 & 3.6 \\
{\small brain} & 163 & 399 & 5 & 4512 & 4142 & 4512 & 4220 & 4142 & 1.3 & 67.6 & 9.7 & 2.9 & 65.7 \\
{\small cost266} & 39 & 80 & 5 & 4507 & 3545 & 4314 & 4050 & 3859 & 2.9 & 55.5 & 2.2 & 2.4 & 11.0 \\
{\small dfn-bwin} & 12 & 50 & 5 & 3657 & 1155 & 3626 & 3060 & 118 & 8.4 & 19.8 & 10.2 & 8.5 & 2.6 \\
{\small dfn-gwin} & 13 & 54 & 5 & 3013 & 1361 & 2750 & 2795 & 979 & 8.8 & 37.7 & 10  & 8.4 & 2.0 \\
{\small di-yuan} & 13 & 49 & 5 & 4653 & 1869 & 4377 & 3850 & 1803 & 9.7 & 33.2 & 9.7 & 7.1 & 1.2 \\
{\small france} & 27 & 60 & 5 & 4502 & 2912 & 4236 & 4340 & 3755 &2.5 & 42.7 & 2.5 & 3.1 & 1.3 \\
{\small geant} & 24 & 48 & 5 & 4633 & 3840 & 4467 & 4435 & 3984 & 1.8 & 64.3 & 3.5 & 2.7 & 2.3 \\
{\small germany50} & 52 & 109 & 5 & 4338 & 3989 & 4148 & 4130 & 4046 & 2.9 & 66.1 & 8.6 & 6.7 & 6.6 \\
{\small giul39} & 41 & 189 & 5 & 4710 & 1488 & 4447 & 4190 & 2118 & 10.5 & 24.3 & 33.7 & 13.0 & 3183 \\
{\small india35} & 37 & 97 & 5 & 4489 & 3421 & 4290 & 4195 & 3737 & 2.4 & 57.1 & 5.5 & 4.1 & 5.4 \\
{\small janos-us-ca} & 41 & 139 & 5 & 4647 & 3093 & 4219 & 4310 & 3100 & 9.3 & 44.7 & 16.1 & 14.7 & 2443 \\
{\small janos-us} & 28 & 97 & 5 & 3891 & 1784 & 3205 & 3205 & 3057 & 3.7 & 26.1 & 9.8 & 3.4 & 7.8 \\
{\small newyork} & 18 & 58 & 5 & 4700 & 3106 & 4607 & 4470 & 3604 & 3.9 & 46.5 & 3.3 & 4.1 & 4.1 \\
{\small nobel-eu} & 30 & 55 & 5 & 4403 & 3367 & 4020 & 3690 & 3582 & 3.9 & 52.4 & 2.7 & 1.5 & 0.8 \\
{\small nobel-germany} & 19 & 39 & 5 & 4691 & 2612 & 4400 & 4400 & 3441 & 2.6 & 39.3 & 1.5 & 1.5 & 1.8 \\
{\small nobel-us} & 16 & 33 & 5 & 4828 & 3293 & 4726 & 4665 & 4621 & 0.8 & 50.9 & 0.5 & 0.3 & 1.3 \\
{\small norway} & 29 & 64 & 5 & 4100 & 2321 & 3817 & 3215 & 3028 & 4.2 & 37.6 & 7.7 & 1.9 & 5.6 \\
{\small pdh} & 13 & 41 & 5 & 4283 & 1422 & 4039 & 3645 & 2062 & 8.6 & 21.0 & 9.5 & 8.9 & 0.8 \\
{\small pioro40} & 42 & 106 & 5 & 4283 & 1422 & 4039 & 3645 & 2062 & 1.5 & 57.2 & 6.4 & 2.2 & 15.1 \\
{\small polska} & 14 & 26 & 3 & 2695 & 1908 & 2672 & 2460 & 1837 & 3.9 & 54.1 & 7.1 & 6.7 & 0.3 \\
{\small sun} & 29 & 115 & 5 & 4552 & 2918 & 4324 & 3930 & 3692 & 4.0 & 45.3 & 17.6 & 5.8 & 53.2 \\
{\small ta1} & 26 & 66 & 5 & 4552 & 2918 & 4324 & 3930 & 3692 & 10.6 & 33.9 & 14.5 & 11.7 & 80.7 \\
{\small ta2} & 67 & 138 & 5 & 38464 & 15776 & 25200 & 15120 & 14023 & 9.3 & 17.1 & 5.4 & 1.3 & 8.0 \\
{\small zib54} & 56 & 108 & 5 & 4786 & 4378 & 4671 & 4523 & 4483 & 1.3 & 70.3 & 3.5 & 2.9 & 11.7 \\
\specialrule{.1em}{.05em}{.05em} 
\end{tabular}
\end{center}
% \vspace{-0.1in}
\caption{Empirical results on SNDlib data set.}
%\vspace{-0.1in}
\label{table:sndlib}
\end{table*}

Table~\ref{table:sndlib} elaborates the comparison results on all the 26 instances of SNDlib data set. For each instance, we show the network details (i.e., the number of nodes and edges, and the adversary's budget), the amount of lost flow for all the five approaches and the runtime for our \emph{RAMF} approach. We also provide the percentage gains in objective value for our approach over four benchmarks. For all the instances, the amount of lost flow for the \emph{MF} approach is significantly higher than the \emph{RAMF} approach. The \emph{RAMF} approach always outperforms all the four benchmark approaches with respect to maximizing the administrator's objective value. On an average, our approach improves the objective value by 4.8\%, 43.7\%, 7.9\% and 4.9\% over \emph{MF}, \emph{OSP}, \emph{RF} and \emph{AAMF}, respectively. In addition, we observe that our approach is computationally attractive for these structured benchmark networks. The runtime for our approach is always bounded by 90 seconds except for two instances (i.e., `giul39' and `janos-us-ca') for which the edge capacities are generated randomly as they are stated as 0 in the SNDlib database.

\subsection{Empirical Results on Large-scale Data Sets}\label{largeResults}
In this section, we present empirical results on a set of synthetic and real-world large scale problem instances. For these large-scale problem instances, we employ heuristics from \cref{heuristics} to efficiently solve the adversary's decision problem. We elucidate the following key performance results:
\begin{enumerate}
\item Solution quality comparison between the two proposed heuristics.
\item Sensitivity results on synthetic networks by varying three tunable input parameters.
\item The runtime analysis of our \emph{RAMF} approach.
\item Performance analysis on benchmark data set from RMFGEN \citep{goldfarb1988computational} networks.
\end{enumerate}

\subsubsection{Performance Comparison between the Proposed Heuristic Approaches} 
We begin by showing the performance comparison between the accelerated greedy approach from \cref{greedyHeuristic} and the network partitioning based heuristic approach from \cref{partitionHeuristic}. We empirically observe that the partitioning based heuristic mostly performs better than the greedy approach, specially at the later stage of the game. However, as both the approaches provide sub-optimal solutions, the greedy approach outperforms the partitioning based heuristic in some cases. To illustrate this behavior, we show the number of lost flows, overall objective value and runtime for both heuristics in each iteration of the game on a problem instance with 50 nodes, 0.4 edge density and the adversary's budget as 10. As the adversary seeks to minimize the administrator's objective value, a better quality solution should provide lower objective value and higher number of lost flows. As shown in Table~\ref{table:GrdPart}, while the partitioning based heuristic mostly outperforms the greedy approach, the solution quality of the greedy approach is better for a few iterations (e.g., 1, 6 and 7). As the partitioning based heuristic does not always dominate the greedy approach, we solve the adversary's decision problem using both heuristics in each iteration of the game and choose the attack with better solution quality. In terms of the computational complexity, as expected, the greedy approach is proven to be much faster than the network partitioning based heuristic. 

\begin{table*}[!htb]
\begin{center}
{\small
\begin{tabular}{>{\centering}m{1.3cm} >{\centering}m{1.25cm} >{\centering}m{1.8cm} >{\centering}m{1.25cm} >{\centering}m{1.8cm} >{\centering}m{1.7cm} >{\centering}m{1.25cm} c}
\hline
\multicolumn{1}{>{\centering}m{1.3cm}|}{\text{Iteration}} & \multicolumn{2}{>{\centering}m{3.05cm}|}{Lost Flow} & \multicolumn{3}{>{\centering}m{5.35cm}|}{Objective Value} & \multicolumn{2}{>{\centering}m{3.35cm}}{Runtime (Sec)}\\
\hline
\multicolumn{1}{>{\centering}m{1.3cm}|}{}& \text{Greedy} & \multicolumn{1}{>{\centering}m{1.8cm}|}{\text{Partitioning}} & {\text{Greedy}} & \text{Partitioning} & \multicolumn{1}{>{\centering}m{1.7cm}|}{\text{Difference}} & \text{Greedy} & \text{Partitioning} \\
\specialrule{.1em}{.05em}{.05em} 
1 & 466 & 447 & 1171.99 & 1190.91 & \textbf{-18.92} & 2.05 & 106.68 \\
2 & 143 & 213 & 1071.76 & 1011.77 & 59.99 & 29.92 & 224.19 \\
3 & 124 & 185 & 1096.4 & 1040.24 & 56.16 & 31.24 & 132.21 \\
4 & 135 & 185 & 1118.11 & 1077.06 & 41.05 & 30.86 & 183.59 \\
5 & 178 & 197 & 1117.71 & 1104.38 & 13.33 & 24.18 & 265.3 \\
6 & 168 & 151 & 1157.89 & 1181.58 & \textbf{-23.69} & 25.66 & 684.01 \\
7 & 187 & 182 & 1152.21 & 1158.04 & \textbf{-5.83} & 26.14 & 505.1 \\
8 & 135 & 165 & 1206.67 & 1187.68 & 18.99 & 31.19 & 278.07 \\
9 & 144 & 173 & 1201.79 & 1179.36 & 22.43 & 32.65 & 240.05 \\
10 & 127 & 138 & 1214.84 & 1211.69 & 3.15 & 32.49 & 822.93 \\
11 & 127 & 135 & 1214.81 & 1214.27 & 0.54 & 32.53 & 523.11 \\
\specialrule{.1em}{.05em}{.05em} 
\end{tabular}}
\end{center}
\caption{Solution quality and runtime comparison between two proposed heuristic approaches.}
\label{table:GrdPart}
\end{table*}

\subsubsection{Sensitivity Results with Different Settings of Input Parameters} 
We now present empirical results on a set of large scale synthetic problem instances. We use the same setting used in \cref{smallResultsSensitive} to generate a set of large directed connected synthetic networks. For all the networks, we randomly draw the capacities for the edges from the range of 1 to 50. The unit costs for routing flows through edges (i.e., $p$) are drawn randomly from the range of 0.01 to 0.1. 
In the default setting of experiments, we use networks with 50 nodes and 0.4 edge density, and the adversary's budget value is set to 10. 

\begin{figure}[!htb]
	\centering
	\begin{subfigure}{0.325\textwidth}
		\includegraphics[width=\textwidth]{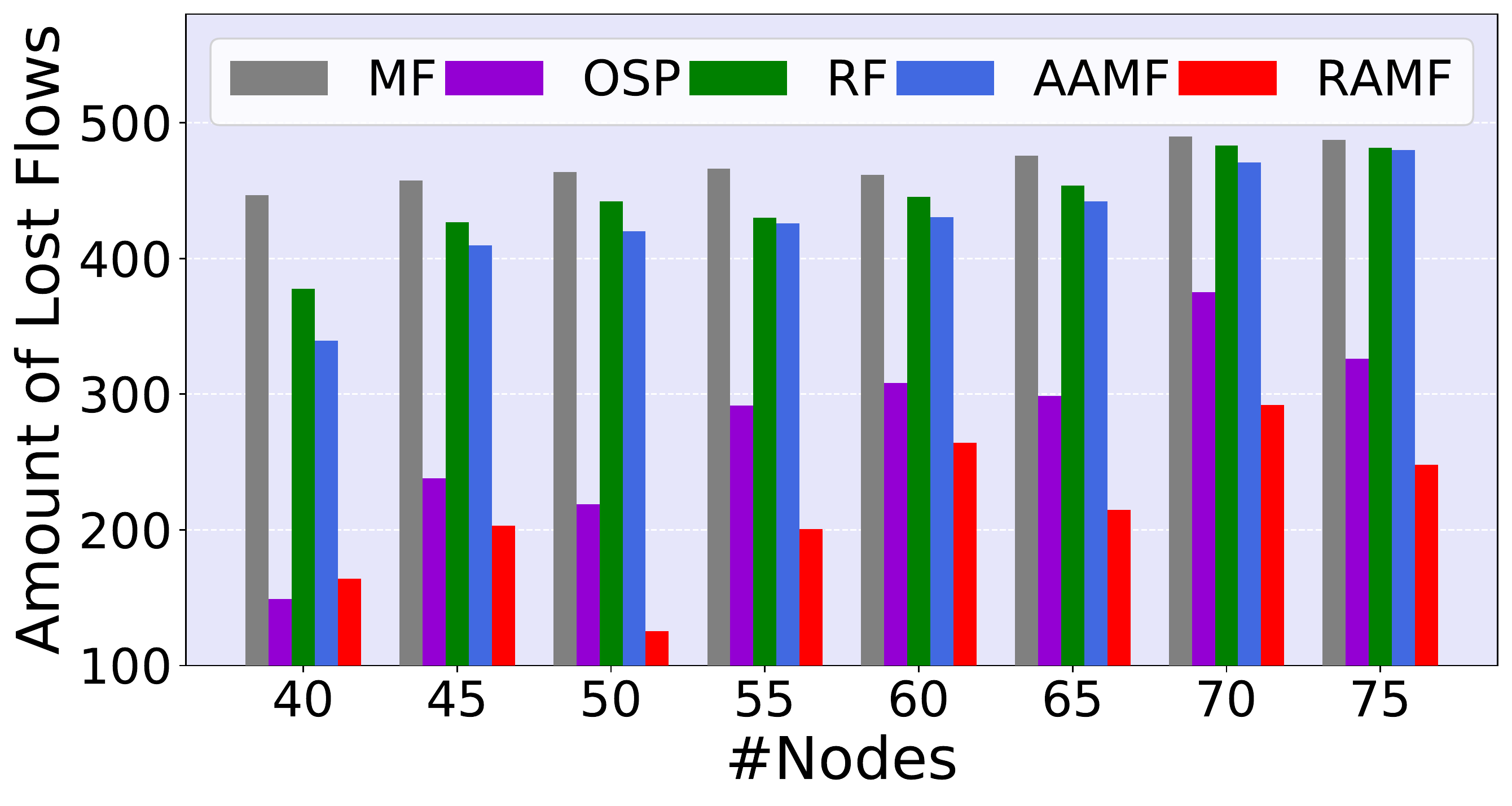} \caption{}
	\end{subfigure} 
	\begin{subfigure}{0.325\textwidth}
		\includegraphics[width=\textwidth]{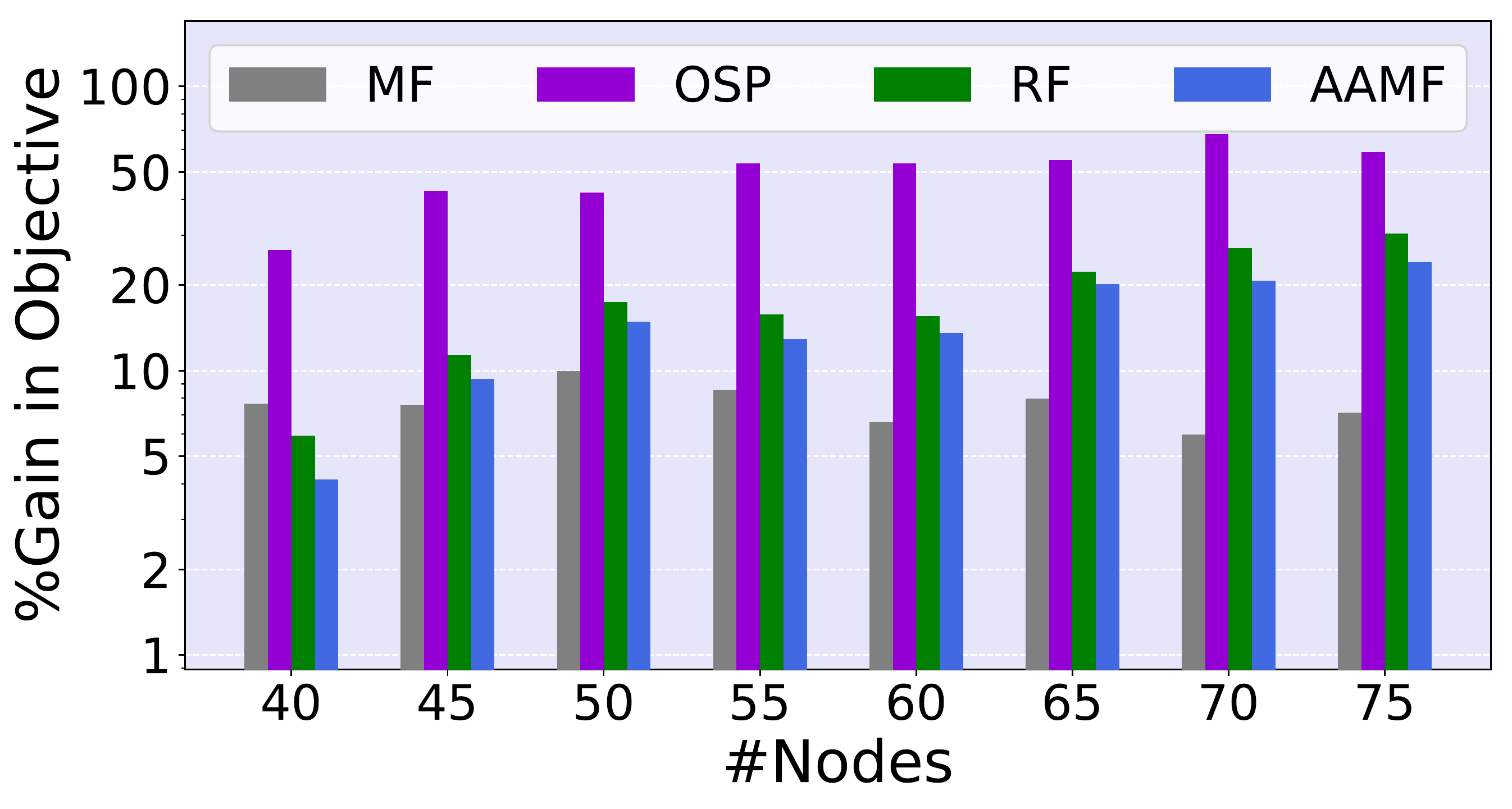} \caption{}
	\end{subfigure} 
	\begin{subfigure}{0.325\textwidth}
		\includegraphics[width=\textwidth]{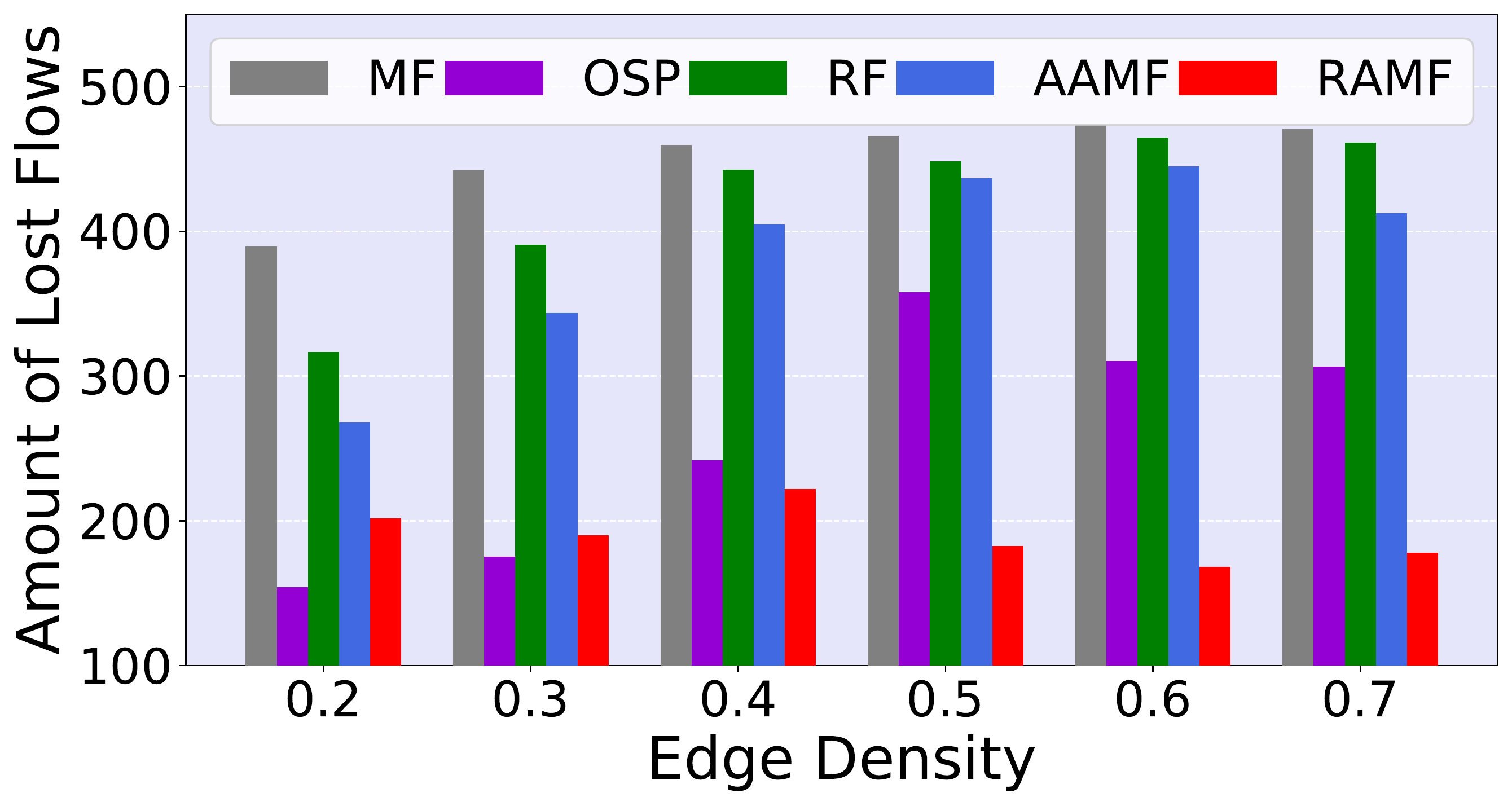} \caption{}
	\end{subfigure} 
	\begin{subfigure}{0.325\textwidth}
		\includegraphics[width=\textwidth]{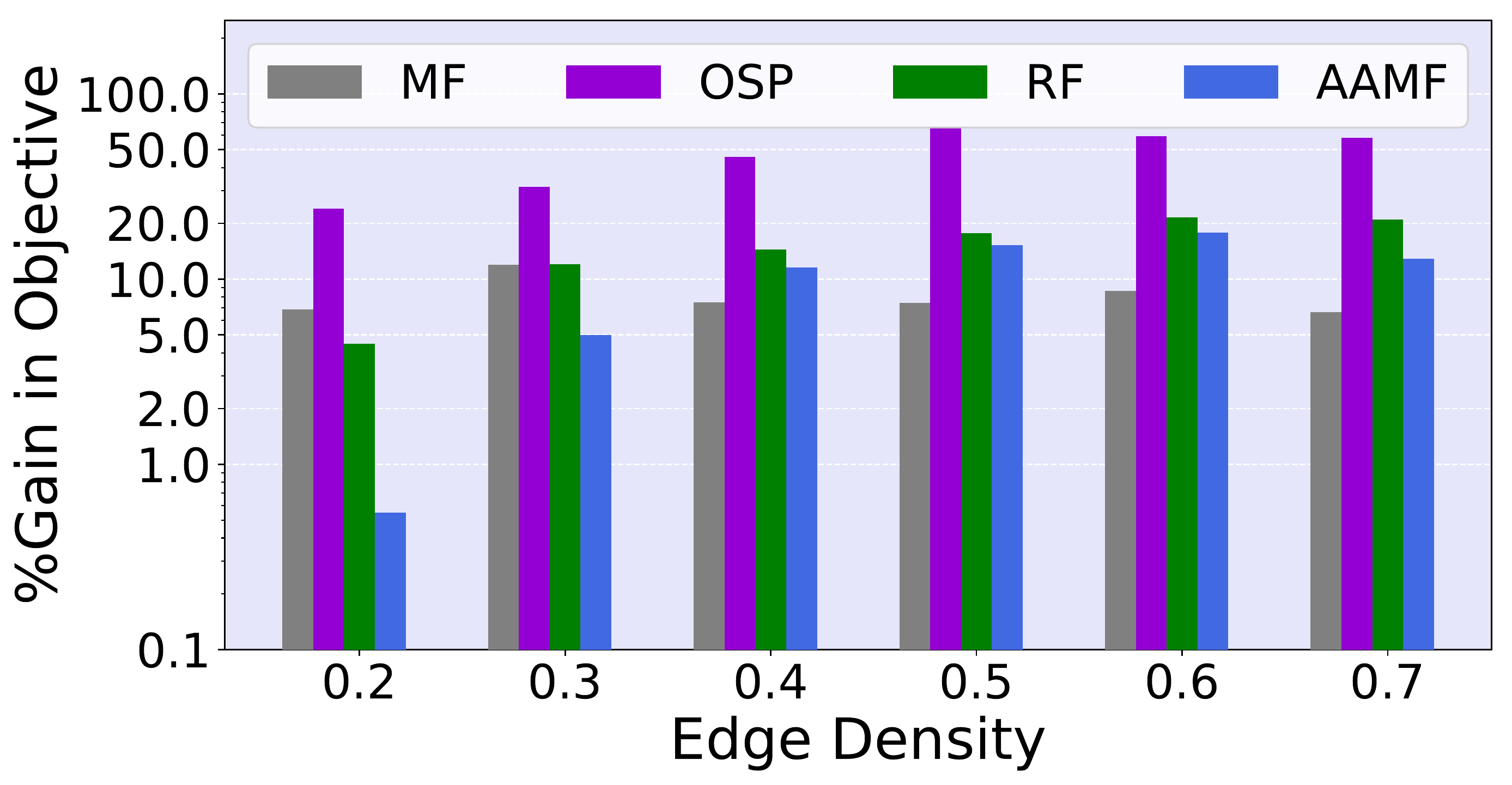} \caption{}
	\end{subfigure} 
	\begin{subfigure}{0.325\textwidth}
		\includegraphics[width=\textwidth]{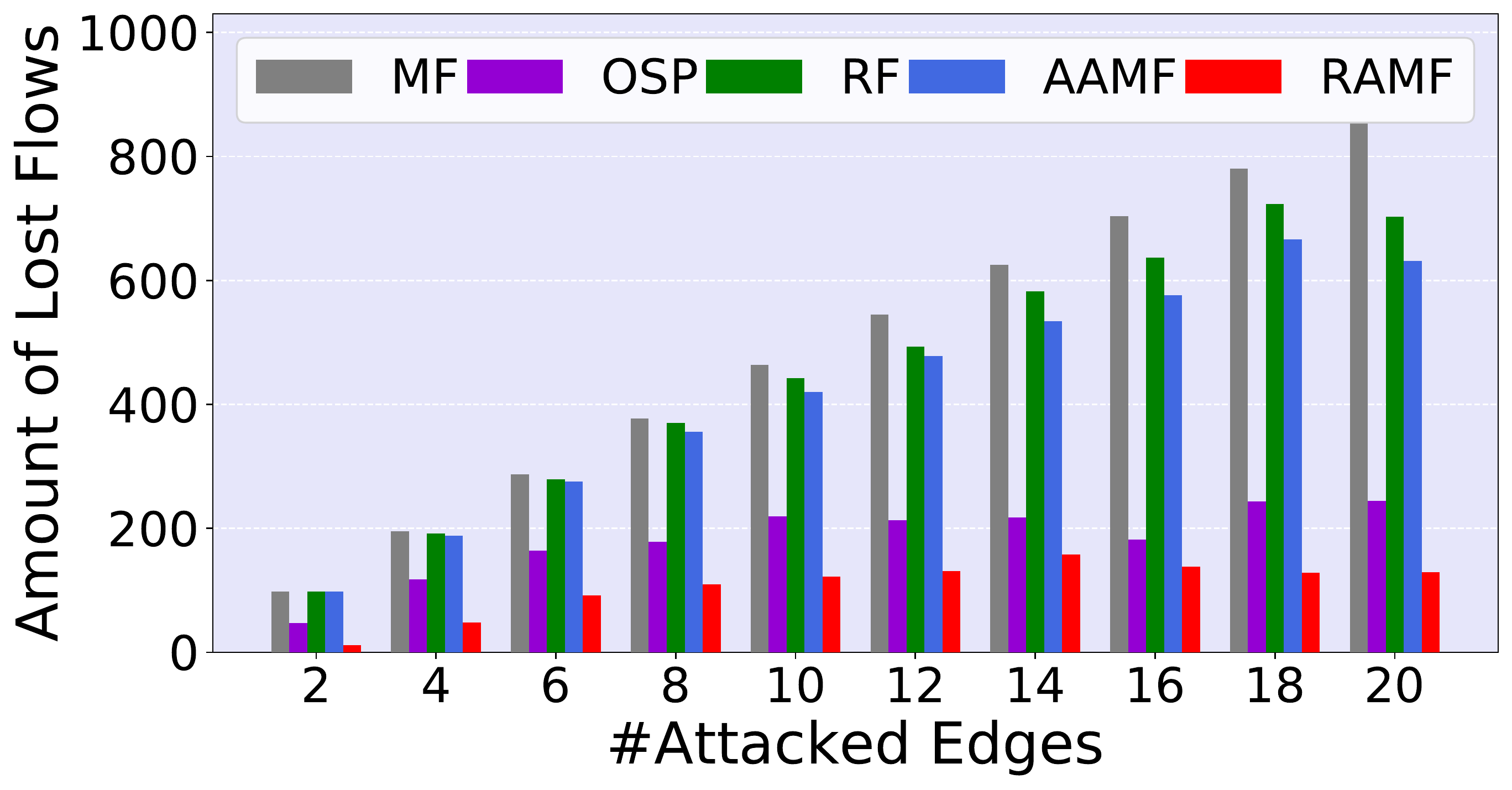} \caption{}
	\end{subfigure} 
	\begin{subfigure}{0.325\textwidth}
		\includegraphics[width=\textwidth]{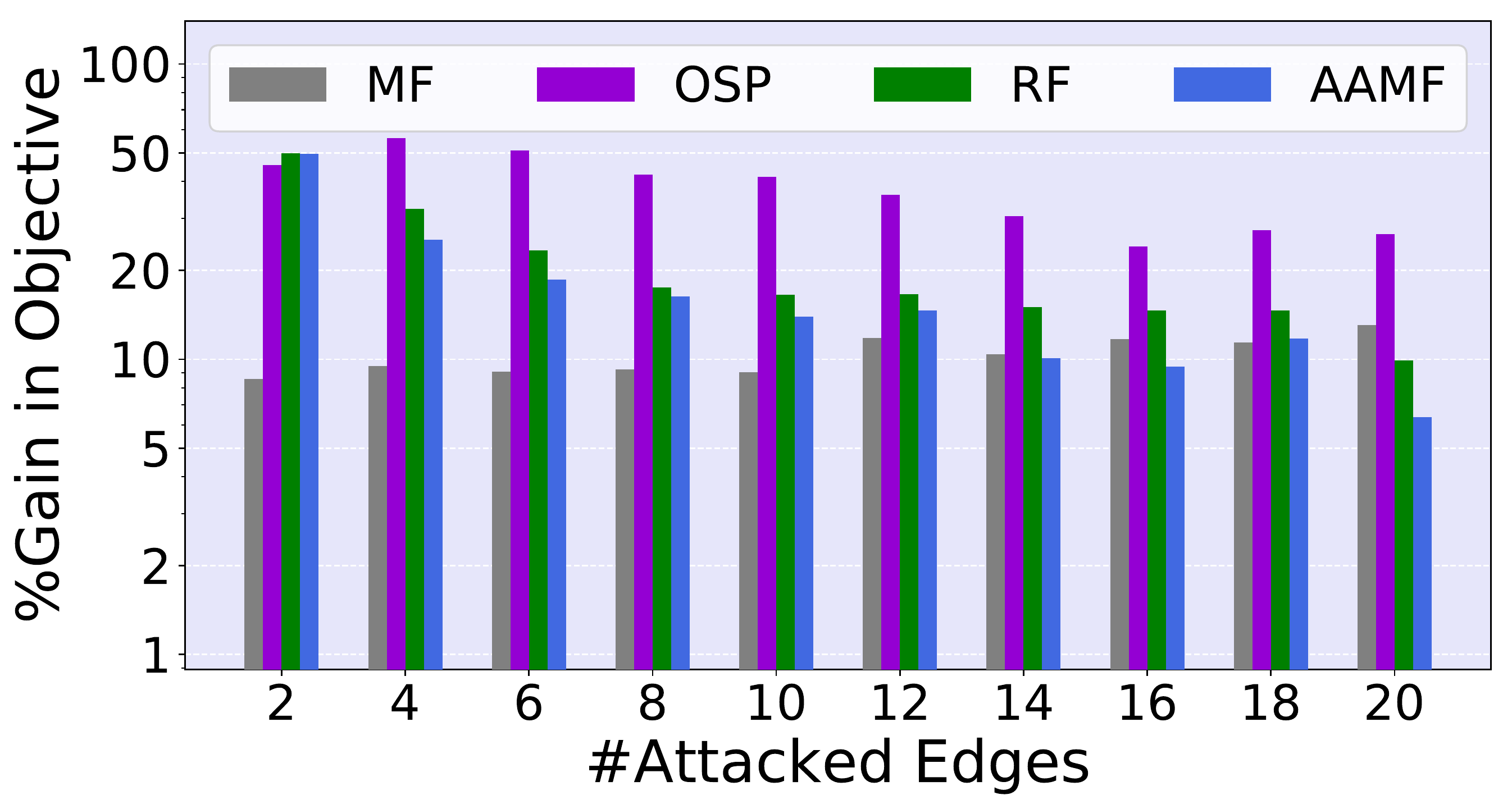} \caption{}
	\end{subfigure} 
	 \caption{Effect of number of nodes on (a) Lost flows and (b) Objective value; Effect of number of edges on (c) Lost flows and (d) Objective value; Effect of the value of $\Gamma$ on (e) Lost flows and (f) Objective value.}
	 \label{fig:largeResults}
	 \vspace{0.1in}
\end{figure}

Figure~\ref{fig:largeResults}(a) demonstrates the net amount of lost flow for our \emph{RAMF} approach and the four benchmark approaches, where we vary the number of nodes from 40 to 75 in the X-axis. The amount of lost flow for the \emph{MF} approach is significantly high in all the settings. The \emph{RF} and \emph{AAMF} approaches perform equally poorly in reducing the number of lost flow. Except for the relatively small problem instances with 40 nodes, our \emph{RAMF} approach always outperforms all the four benchmark approaches in terms of reducing the number of lost flow.
Figure~\ref{fig:largeResults}(b) delineates the percentage gain in the objective value for our \emph{RAMF} approach against the four benchmarks in a logarithmic scale. As clearly shown, the percentage gains in objective for our approach over all the benchmarks are always positive. The average percentage gains in the objective value for our approach over the \emph{MF}, \emph{OSP}, \emph{RF} and \emph{AAMF} approaches are 7.7\%, 50.1\%, 18.2\%  and 14.97\%, respectively. 

Figure~\ref{fig:largeResults}(c) shows the net amount of lost flows for all the five approaches, where we vary the edge density from 0.2 to 0.7 in the X-axis. The lost flows for the \emph{MF}, \emph{RF}, \emph{AAMF} approaches are always significantly high. Except for edge density 0.2 and 0.3, our \emph{RAMF} approach significantly outperforms the \emph{OSP} approach in terms of reducing the lost flows. 
Figure~\ref{fig:largeResults}(d) delineates that the \emph{AAMF} approach always provides a better quality solution over the \emph{RF} approach. The \emph{MF} approach outperforms the \emph{RF} and \emph{AAMF} approaches on larger problem instances with edge density 0.4 and beyond. As expected, the \emph{OSP} approach performs poorly in maximizing the objective value in all the cases. On an average, the percentage gains in objective for our \emph{RAMF} approach over the \emph{MF}, \emph{OSP}, \emph{RF} and \emph{AAMF} approaches are 8.2\%, 47.3\%, 15.2\% and 10.5\%, respectively. 

Figure~\ref{fig:largeResults}(e) exhibits the net amount of lost flow for all the approaches, where we vary the budget of the adversary from 2 to 20 in the X-axis. The amount of lost flow for \emph{MF} approach increases monotonically from 100 to almost 900 as we increase the adversary's budget, whereas the number of lost flow for our \emph{RAMF} approach is always lower than all the benchmarks and is always bounded by 160. Moreover, we observe that the number of lost flow for our \emph{RAMF} approach remains steady when the adversary's budget value goes beyond 14. 
Although Figure~\ref{fig:largeResults}(f) demonstrates that the gains in objective for our approach over the benchmark approaches almost always reduce monotonically with the increasing value of $\Gamma$, the net difference between the objective values increases monotonically with the value of $\Gamma$, which indicates that the performance of our approach improves gradually if the adversary becomes stronger. On an average, the percentage gains in the objective of our \emph{RAMF} approach over the \emph{MF}, \emph{OSP}, \emph{RF} and \emph{AAMF} approaches are 10.4\%, 38.1\%, 21.1\% and 17.6\%, respectively.

In a nutshell, we observe a consistent pattern that our \emph{RAMF} approach reduces the lost flow significantly over \emph{MF}, \emph{RF} and \emph{AAMF} approaches. On the other hand, the percentage gain in the objective value for the \emph{RAMF} approach over the \emph{OSP} approach is always significantly high. Therefore, we can conclude that among five approaches, only our \emph{RAMF} approach is able to maintain the right trade-off between the two performance metrics. Moreover, the results on the large scale problem instances replicate the similar trend observed for the optimal results on the small problem instances presented in \cref{smallResults}.

\subsubsection{Runtime Performance} 
We now demonstrate the runtime performance of our \emph{RAMF} approach for different settings of network size and adversary's budget value. As the number of iterations required to converge the game may vary randomly for different problems, we show the average runtime for one iteration of the game.
\begin{wrapfigure}{r}{0.5\linewidth}
	\centering
	\includegraphics[width=0.5\textwidth]{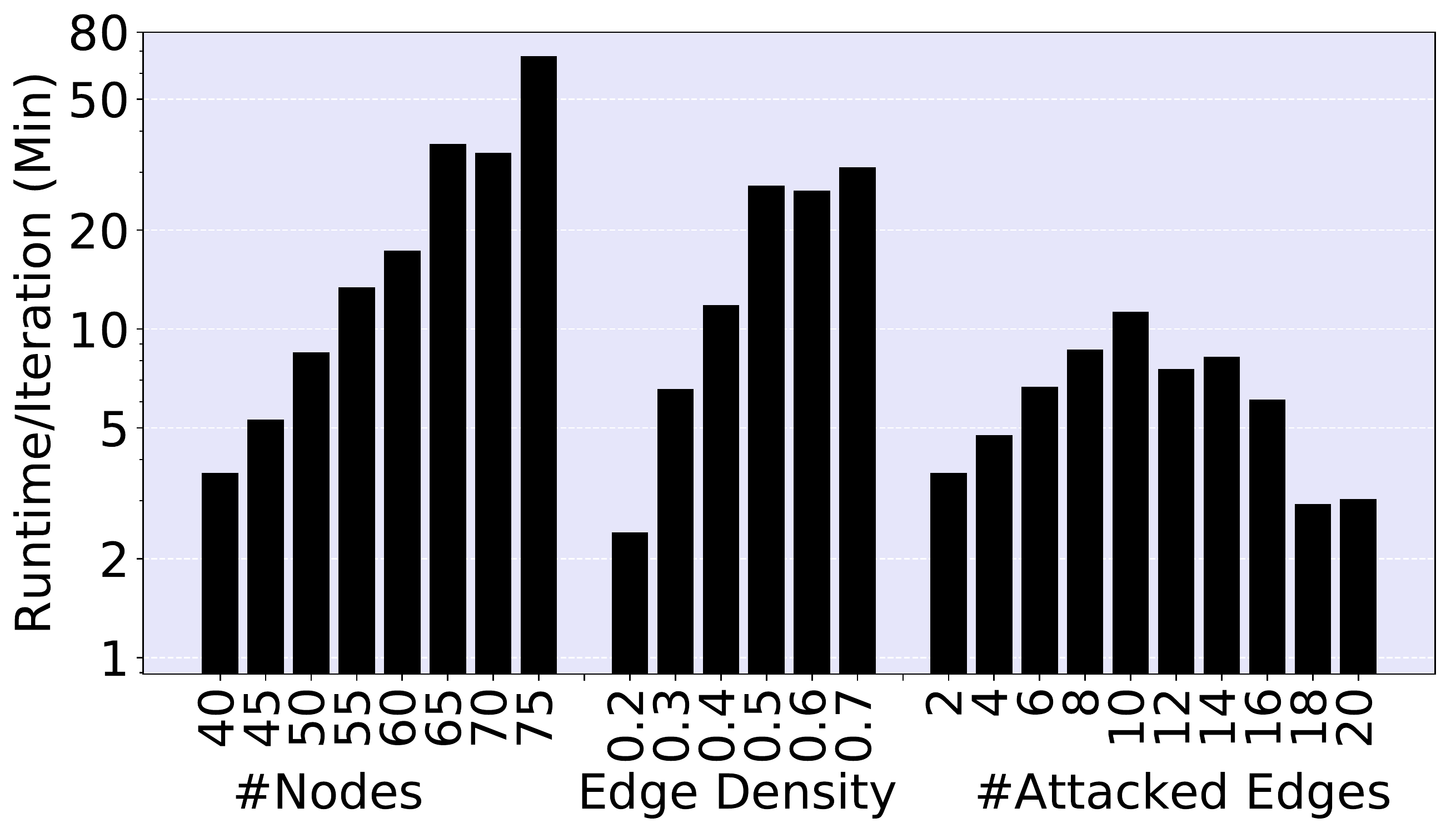} 
	 \caption{Runtime results for large problem instances.}
	 \label{fig:largeRuntime}
\end{wrapfigure}
Figure~\ref{fig:largeRuntime} presents the runtime for the \emph{RAMF} approach in minutes in a logarithmic scale. We observe that the runtime almost always increases monotonically with the network size (both in terms of the number of nodes and edges). The runtime increases monotonically with the adversary's budget until $\Gamma=10$. However, as the sub-problems of the network partitioning based heuristic approach have relatively small number of edges and the optimization problem tries to identify $\frac{\Gamma}{2}$ potential edges from a small set of edges, the combinatorial space and the complexity of the sub-problems reduces once the budget value becomes large. Therefore, the runtime starts to decrease as the value of $\Gamma$ goes beyond 10.

\subsubsection{Performance Analysis on RMFGEN Data Set}
In the last thread of results, we provide the performance comparison between different approaches on instances from RMFGEN networks \citep{goldfarb1988computational}. RMFGEN networks are widely used for validating large scale network flow solutions. We employ 7 moderately large RMFGEN networks for the experiment\footnote{The data set is collected from http://elib.zib.de/pub/mp-testdata/maxflow/index.html}. As some of the instances are undirected networks, we convert them into directed networks by randomly adding edges until it evolves into a connected (i.e., every node has at least one incoming and one outgoing edge) network. As the capacities of edges, $U$ are mentioned as continuous value in the database, we generate the capacity values randomly from the range of 10 to 50. The unit routing costs for the edges, $p$ are randomly drawn from the range of 0.01 to 0.1. Finally, we set the adversary's budget value to 10 for all the problem instances.

\begin{table*}[!htb]
\footnotesize
\begin{center}
\begin{tabular}{>{}m{1.1cm} >{\centering}m{0.3cm} >{\centering}m{0.4cm} >{\centering}m{0.25cm} >{\centering}m{0.4cm} >{\centering}m{0.4cm} >{\centering}m{0.4cm} >{\centering}m{0.65cm} >{\centering}m{0.8cm} >{\centering}m{0.7cm} >{\centering}m{0.7cm} >{\centering}m{0.7cm} >{\centering}m{0.7cm} >{\centering}m{0.8cm} c}
\hline
\multicolumn{4}{>{\centering}m{3.25cm}|}{Problem Instance} & \multicolumn{5}{>{}m{3.5cm}|}{Amount of Lost Flows} & \multicolumn{5}{>{\centering}m{5cm}|}{Objective Value} & \multicolumn{1}{>{\centering}m{1.2cm}}{Runtime/ Iteration} \\
\hline
{\footnotesize Name} & $|{\mathcal V}|$ & $|{\mathcal E}|$ & \multicolumn{1}{>{\centering}m{0.25cm}}{$\Gamma$} & {\footnotesize MF} & {\footnotesize OSP} & {\footnotesize RF} & { AAMF} & \multicolumn{1}{>{\centering}m{0.8cm}}{{ RAMF}} & {\footnotesize MF} & {\footnotesize OSP} & {\footnotesize RF} & \multicolumn{1}{>{\centering}m{0.8cm}}{{\footnotesize AAMF}} &  \multicolumn{1}{>{\centering}m{0.8cm}}{{\footnotesize RAMF}} & {\footnotesize RAMF} \\
\specialrule{.1em}{.05em}{.05em} 
{ elist96} & 96 & 348 & 10 & 357 & 169 & 276 & 269 & 145 & 155.5 & 61.2 & 170.9 & 163.8 & 216.7 & 18.30\\
{ elist96d} & 96 & 539 & 10 & 439 & 343 & 378 & 376 & 213 & 452.9 & 307.0 & 323.6 & 365.4 & 508.2 & 11.26 \\
{ elist160} & 160 & 586 & 10 & 438 & 266 & 370 & 370 & 310 & 408.7 & 255.7 & 345.3 & 345.4 & 449.1 & 28.07\\
{ elist160d} & 160 & 933 & 10 & 481 & 377 & 456 & 454 & 345 & 1602.1 & 1380.1 & 1356.1 & 1467.8 & 1647.7 & 54.14\\
{ elist200} & 200 & 818 & 10 & 463 & 344 & 408 & 405 & 352 & 969.3 & 737.2 & 840.8 & 904.8 & 1010.9 & 25.37 \\
{ elist200d} & 200 & 1370 & 10 & 476 & 413 & 458 & 458 & 452 & 2255.4 & 1938.3 & 1823.5 & 1750.6 & 2272.8 & 35.64 \\ 
{ elist500} & 500 & 2042 & 10 & 477 & 422 & 453 & 440 & 430 & 3106.8 & 2799.6 & 2694.4 & 2848.2 & 3126.1 & 56.11  \\
\specialrule{.1em}{.05em}{.05em} 
\end{tabular}
\end{center}
%\vspace{-0.1in}
\caption{Empirical results on RMFGEN data set.}
%\vspace{-0.1in}
\label{table:rmfgen}
\end{table*}

Table~\ref{table:rmfgen} elaborates the performance comparison results on the instances of RMFGEN networks. For each instance, we show the network details (i.e., the number of nodes and edges, and the adversary's budget value), the amount of lost flow and the objective values for all the five approaches and the average runtime (in minutes) per iteration for the \emph{RAMF} approach. The net amount of lost flows for \emph{MF}, \emph{RF} and \emph{AAMF} approaches is significantly higher than our \emph{RAMF} approach for all the instances. For all the problem instances, the \emph{RAMF} approach also provides higher objective value over all the benchmark approaches. On an average, our approach improves the objective value by 8\%, 50.1\%, 47.9\%, 39.6\% over \emph{MF}, \emph{OSP}, \emph{RF} and \emph{AAMF} approaches, respectively. Most importantly, we observe that our proposed heuristic approaches can scale gracefully to solve these large problem instances while providing a significant performance gain over the benchmark approaches.

\section{Concluding Remarks} \label{conclusion}
To evaluate the resilience and sustainability of modern critical infrastructure networks, we propose a robust and adaptive network flow model by assuming that the network parameters are deterministic but the network structure (e.g., edges) is vulnerable to adversarial attacks or failures. To compute a robust and adaptive network flow strategy, we introduce a novel scenario generation approach based on a two-player iterative game between the network administrator and an adversary. In each iteration of the game, the adversary identifies an optimal attack to disrupt the flow strategy generated by the administrator in the current iteration and the administrator computes a robust flow strategy by considering a set of attacks revealed by the adversary in previous iterations. As the computational complexity of the adversary's decision problem increases significantly with network size, we propose two novel heuristics, one leverages an accelerated greedy approach and other employs a network partitioning based optimization approach, to speed up the solution process. The empirical results on multiple synthetic and real-world benchmark data sets demonstrate that our proposed approach scales gracefully to the large-scale problem instances and improves the operational efficiency of the network by reducing the expected lost flow. 

In future, this work can be extended in the following two directions: (a) Develop faster heuristics for solving the decision problem of both the adversary and the administrator, so as to scale up the solution process to massive real-world urban networks consisting of tens of thousands edges; and (b) Incorporate precise domain constraints in the optimization models of both the players to cater to specific real-world application domain. For example, in the context of urban transportation, a detailed traffic model with congestion effects would make the model more realistic. However, incorporating precise traffic details (e.g., representing the routing cost as a latency function of traffic to capture the congestion effects) would make our solution approach computationally intractable due to non-linearity in objective function and therefore, efficient approximation methods need to be designed by analyzing the properties of the specific problem domain.

\section*{Acknowledgements}
This work was partially supported by the Singapore National Research Foundation through the Singapore-MIT Alliance for Research and Technology (SMART) Centre for Future Urban Mobility (FM) and National Research Foundation, Prime Ministers Office, Singapore under its International Research Centres in Singapore Funding Initiative.

%\section*{Appendix}

\appendix

\section{Robust Flow Solution} \label{sec:appendixA}
In this section, we provide the details of the robust flow (RF) solution \citep{bertsimas2013robust} that is used as a benchmark approach. 
For this benchmark approach, our goal is to proactively compute a robust flow solution by assuming that the entire flow of an attacked edge is lost. For a fair comparison with other approaches, we modified the robust flow solution proposed by \cite{bertsimas2013robust} to ensure that a maximum of $\Gamma$ edges in the network can be attacked. Let ${x}$ denote the resulting robust flow scenario and ${\cal L}_{\Gamma}$ denote the worst-case lost flow value if the adversary is restricted to attack a maximum of $\Gamma$ edges. The optimization model \eqref{eq:robustflow} compactly delineates the details of our robust flow solution, where the inner optimization model computes the value of ${\cal L}_{\Gamma}$ as the sum of the first $\Gamma$ biggest edge flow values.

\begin{equation}
	\label{eq:robustflow}
	\begin{aligned}
		\max_{{x}} \hspace{0.1in} & x_{(t,s)} - {\cal L}_{\Gamma} &  \hspace{0.05in} \textbf{where,  } &\hspace{0.2in}  {\cal L}_{\Gamma} =  & \max_{{\mu}} & \sum_e x_e \mu_e \hspace{0.2in}  & \\
		\text{s.t.} \hspace{0.1in} & \sum_{e\in \delta^+_v} x_e - \sum_{e\in \delta^-_v} x_{e} = 0, & \hspace{0.05in}  \forall v \in {\mathcal V}\setminus \{s\} & &\hspace{0.05in} \text{s.t.} & \sum_{e}  \mu_e \leq \Gamma \hspace{0.1in} &  \\
		 & 0 \leq x_e \leq U_e, & \forall e \in {\mathcal E} & & & 0 \leq \mu_e \leq 1, & \hspace{0.05in} \forall e \in {\mathcal E}
	\end{aligned}
\end{equation}

The two components of problem \eqref{eq:robustflow} can be combined by taking the dual of the inner problem ${\cal L}_{\Gamma}$. To achieve this goal, we first compute the \emph{Lagrangian} function \eqref{eq:robustflowLagrange} by introducing the price variables $\theta$ and $\zeta$. From this \emph{Lagrangian} function, we construct the dual problem \eqref{eq:robustflowDual}.
\begin{equation}
	\label{eq:robustflowLagrange}
	\begin{aligned}
		\min_{{\zeta, \theta}} -\sum_e x_e \mu_e + \zeta (\sum_e \mu_e - \Gamma) +\sum_e \theta_e (\mu_e-1)
	\end{aligned}
\end{equation}

\begin{equation}
	\label{eq:robustflowDual}
	\begin{aligned}
		\max_{{\zeta, \theta}} \hspace{0.1in} & - \sum_e \theta_e -\zeta \Gamma & \\
		\text{s.t.} \hspace{0.1in}  & \theta_e +\zeta \geq x_e, & \forall e \in {\cal E} \\
		& \theta_e \geq 0, \zeta \geq 0 & 
	\end{aligned}
\end{equation}

Putting the optimization models \eqref{eq:robustflow} and \eqref{eq:robustflowDual} together, we construct the linear optimization model \eqref{eq:robustflowFinal} that is used to solve the robust maximum flow problem.
\begin{equation}
	\label{eq:robustflowFinal}
	\begin{aligned}
		\max_{{x,\theta,\zeta}} \hspace{0.1in} & x_{(t,s)} - \sum_e \theta_e -\zeta \Gamma & \\
		\text{s.t.} \hspace{0.1in} & \sum_{e\in \delta^+_v} x_e - \sum_{e\in \delta^-_v} x_{e} = 0, & \hspace{0.05in}  \forall v \in {\mathcal V}\setminus \{s\} \\
		& \theta_e +\zeta \geq x_e, & \forall e \in {\cal E} \\
		& 0 \leq x_e \leq U_e, & \forall e \in {\cal E} \\
		& \theta_e \geq 0, \zeta \geq 0 & 
	\end{aligned}
\end{equation}

\section{Approximate Adaptive Maximum Flow Solution} \label{sec:appendixB}
In this section, we provide the details of another benchmark approach that computes an approximate adaptive maximum flow (AAMF) solution by assuming that the flow can be adjusted after the edge failure occurred. \cite{bertsimas2013robust} show that the adaptive maximum flow problem is strongly NP-hard. Therefore, they propose a scalable linear optimization model to approximately solve the problem.
Let ${x}$ denote the resulting approximate adaptive maximum flow solution and $\theta$ denote the largest edge flow value. In addition, let us define an $s-t$ cut for the network as a subset $S \in {\cal V}$ of nodes with $s\in S$ and $t\in V\setminus S$. We say that a node $v$ is on the $s$-side if $v\in S$ and on the $t$-side if $v\in {\cal V}\setminus S$. Let $\delta^+(S)$ denote the set of outgoing edges $e=(v,w)$ from the $S$ side such that $v\in S$ and $w\in {\cal V}\setminus S$ and $\delta^-(S)$ represent the set of incoming edges $e=(v,w)$ to the $S$ side with $v\in {\cal V}\setminus S$ and $w\in S$. Then, the capacity of the $s-t$ cut is defined as: $Cap(S) = \sum_{e\in \delta^+(S)} U_e$. Let us assume that $S$ denotes the $s-t$ cut with minimum capacity value for the network of interest (i.e., a min-cut solution). The linear optimization model \eqref{eq:adaptiveMaxFlow} provides details of the approximate solution proposed in \cite{bertsimas2013robust}.

\begin{equation}
	\label{eq:adaptiveMaxFlow}
	\begin{aligned}
		\max_{{x,\theta}} \hspace{0.1in} & \sum_{e\in \delta^+(S)} x_e - \sum_{e\in \delta^-(S)} x_e -\theta \Gamma & \\
		\text{s.t.} \hspace{0.1in} & \sum_{e\in \delta^+_v} x_e - \sum_{e\in \delta^-_v} x_{e} = 0, & \hspace{0.05in}  \forall v \in {\mathcal V}\setminus \{s,t\} \\
		& x_e \leq \theta, & \forall e \in {\cal E} \\
		& 0 \leq x_e \leq U_e, & \forall e \in {\cal E}
	\end{aligned}
\end{equation}

Let $x^*$ be a flow with maximum robust flow value, such that $\beta \text{Val}(x^*) \leq \text{RVal}(x^*)$ for some $\beta \in (0,1]$, where $\text{Val}(x^*)$ represents the objective value for the administrator if no attack is executed in the network, and $\text{RVal}(x^*)$ denotes the robust flow value of $x^*$. Further, suppose $\bar{x}, \bar{\theta}$ be the optimal solution for the optimization problem \eqref{eq:adaptiveMaxFlow}. Let $\rho$ denote the value of $(1-(1-\frac{1}{n})^n)$, where $n$ represents the number of nodes in the network. Then, \cite{bertsimas2013robust} show that $\bar{x}$ is a $1- (((1-\rho)/\rho) ((1-\beta)/\beta))$-approximation of $x^*$, i.e., 
$$ \text{RVal}(\bar{x}) \geq \Big (1- \frac{1-\rho}{\rho} \frac{1-\beta}{\beta} \Big ) \text{RVal}(x^*)$$
In addition, \cite{bertsimas2013robust} show that the resulting flow from problem \eqref{eq:adaptiveMaxFlow} provides an $\alpha$-approximation to the optimal adaptive maximum flow solution. Let $x^*$ be an adaptive maximum flow such that $\beta \text{Val}(x^*) \leq \text{RVal}(x^*)$ for some $\beta \in (0,1]$, and $\bar{x}, \bar{\theta}$ be the optimal solution for the optimization problem \eqref{eq:adaptiveMaxFlow}. Then, adaptive value of $\bar{x}$, $\text{AVal}(\bar{x})$ yields a $\beta(1- (((1-\rho)/\rho) ((1-\beta)/\beta)))$-approximation for the adaptive value of $x^*$, i.e.,
$$ \text{AVal}(\bar{x}) \geq \beta\Big (1- \frac{1-\rho}{\rho} \frac{1-\beta}{\beta} \Big ) \text{AVal}(x^*)$$

\bibliographystyle{ormsv080} 
\bibliography{referencelog}

\end{document}